\pdfoutput=1
\documentclass[PAPER, atlasdraft=false, UKenglish, cernpreprint, texmf, orcidlogo]{atlasdoc}

\usepackage[subcaption=true]{atlaspackage}
\usepackage{atlasbiblatex}

\usepackage{atlasphysics}
\graphicspath{{logos/}{figures/}}

\usepackage{ANA-HMBS-2024-65-PAPER-defs}
\usepackage{atlasbsm}
\usepackage{multirow}
\usepackage{adjustbox}
\usepackage{comment}
\usepackage{cancel}

\AtlasTitle{Search for higgsinos in compressed mass spectra using low-momentum tracks in $pp$ collisions at $\sqrt{s}=13$ TeV with the ATLAS detector}

\AtlasAbstract{%
This paper presents two searches for the electroweak production of higgsinos with compressed mass spectra using 140~fb$^{-1}$ of $\sqrt{s}=13$ TeV proton--proton
collision data collected by the ATLAS experiment at the Large Hadron Collider.
Events are required to feature an energetic jet, large missing transverse momentum, and at least one low-momentum charged particle that serves as a candidate
higgsino decay product.
In the first search, targeting higgsino mass splittings in the range of 0.3--1~GeV, the higgsinos are expected to predominantly decay into pions that are identified as low-momentum
charged particles with large transverse impact parameters due to the long higgsino lifetime ($c\tau\approx\mathcal{O}(0.1\text{--}10~\text{mm})$), and neural networks are used to discriminate between signal and background processes.
The second search targets larger mass splittings in the range of 1--3~GeV, where the higgsinos are expected to decay promptly into low-momentum leptons, one of
which is identified by dedicated low-momentum electron or muon taggers based on neural networks utilising tracking and calorimeter information.
No significant excess above the Standard Model prediction is observed in either search and the results are interpreted within simplified models, to set lower limits on the masses of the higgsino-like charginos and neutralinos.
Together, these searches exclude chargino masses below 126 GeV at 95\% confidence level for mass splittings between the chargino and lightest neutralino in the range of 0.3--2~GeV. This represents the first ATLAS constraints in a portion of this parameter space and surpasses the limits previously set by other experiments.
}

\AtlasRefCode{HMBS-2024-65}

\PreprintIdNumber{CERN-EP-2025-246}

\AtlasJournal{JHEP}
\AtlasJournalRef{JHEP 06 (2026) 094}
\AtlasDOI{10.1007/JHEP06(2026)094}

\AtlasCoverEgroupAnalysisTeam{atlas-hmbs-2024-65-analysis-team@cern.ch}

\hypersetup{pdftitle={ATLAS document},pdfauthor={The ATLAS Collaboration}}

\begin{document}

\maketitle
\section{Introduction}
%

Supersymmetry (SUSY)~\cite{Golfand:1971iw,Volkov:1973ix,Wess:1974tw,Wess:1974jb,Ferrara:1974pu,Salam:1974ig} proposes a symmetry between bosons and fermions, thus predicting new particles that have identical quantum numbers as their Standard Model (SM) partners except for their spin. Specifically, SUSY introduces bosonic superpartners of the SM fermions and fermionic superpartners of the SM bosons.
In SUSY models with \(R\)-parity conservation~\cite{Farrar:1978xj}, the lightest SUSY particle (LSP) must be stable and, if electrically neutral and weakly interacting, can be a viable dark matter candidate~\cite{Goldberg:1983nd,Ellis:1983ew}.
In addition, SUSY can provide a necessary condition gauge coupling unification~\cite{Sakai:1981gr, Dimopoulos:1981yj, Ibanez:1981yh, Dimopoulos:1981zb} and a solution to the fine-tuning problem of the Higgs boson mass~\cite{Barbieri:1987fn, deCarlos:1993yy}, making it a leading candidate for physics beyond the SM.

In the minimal supersymmetric extension of the SM (MSSM)~\cite{Fayet:1976et,Fayet:1977yc}, the bino, wino, and higgsino fields are the fermionic partners of the \(\mathrm{SU(2)}\times\mathrm{U(1)}\) gauge fields in the SM and the two complex scalar Higgs doublets, respectively.
The bino, wino, and higgsino fields are collectively referred to as `electroweakinos,' which mix to form mass eigenstates known as neutralinos \(\ninoi\,(i=1,2,3,4)\) and charginos \(\chinojpm\,(j=1,2)\) with the subscripts indicating increasing mass. The masses of the bino, wino, and higgsino fields are parameterised in terms of $M_{1}$, $M_{2}$, and $\mu$, respectively. For large values of $\tan\beta$, representing the ratio between the vacuum expectation values of the two Higgs doublets, the phenomenology of the electroweakinos is driven by these three mass parameters.

This paper presents searches targeting SUSY scenarios in which the masses of the higgsino states are much smaller compared with the masses of the bino and wino states (i.e. $|\mu|\ll M_{1},M_{2}$). Such scenarios are motivated by the role of $\mu$ in naturalness arguments~\cite{Barbieri:1987fn,deCarlos:1993yy}, which favour values of $|\mu|$ near the weak scale~\cite{Barbieri:2009ev,Baer:2011ec,Papucci:2011wy,Baer:2012up}. In these scenarios, the lightest SUSY particles are a triplet of higgsino-like states ($\ninoone,\chinoonepm,\ninotwo$) having `compressed mass spectra', with minimal mass splittings of about 300~\MeV due to radiative corrections. Larger mass splitting values of a few \GeV can be generated depending on the values of $M_{1}$ and $M_{2}$ relative to $\mu$. Throughout this paper, the LSP is assumed to be the lightest neutralino (\ninoone) and the higgsino-like mass eigenstates are referred to as `higgsinos' for simplicity.

Various ATLAS and CMS searches were conducted to explore this regime of compressed mass spectra, primarily with the use of `tracks' that represent the reconstructed trajectories of charged particles in the inner detector.
Highly compressed scenarios with $\Delta m(\chinoonepm,\ninoone)\lesssim300\textrm{--}400~\MeV$ were searched for using the `disappearing track' signature that targets the long-lived ($c\tau>\mathcal{O}(10~\textrm{mm})$) \chinoonepm.
In these scenarios, a significant fraction of the produced $\chinoonepm$ particles can traverse several silicon layers of the inner detector before decaying into an invisible $\ninoone$ and a low-momentum (`soft') charged pion (i.e. $\chinoonepm\rightarrow\ninoone\pi^{\pm}$) that is not reconstructed, leading to the distinct signature of a charged particle track disappearing within the inner detector~\cite{SUSY-2016-06,SUSY-2018-19,CMS-EXO-16-044,CMS-EXO-19-010}.
Scenarios with $\Delta m(\chinoonepm,\ninoone)\approx300\textrm{--}900~\MeV$, and thus shorter $\chinoonepm$ lifetimes ($c\tau\approx\mathcal{O}(0.1\text{--}10~\text{mm})$),
were searched for using a `mildly displaced track' signature~\cite{Fukuda:2020}, in which the $\chinoonepm$ decay occurs before reaching the tracking detector and instead, the charged pion is reconstructed as a soft track with a large transverse impact parameter~\cite{SUSY-2020-04}, denoted by $d_0$ and defined as the shortest distance between the track and the beam line in the transverse plane.
Scenarios with mass splittings larger than approximately $1.5~\GeV$, corresponding to $c\tau<\mathcal{O}(0.1~\textrm{mm})$, were explored by searches targeting low-momentum prompt lepton\footnote{The term `lepton' in this paper refers to electrons or muons and does not include neutrinos and $\tau$-leptons unless otherwise specified.} signatures from $\ninotwo\rightarrow\ninoone Z^{*}\rightarrow\ninoone\ell^+\ell^-$~\cite{SUSY-2016-25, SUSY-2018-16,SUSY-2019-09,CMS-SUS-19-012,CMS-SUS-24-003}.
Despite these efforts, the strongest constraints on scenarios with $\Delta m(\chinoonepm,\ninoone)\approx0.9\textrm{--}1.2~\GeV$ are still set by the LEP experiments~\cite{LEPlimits}.
Additionally, ATLAS has performed a scan of the electroweak sector of the phenomenological MSSM~\cite{Djouadi:1998di,Berger:2008cq} that highlights the continued viability of higgsinos in this mass regime after accounting for constraints from previous Run-2 searches, flavour physics observables, precision electroweak measurements, and the observed dark matter relic density~\cite{SUSY-2020-15}.

The searches in this paper use $140~\mathrm{fb^{-1}}$ of proton--proton ($pp$) collision data at a centre-of-mass energy $\sqrt{s}=13~\TeV$, collected by the ATLAS experiment~\cite{PERF-2007-01} from 2015 to 2018 at the CERN Large Hadron Collider (LHC).
The pair production of higgsinos via electroweak interactions is assumed, with subsequent decays into a stable \ninoone, the LSP, and SM particles.
Simplified SUSY models~\cite{Alwall:2008ve,Alwall:2008ag,Alves:2011wf} are used to optimise the searches and interpret the results.

Two searches are designed to identify at least one of the higgsino decay products as a low-momentum track using machine learning techniques, which are adopted to search a range of mass splitting scenarios of the higgsinos.
The first search targets smaller mass splittings ($\Delta m(\chinoonepm,\ninoone)\approx0.3\textrm{--}1~\GeV$) using event signatures similar to the one employed in Ref.~\cite{SUSY-2020-04}, where soft pion tracks with large $d_0$
are identified as the long-lived higgsino decay product candidates.
This signature is also relevant for SUSY models where $M_{2} \ll M_{1}, |\mu|$, such that the lightest particles are the wino-like \chinoonepm and \ninoone with $\Delta m(\chinoonepm,\ninoone)$ being on the order of a few hundred \MeV~\cite{Ibe:2012sx}.
In contrast to this previous search, the sensitivity is enhanced using two dedicated neural-network (NN) classifiers: one exploiting the kinematic information of individual tracks (referred to as the `track-level' NN) and the other exploiting global kinematic observables of the event (referred to as the `event-level' NN) to provide discrimination between signal and SM background processes.
The second search targets slightly larger mass splittings ($\Delta m(\chinoonepm,\ninoone)\approx1\textrm{--}3~\GeV$) by using NN-based soft-electron and -muon identification algorithms, or `taggers', to identify the leptons from the $\ninotwo \to Z^*(\to \ell^+ \ell^-)\ninoone$ decays.
These advanced soft-lepton taggers extend the identification of electrons and muons to lower transverse momenta than what is achievable with the existing ATLAS algorithms, granting the ability to search for such new physics in this regime of mass splittings.

The paper is structured as follows: Section~\ref{sec:strategy} highlights the analysis strategies of the two searches in further detail. Section~\ref{sec:detector} provides a brief overview of the ATLAS detector. Details of the data and simulated event samples used in the analyses are given in Section~\ref{sec:samples}. The object reconstruction and identification methods are presented in Section~\ref{sec:reconstruction}. The event selection strategies and signal region definitions are outlined in Section~\ref{sec:selection}. The background estimation methods are discussed in Section~\ref{sec:background}, followed by a summary of the systematic uncertainty in Section~\ref{sec:systematics}. Finally, the search results and their statistical interpretations are presented in Section~\ref{sec:result}, followed by the conclusion in Section~\ref{sec:conclusion}.


%

%
\section{Analysis strategy}
\label{sec:strategy}

Two search strategies are employed in this paper to search for the direct production of higgsinos that feature small mass splittings between the \ninoone LSP and the slightly heavier \ninotwo and \chinoonepm particles. They are referred to as the `displaced track' and the `one-lepton and one-track ($1\ell1$T)' searches.
In this regime, the transverse momenta (\pt) of the SM particles arising from the \ninotwo and \chinoonepm decays are well below the ATLAS trigger thresholds.
Additionally, the signal processes are expected to result in two undetected \ninoone particles that are typically back-to-back in the transverse plane, therefore producing
only a small amount of missing transverse momentum \ptmiss (with magnitude denoted by \met), defined as the negative vector sum of the transverse momenta of all reconstructed particles in the event. In such cases, the missing transverse momentum triggers remain inefficient.
Both of the searches therefore target events in which the higgsinos are produced in association with a high-\pt jet from initial-state radiation (ISR), which provides a boost
to the higgsinos in the direction opposite to that of the ISR jet in the transverse plane.
While reducing the signal acceptance, the boost from the ISR jet serves to increase the \pt of the two \ninoone particles and align their directions in the transverse plane, thereby increasing the \met in the event.
The large \met then allows the \met trigger to be highly efficient and provides important discrimination between the signal and SM background processes.

In addition to this common selection, each search attempts to reconstruct the soft SM particles from the \ninotwo and \chinoonepm decays to further reduce the large SM backgrounds.
In contrast to previous ATLAS searches for higgsino-like LSPs with compressed mass spectra, the identification of these soft decay products, and the discrimination between signal and SM background processes, are performed using several NN classifiers.

\begin{figure}[tbp]
\centering
\subfloat[]{
\includegraphics[width=0.4\textwidth]{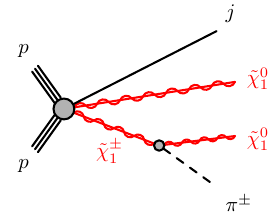}
\label{fig:feynman_diagram_DT}
}
\subfloat[]{
\includegraphics[width=0.4\textwidth]{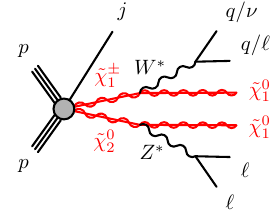}
\label{fig:feynman_diagram_1L1T}
}
\caption{Example diagrams for the targeted signal processes featuring a jet ($j$) from initial-state radiation for (a) the displaced track search and (b) the 1$\ell$1T search.
While only the $\ninoone\chinoonepm$ ($\ninotwo\chinoonepm$) diagram is shown for the displaced track (1$\ell$1T) search, the additional production processes described in Section~\ref{sec:samples} are considered as well.}
\end{figure}

The first, `displaced track' search, targets models in which $\Delta m(\chinoonepm,\ninoone)\approx0.3\textrm{--}1~\GeV$, where the dominant decay mode
of the charged higgsino is expected to be $\chinoonepm \to \pi^\pm \ninoone$~\cite{pasechnik2009neutralino} and the decay length of this particle is $\mathcal{O}(0.1\text{--}10~\text{mm})$, i.e. well before reaching the ATLAS tracking detector.
The charged pions from these decays can be identified as low-transverse-momentum tracks, typically having $\pt<5~\GeV$, with moderately large $|d_0|$ relative to the beam line.
An example diagram showing $\chinoonepm\ninoone$ production in association with an ISR jet is shown in Figure~\ref{fig:feynman_diagram_DT}, where the \chinoonepm decays into a charged pion and the \ninoone LSP.
The displaced track search leverages two fully connected NNs to discriminate between the signal and SM background processes.
The first network uses event-level information such as the jet kinematics and missing transverse momentum, while the second network uses tracking information to identify tracks that are consistent with arising from the decay of a moderately long-lived \chinoonepm, providing additional discriminating power.
The output scores of these networks are used to define the analysis regions, as described in Section~\ref{sec:selection_DT}.
The estimate of the SM backgrounds, which are largely represented by $Z(\to\nu\bar{\nu})+\text{jets}$ and $W^\pm(\to\tau^\pm\nu_\tau)+\text{jets}$ processes, is performed using simulation, with the normalisations of the dominant backgrounds being constrained in control regions, as described in Section~\ref{sec:background_DT}.

The second, `$1\ell1$T' search, is designed to have sensitivity to models with larger mass splittings where $1~\GeV \lesssim \Delta m(\chinoonepm, \ninoone) \lesssim 3~\GeV$, under the assumption that $\Delta m(\ninotwo, \ninoone)=2\Delta m(\chinoonepm, \ninoone)$.
Here, the \chinoonepm and \ninotwo are assumed to decay promptly into off-shell SM gauge bosons and the \ninoone LSP: $\chinoonepm \to W^{\pm *} \ninoone$ and $\ninotwo \to Z^{*} \ninoone$.
The search targets events featuring two electrons or muons with opposite electric charges from the $Z^*$ decay, where one of the leptons is identified using the standard ATLAS identification algorithms, defined in Section~\ref{sec:reconstruction}, and the other
is identified using dedicated soft-electron and soft-muon taggers based on NNs that are developed for this search.
These taggers use charged particle tracks and energy deposits in the electromagnetic calorimeter to significantly increase the identification efficiency for electrons and muons with $\pt\in[0.5, 5]~\GeV$ relative to standard identification methods,
as described in Section~\ref{sec:lowPT_ID}.
Figure~\ref{fig:feynman_diagram_1L1T} shows an example diagram of $\chinoonepm\ninotwo$ production in association with an ISR jet where the heavier higgsinos decay into an off-shell $W/Z$ boson and the \ninoone LSP\@.
To further enhance the sensitivity to higgsino production, the search employs a parameterised NN (pNN)~\cite{Baldi:2016fzo} to provide discrimination between the signal and SM background processes. The parameterisation is performed
relative to $\Delta m(\ninotwo, \ninoone)$, the parameter of the benchmark scenarios that has a significant impact on the signal event kinematics.
The details of the event selection are given in Section~\ref{sec:selection_1L1T}.
The dominant SM backgrounds in which the lepton-tagged track is not a real lepton, dominated by $W^\pm+\text{jets}$ processes, are determined by data-driven methods, while the residual SM backgrounds featuring a real lepton-tagged track
are estimated by using simulation, as described in Section~\ref{sec:background_1L1T}.

%

%
\section{ATLAS detector}
\label{sec:detector}

\newcommand{\AtlasCoordFootnote}{%
ATLAS uses a right-handed coordinate system with its origin at the nominal interaction point (IP)
in the centre of the detector and the \(z\)-axis along the beam pipe.
The \(x\)-axis points from the IP to the centre of the LHC ring,
and the \(y\)-axis points upwards.
Polar coordinates \((r,\phi)\) are used in the transverse plane,
\(\phi\) being the azimuthal angle around the \(z\)-axis.
The pseudorapidity is defined in terms of the polar angle \(\theta\) as \(\eta = -\ln \tan(\theta/2)\) and is equal to the rapidity
$ y = \frac{1}{2} \ln \left( \frac{E + p_z}{E - p_z} \right) $ in the relativistic limit.
Angular distance is measured in units of \(\Delta R \equiv \sqrt{(\Delta y)^{2} + (\Delta\phi)^{2}}\).}

The ATLAS detector~\cite{PERF-2007-01} at the LHC covers nearly the entire solid angle around the collision point.\footnote{\AtlasCoordFootnote}
It consists of an inner tracking detector surrounded by a thin superconducting solenoid, electromagnetic and hadronic calorimeters,
and a muon spectrometer incorporating three large superconducting air-core toroidal magnets.

The inner-detector system is immersed in a \qty{2}{\tesla} axial magnetic field
and provides charged-particle tracking in the range of  \(|\eta| < 2.5\). It consists of a high-granularity silicon pixel detector, a silicon microstrip detector (SCT), and a transition radiation straw-tube tracker (TRT), each spanning a sensitive radial distance from the interaction point of 33--150~mm, 299--560~mm, and 563-1066~mm, respectively~\cite{PERF-2015-07}. The barrel (each endcap) consists of four (three) pixel layers, four (nine) double-layers of single-sided silicon microstrips, and 73 (160) layers of TRT straws. The pixel detector covers the vertex region and typically provides four measurements per track, the first hit generally being in the insertable B-layer (IBL) installed before Run~2~\cite{ATLAS-TDR-19,PIX-2018-001}. The smaller radius and the reduced pixel size of IBL results in improvements of both the transverse and longitudinal impact parameter resolutions. The $d_0$ resolution reaches $\sim0.09$ mm for tracks with $\pt=1~\GeV$~\cite{ATL-IDTR-2018-008}. The SCT usually provides eight measurements, each corresponding to a single side of the double-layered strips, per track. The TRT enables radially extended track reconstruction up to \(|\eta| = 2.0\). As electrons tend to have higher relativistic gamma factors and thus higher probabilities of inducing transition radiation compared with other particles, the TRT is also capable of providing electron identification information based on the fraction of hits (typically 30 in total) above a higher energy-deposit threshold.

The calorimeter system covers the pseudorapidity range \(|\eta| < 4.9\).
Within the region \(|\eta|< 3.2\), electromagnetic calorimetry is provided by barrel and
endcap high-granularity lead/liquid-argon (LAr) calorimeters,
with an additional thin LAr presampler covering \(|\eta| < 1.8\)
to correct for energy loss in material upstream of the calorimeters.
Hadronic calorimetry is provided by the steel/scintillator-tile calorimeter,
segmented into three barrel structures within \(|\eta| < 1.7\), and two copper/LAr hadronic endcap calorimeters.
The solid angle coverage is completed with forward copper/LAr and tungsten/LAr calorimeter modules
optimised for electromagnetic and hadronic energy measurements, respectively. The electromagnetic calorimeters consist of three main layers and the presampler layer. The hadronic calorimeters consist of three (four) layers in the barrel (endcap) regions, providing sufficient containment of hadronic showers.

The muon spectrometer comprises separate trigger and
high-precision tracking chambers measuring the deflection of muons in a magnetic field generated by the superconducting air-core toroidal magnets.
The field integral of the toroids ranges between \num{2.0} and \qty{6.0}{\tesla\metre}
across most of the detector.
Three layers of precision chambers, each consisting of layers of drift tubes, cover the region \(|\eta| < 2.7\),
complemented by cathode-strip chambers in the forward region, where the background is highest.
The muon trigger system covers the range \(|\eta| < 2.4\) with resistive-plate chambers in the barrel, and thin-gap chambers in the endcap regions.

The luminosity is measured mainly by the LUCID--2~\cite{LUCID2} detector that records Cherenkov light produced in the quartz windows of photomultipliers located close to the beampipe.

Events are selected by the first-level trigger system implemented in custom hardware,
followed by selections made by algorithms implemented in software in the high-level trigger~\cite{TRIG-2016-01}.
The first-level trigger accepts events from the \qty{40}{\MHz} bunch crossings at a rate below \qty{100}{\kHz},
which the high-level trigger further reduces to record complete events to disk at about \qty{1.25}{\kHz}.

A software suite~\cite{SOFT-2022-02} is used in the trigger and data acquisition systems of the experiment, in detector operations, in data simulation, and in the reconstruction and analysis of real and simulated data.


%

%
\section{Data and simulated event samples}
\label{sec:samples}

%
The considered data sample amounts to 140~\ifb of $\sqrt{s}=13$~\TeV{} $pp$ collision data recorded by the ATLAS detector from 2015--2018 with an uncertainty of 0.83\% in the integrated luminosity~\cite{DAPR-2021-01}.
The average number of interactions per bunch-crossing in the data sample is 33.7~\cite{DAPR-2018-01}.
The data are selected using missing transverse momentum triggers~\cite{TRIG-2019-01} that use different trigger thresholds depending on the data-taking periods, resulting in an efficiency of more than $95\%$
for events with offline-calibrated $\met$ greater than $200~\GeV$.
Additionally, single-electron and single-muon triggers~\cite{TRIG-2018-05,TRIG-2018-01} are employed to collect a sample of events that are used to constrain SM backgrounds involving the production of $W^\pm$ or $Z$
bosons, as described in Section~\ref{sec:background_DT}.

Monte Carlo (MC) simulation samples are produced for the signal and SM background processes to estimate their event yields and associated systematic uncertainty.
The SM backgrounds in the searches are determined from both the data-driven methods and  the MC simulation, as described in Section~\ref{sec:background}.
While the details of the signal and SM background MC simulation are given below, both of the sets of samples undergo the same procedure to include the effects of multiple $pp$ interactions
in the same and neighbouring bunch crossings (pile-up).
In each case, the simulated hard-scattering events are overlaid with inelastic $pp$ collision events generated with \PYTHIA[8.186]~\cite{Sjostrand:2007gs} using the \NNPDF[2.3lo] set of parton distribution
functions (PDF)~\cite{Ball:2012cx} and the A3 set of tuned parameters~\cite{ATL-PHYS-PUB-2016-017}.
Weights are then applied to the simulated events to match the pile-up distribution observed in the data and the detector response is simulated using the ATLAS simulation framework~\cite{SOFT-2010-01}
based on \textsc{Geant4}~\cite{Agostinelli:2002hh}. The full detector simulation is used for the SM background samples, while the signal samples use either this full detector simulation or the ATLAS fast
simulation that relies on a parameterisation of the calorimeters, as detailed below.

Signal MC samples consisting of direct higgsino production are simulated for the displaced track and 1$\ell$1T searches to optimise the event selections and interpret the results in the context of the simplified models under study.
Since the 1$\ell$1T search targets the soft leptons from the $\ninotwo \to Z^*(\to \ell^+ \ell^-) \ninoone$ decays, the $\ninotwo\chinoonepm$ and $\ninotwo\ninoone$ production modes are simulated, while the
samples used for the displaced track search also include the $\ninoone\chinoonepm$ and $\chinoonep\chinoonem$ processes.
In each case, the $\ninoone\ninoone$ and $\ninotwo\ninotwo$ production modes are ignored due to their negligible production cross-sections when the $\ninoone$ and $\ninotwo$ are higgsino-like states.
In the simplified model considered here, the $\chinoonepm$ mass is set to be halfway between the neutralino masses, such that $\Delta m(\ninotwo, \ninoone) = 2 \Delta m(\chinoonepm, \ninoone)$.
The MC samples are generated at leading-order (LO) with \madgraph~2.9.5~\cite{Alwall:2014hca} for the displaced track search and version 2.9.3 for the 1$\ell$1T search using the NNPDF2.3LO~\cite{Ball:2012cx} PDF set and
including up to two additional partons in the matrix element calculation. The simulated events for the displaced track search and the 1$\ell$1T search are then interfaced with \pythia~8.306~\cite{Bierlich:2022pfr} and \pythia~8.245~\cite{Sjostrand:2014zea}, respectively,
to model the parton shower, hadronisation, and underlying event using the A14 set of tuned parameters~\cite{ATL-PHYS-PUB-2014-021}.
Matching between the matrix element and parton shower jets is performed using the CKKW-L scheme~\cite{Lonnblad:2011xx} with the matching scale set to $15~\GeV$.
At least one parton in the final state is required to have a transverse momentum greater than $50~\GeV$ to enforce the ISR topology targeted by both of the searches.
The decays of $b$- and $c$-hadrons are modelled by \EVTGEN 1.7.0~\cite{Lange:2001uf}.

For the signal samples used by the 1$\ell$1T search, the branching fractions for the $\ninotwo \to Z^* \ninoone$ and $\chinoonepm \to W^{\pm *} \ninoone$ are set to 100\% and the higgsinos are decayed with \madspin~\cite{Artoisenet:2012st} to account for spin
correlations effects. The leptonic branching fractions of the $\ninotwo \to Z^* \ninoone$, typically 5\% each for $\ninotwo \to e^{+}e^{-} \ninoone$ and $\ninotwo \to \mu^{+}\mu^{-} \ninoone$, are calculated with \textsc{SUSY-HIT}~\cite{Djouadi:2006bz} for different $\Delta m(\ninotwo,\ninoone)$ values and incorporated in the signal yield calculation. The higgsino branching fractions in the samples used by the displaced track search are calculated following the prescription in Ref.~\cite{pasechnik2009neutralino}, which accounts for the mass splitting between the heavier \ninotwo/\chinoonepm
and the \ninoone.
For $\Delta m(\chinoonepm,\ninoone) = 0.5~\GeV$, the branching fraction for $\chinoonepm \to \pi^\pm \ninoone$ is approximately $80\%$, with the remaining $20\%$ corresponding to $\chinoonepm \to e^\pm/\mu^\pm \nu \ninoone$.
In the simplified model, this corresponds to $\Delta m(\ninotwo,\ninoone) = 1~\GeV$, for which the dominant branching fractions are approximately $30\%$ for each of the $\ninotwo \to \pi^0 \ninoone$, $\ninotwo \to \pi^+\pi^- \ninoone$, and
$\ninotwo \to \nu\bar{\nu} \ninoone$ processes.
In addition to the branching fractions, the mass splitting between the higgsinos determines the $\chinoonepm$ and $\ninotwo$ lifetimes, where smaller mass splittings result in longer lifetimes and flight lengths.
The higgsino decays for these samples are performed using \pythia{} and the higgsino lifetimes are determined by the mass splitting according to the prescription in Ref.~\cite{pasechnik2009neutralino}.

The higgsino production cross-sections and associated uncertainty are calculated using \textsc{Resummino} 2.0.1 at NLO+NLL precision~\cite{Beenakker:1999xh,Debove:2010kf,Fuks:2012qx,Fuks:2013vua,Fiaschi:2018hgm,Bozzi:2007qr,Fuks:2013lya,Fiaschi:2018xdm,Fiaschi:2023tkq}
using an envelope of the CTEQ6.6~\cite{Pumplin:2002vw} and MSTW2008nlo90cl~\cite{Martin:2009iq} PDF sets. The uncertainty originates primarily from QCD renormalisation and factorisation scales.
Taking an example with $m(\ninotwo) = 102~\GeV$, $m(\chinoonepm) = 101~\GeV$, and $m(\ninoone) = 100~\GeV$, the cross-section values are $5.3 \pm 0.2$~pb, $3.2 \pm 0.1$~pb, $5.5 \pm 0.2$~pb and $2.9 \pm 0.1$~pb for $\ninotwo\chinoonepm$,
$\ninotwo\ninoone$, $\ninoone\chinoonepm$ and $\chinoonep\chinoonem$ production, respectively.
The samples used for the 1$\ell$1T search use the full detector simulation, while those used for the
displaced track search use the ATLAS fast simulation as its analysis signature is less dependent to calorimeter responses.

The SM background processes are simulated using a variety of MC generators, with Table~\ref{tab:mc_backgr} providing a summary of the matrix element and parton shower configurations as well as the accuracy of the cross-section calculation used
for each of the background samples.
The dominant sources of SM backgrounds for both the displaced track and the 1$\ell$1T searches arise from vector boson production with subsequent leptonic decays, in association with at least one high-\pt jet.
These processes are simulated using \SHERPA 2.2.11 with the \NNPDF[3.0nnlo] PDF set~\cite{Ball:2014uwa}, where the matrix element calculation includes up to two additional partons at NLO accuracy and up to five additional partons at LO accuracy.
The $Z^{(*)}/\gamma^*$+jets samples provide coverage of the dilepton invariant mass down to $10~\GeV$.
Minor backgrounds arise from the production of diboson pairs ($VV=WW,ZZ,WZ$), top quarks ($t\bar{t}$, single-$t$ in the $s$, $t$, and $tW$ channels), multijets, and $\gamma+\text{jets}$.
The diboson samples are modelled with either \SHERPA version 2.2.1 or 2.2.2, depending on the specific process, at NLO accuracy with the \NNPDF[3.0nnlo] PDF set.
The $t\bar{t}$, single-$t$ and $tW$ backgrounds are simulated using \POWHEGBOX[v2] for the matrix element and \PYTHIA[8.230] for the parton shower, using the \NNPDF[3.0nlo] and \NNPDF[2.3lo] PDF sets, respectively.
The diagram removal scheme~\cite{Frixione:2008yi} is used to account for interference effects between the $t\bar{t}$ and $tW$ processes.
Multijet samples are modelled at LO using \PYTHIA[8.230] and the \NNPDF[2.3lo] PDF set for both the matrix element and the parton shower.
Finally, the $\gamma+\text{jets}$ processes are modelled using \SHERPA 2.2.2 at NLO accuracy with the \NNPDF[3.0nnlo] PDF set.
Backgrounds that are simulated using \SHERPA model the decays of $b$- and $c$-hadrons internally, while the other samples make use of \EVTGEN 1.6.0 or 1.2.0~\cite{Lange:2001uf}.
Backgrounds such as triboson production and other rare processes involving top quarks or Higgs bosons are found to be negligible in the searches.

%
\begin{table}[tbp]
\fontsize{6.5pt}{8pt}\selectfont
\centering
\caption{\label{tab:mc_backgr}
A summary of the configurations used to generate the MC samples for the SM background processes. Starting from the second column, the table provides the generators used for the matrix element (ME), the generators used for the parton shower (PS),
the order in $\alpha_\text{s}$ used to determine the production cross-section, the tune for the underlying event, the PDF set used for the matrix element calculation, and the PDF set used for the parton shower.
The symbol $V$ is used to denote the $W$ and $Z$ vector bosons.}
\begin{tabular}{p{0.087\linewidth}p{0.190\linewidth}p{0.152\linewidth}p{0.133\linewidth}p{0.053\linewidth}p{0.104\linewidth}p{0.09\linewidth}}

\toprule

Physics process          			& Matrix element 	& Parton shower 	&  Cross-section 		& Tune 			& PDF (ME) & PDF (PS)\\
\midrule

$V+\text{jets}$                                 & \SHERPA[2.2.11]~\cite{Bothmann:2019yzt}                      & \SHERPA[2.2.11]               & NLO~\cite{Hoeche:2011fd,Hoeche:2012yf,Catani:2001cc,Hoeche:2009rj}           & Default~\cite{ATL-PHYS-PUB-2017-005} & \NNPDF[3.0nnlo]~\cite{Ball:2014uwa} & \NNPDF[3.0nnlo] \\

Diboson ($VV$)                      &  \SHERPA[2.2.1],\,2.2.2       &       \SHERPA[2.2.1],\,2.2.2~\cite{Gleisberg:2008fv,Schumann:2007mg}
& NLO              &  Default
&\NNPDF[3.0nnlo]   & \NNPDF[3.0nnlo] \\

$t\bar{t}$                                 & \POWHEGBOX[v2]~\cite{Frixione:2007nw,Nason:2004rx,Frixione:2007vw,Alioli:2010xd}                      & \PYTHIA[8.230]~\cite{Sjostrand:2014zea}               & NNLO+NNLL~\cite{Beneke:2011mq,Cacciari:2011hy,Baernreuther:2012ws,Czakon:2012zr,Czakon:2012pz,Czakon:2013goa,Czakon:2011xx}           & A14~\cite{ATL-PHYS-PUB-2014-021} & \NNPDF[3.0nlo]~\cite{Ball:2014uwa} & \NNPDF[2.3lo]~\cite{Ball:2012cx}\\

Single-top                                 & \POWHEGBOX[v2]~\cite{Alioli:2009je,Nason:2004rx,Frixione:2007vw,Alioli:2010xd,Re:2010bp}    & \PYTHIA[8.230]        & NLO+NNLL~\cite{Kidonakis:2010ux,Kidonakis:2013zqa}     & A14                   & \NNPDF[3.0nlo]%
& \NNPDF[2.3lo]%
\\

Multijets                                 & \PYTHIA[8.230]                      & \PYTHIA[8.230]               & LO           & A14 & \NNPDF[2.3lo] & \NNPDF[2.3lo] \\

$\gamma+\text{jets}$                                 & \SHERPA[2.2.2]                      & \SHERPA[2.2.2]               & NLO           & Default & \NNPDF[3.0nnlo] & \NNPDF[3.0nnlo] \\

\bottomrule

\end{tabular}

\end{table}



%

%
\section{Event reconstruction}
\label{sec:reconstruction}

The trajectories of charged particles are reconstructed from the inner detector hits as tracks using clusterisation and combinatorial track finding algorithms~\cite{PERF-2015-08}. The kinematic variables of the track such as \pt, $\eta$ and $\phi$ are defined at the perigee, the point of closest approach of the track to the beam line. Tracks with \pt down to 500~\MeV are reconstructed. All events are required to have at least one reconstructed $pp$ interaction vertex with a minimum of two associated tracks. In events with multiple vertices, the primary vertex is defined as the vertex having the highest $\sum\pt^2$ of associated tracks. A set of basic quality criteria~\cite{DAPR-2012-01} is applied to reject events with detector noise or non-collision backgrounds.

Leptons, jets, and tracks are `preselected' using loose identification criteria and must satisfy more restrictive `tight' identification requirements to be selected for the search regions described in Section~\ref{sec:selection}. Preselected leptons and jets are primarily used to resolve ambiguities between tracks and energy clusters associated with multiple lepton and jet candidates.

Isolation criteria are used in the definition of tight leptons and are based on tracking information, calorimeter energy clusters, or both. Isolation energies are computed as a $\sum\pt$ of nearby activity, excluding the contributions from nearby leptons. Requiring small isolation energies helps to reduce contributions from semileptonic heavy-flavour hadron decays and jets misidentified as prompt leptons. The isolation requirements are based on those described in Refs.~\cite{EGAM-2021-01} and~\cite{MUON-2018-03}. The isolation variable $\pt^{\textrm{cone}X}$ is defined as the $\sum\pt$ of all tracks satisfying $\Delta R<X/100$ and $\pt > 1~\GeV$. For the 1$\ell$1T search, this variable is adjusted by subtracting the \pt of the paired lepton or track as described later.

Electrons are required to have $\pt > 4.5~\GeV$ and $|\eta| < 2.47$. Preselected electrons are further required to satisfy the calorimeter- and tracking-based \emph{Loose} likelihood identification~\cite{EGAM-2021-01} and to have a longitudinal impact parameter $z_0$ relative to the primary vertex satisfying $|z_0 \sin \theta| < 0.5$~mm. Tight electrons must satisfy the \emph{Medium} identification criterion~\cite{EGAM-2021-01}, having typical selection efficiency of over 80\% for $\pt>10~\GeV$ and 70--80\% for $\pt=4.5\textrm{--}10~\GeV$, and be associated with the primary vertex, with the significance of the transverse impact parameter relative to the beam position satisfying \mbox{$S(d_0)\equiv|d_0|/\sigma(d_0) < 5$}, where $\sigma(d_0)$ is the uncertainty in $d_0$. For the displaced track search, tight electrons are further selected using the \emph{Loose\_VarRad} isolation working point~\cite{EGAM-2021-01}, whereas for the 1$\ell$1T search, the
\emph{TightTrackOnly\_VarRad} isolation working point~\cite{EGAM-2021-01} is required with an adjustment to subtract the contribution of the signal candidate track. This adjustment is performed to avoid vetoing signal processes in which the two leptons originating from the \ninotwo are nearby.

Muons are required to satisfy $\pt > 3~\GeV$ and $|\eta|<2.5$. Preselected muons are identified using the \emph{Medium} criterion~\cite{MUON-2018-03}, having typical selection efficiency of over 90\% for $\pt>5~\GeV$ and 50--90\% for $\pt=3\textrm{--}5~\GeV$, and are required to satisfy $|z_0 \sin \theta| < 0.5$~mm. Tight muons must satisfy the promptness requirement of \mbox{$|d_0|/\sigma(d_0) < 3$}. For the displaced track search, tight muons are further selected using the \emph{Pflow\_Loose} isolation working point~\cite{MUON-2018-03}, whereas for the 1$\ell$1T search, the \emph{TightTrackOnly} isolation working point~\cite{MUON-2018-03} is required with an adjustment to subtract the contribution of the signal candidate track, similar to the electrons.

Jets are reconstructed from particle flow objects~\cite{PERF-2015-09} using the anti-$k_t$ algorithm~\cite{Cacciari:2008gp,Fastjet} with radius parameter $R = 0.4$. Preselected jets are required to have $\pt>20$~\GeV and $|\eta| < 4.5$ after being calibrated according to the procedure in Ref.~\cite{JETM-2018-05}. For suppressing jets originating from pile-up, jets with $\pt < 60~\GeV$ and $|\eta| < 2.5$ are required to satisfy the \emph{Tight} working point of the jet vertex tagger~\cite{PERF-2014-03}, using information from the tracks associated with the jet. From the sample of preselected jets, tight jets are further selected when they satisfy $|\eta| < 2.8$, which ensures that the jets are within the acceptance of the high-granularity endcap calorimeters.

Jets originating from the hadronisation of $b$-quarks, referred to as $b$-tagged jets, are identified from preselected jets within $|\eta| < 2.5$ using the \emph{DL1r} algorithm~\cite{FTAG-2019-07}.
The $\pt > 20~\GeV$ requirement is maintained to maximise the rejection of the \ttbar background in analysis regions that veto events containing $b$-tagged jets. The $b$-tagging algorithm working point is chosen so that $b$-jets from simulated $t\bar{t}$ events are identified with an 85\% efficiency,
with corresponding rejection factors of approximately 2.9 and 40 for jets containing $c$-hadrons and jets arising from light-flavour quarks and gluons, respectively~\cite{FTAG-2020-08,FTAG-2019-02}.
To resolve ambiguities between nearby reconstructed leptons and jets, the following procedures are performed. First, preselected muons that share an inner detector track with a preselected electron candidate are rejected. Next, non-$b$-tagged preselected jets within $\Delta R<0.2$ of a preselected electron are discarded. Preselected electrons satisfying $\Delta R<\min(0.4, 0.04+10~\GeV / \pt(e))$ relative to a preselected jet are then removed. Non-$b$-tagged preselected jets within $\Delta R<0.2$ of a preselected muon are then rejected. Finally, preselected muons satisfying $\Delta R<\min(0.4, 0.04+10~\GeV / \pt(\mu))$ with respect to a preselected jet are removed.

The missing transverse momentum is defined as the negative vector sum of the transverse momenta of all preselected objects (leptons, jets, and photons~\cite{EGAM-2021-01}) and an additional soft term that is constructed from all tracks that are not associated with any lepton or jet but are associated with the primary vertex.
In this way, \met is adjusted for the best calibration of leptons and jets, while maintaining pile-up independence in the soft term~\cite{JETM-2020-03}.

Scale factors are applied to the efficiencies of reconstructed electrons~\cite{EGAM-2021-01}, muons~\cite{MUON-2018-03} and $b$-tagged jets~\cite{FTAG-2018-01} in the simulated samples to match the reconstruction efficiencies in data.
The scale factors for $b$-tagged jets account for the differences between data and simulated samples in the identification efficiencies for jets including $b$-hadron decays and for the misidentification rates of jets initiated from charm quarks, light-flavour quarks, or gluons. In addition, scale factors are used to reweight simulated samples to properly model the trigger efficiency observed in data.

\subsection{Low-\pt lepton-track identification}
\label{sec:lowPT_ID}

In the 1$\ell$1T search, low-\pt lepton-track identification (ID) methods are used to identify leptons eluding standard identification methods and to gain sensitivity to unexplored low-$\Delta m$ regions. Tracks with $\pt>500~\MeV$ and $|\eta|<2.5$ are preselected using the \emph{Tight Primary} working point defined in Ref.~\cite{ATL-PHYS-PUB-2015-051}. In low-\pt regimes with $\pt<5~\GeV$, tracks are dominated by hadrons (mostly charged pions) and fake combinatorial tracks. The target of the low-\pt lepton-track ID is to distinguish lepton-tracks from such background tracks by exploiting information mainly from detectors situated closer to the interaction point, compared to those used in standard methods. A low-\pt electron-track ID and a low-\pt muon-track ID are developed separately with independent optimisations.

The low-\pt electron-track candidate is required to have $\pt\in[0.5, 5]~\GeV$ and $|\eta|<2.5$. Tracks are extrapolated to a matching calorimeter energy cluster based on its position and energy deposit when available, however, the existence of a matched cluster is not always required.
The matching for low-\pt electron-tracks is performed assuming a constant magnetic field in the volume of the inner detector and using the perigee parameters of the track.
A fully connected NN model with two 128-node hidden layers is trained using single particle MC events to separate low-\pt electron-tracks from dominant low-\pt hadronic tracks using silicon tracking information including track \pt and track $\eta$, along with optional TRT and electromagnetic calorimeter energy cluster information. Tracking fit errors, hit information in the TRT detector, and energy deposits in the inner layers of the electromagnetic calorimeter show notable differences between electron and hadronic tracks due to the effects of bremsstrahlung. An identification working point for the 1$\ell$1T search is dynamically defined by adopting another NN using track \pt, track $|\eta|$, and the score from the previous network as inputs. By setting a threshold on the output score of this network, a working point optimised within the 3D space of the inputs is chosen. The latter network is trained with signal higgsino and background $W^\pm+$jets MC events satisfying the event-selection criteria defined later in Section~\ref{sec:selection}.

The low-\pt muon-track candidate is required to have $\pt\in[1, 5]~\GeV$ and $|\eta|<1.6$. A large fraction of muons with \pt below 3~\GeV make a full-stop in the calorimeters and do not reach the muon spectrometer. Thus, the low-\pt muon track is identified using information only from the inner detector and calorimeters. As opposed to the case of electron-tracks, muon candidate tracks are always required to have a matching calorimeter energy cluster. A fully connected NN model with six 64-node hidden layers is trained using $J/\psi\rightarrow\mu^{+}\mu^{-}$ MC events to separate low-\pt muon-tracks from dominant low-\pt hadronic tracks using track \pt, track $\eta$, and energy deposits in each calorimeter layer. Low-\pt muons tend to act as minimum ionising particles and deposit less energy in the calorimeters compared with low-\pt hadrons, allowing their identification without the use of muon detectors. Similarly to the electron-track case, a working point for the 1$\ell$1T search is dynamically defined with another NN using track \pt, track $\eta$, and the previous network score, with the training performed with signal higgsino and background $W^\pm+$jets MC events that satisfy the 1$\ell$1T event-selection criteria.

Figure~\ref{fig:lowptlep_eff} shows the signal and background track efficiencies of the low-\pt lepton-track ID methods as a function of track \pt. Both of the ID selections are optimised to maximise the final sensitivity to signal higgsino models. Both  of the lepton-track ID methods have tighter criteria in very low-\pt regimes of $\pt<1.5~\GeV$, where contributions from hadronic and fake tracks become significant. The electron-track ID method has tighter criteria in terms of efficiency to achieve sufficient background suppression in the target region.
For the low-\pt electron-track ID, the signal efficiency ranges in 0.6--70\%, while the fake rate is kept within 0.003--3\%. For the low-\pt muon-track ID, the signal efficiency ranges in 15--75\%, while the fake rate is kept within 0.3--2\%.

\begin{figure}[tbp]
\centering
\subfloat[]{
\includegraphics[width=0.495\textwidth]{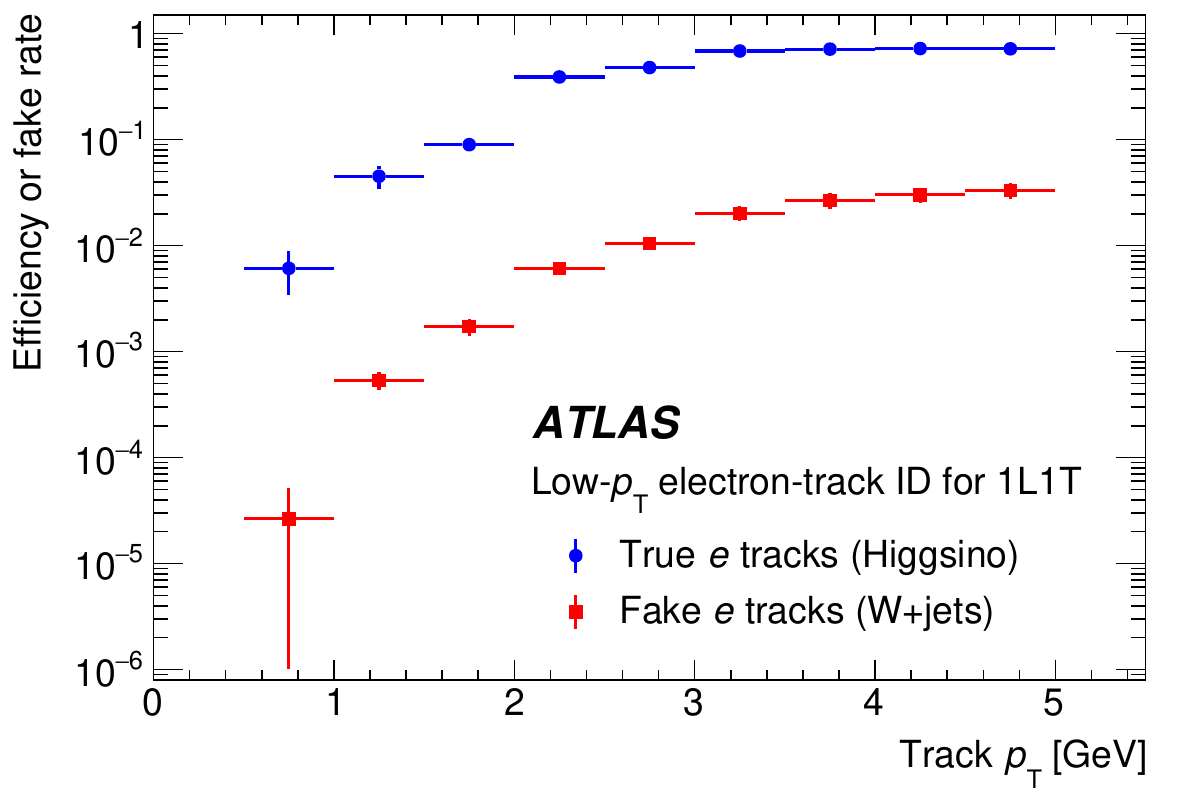}
}
\subfloat[]{
\includegraphics[width=0.495\textwidth]{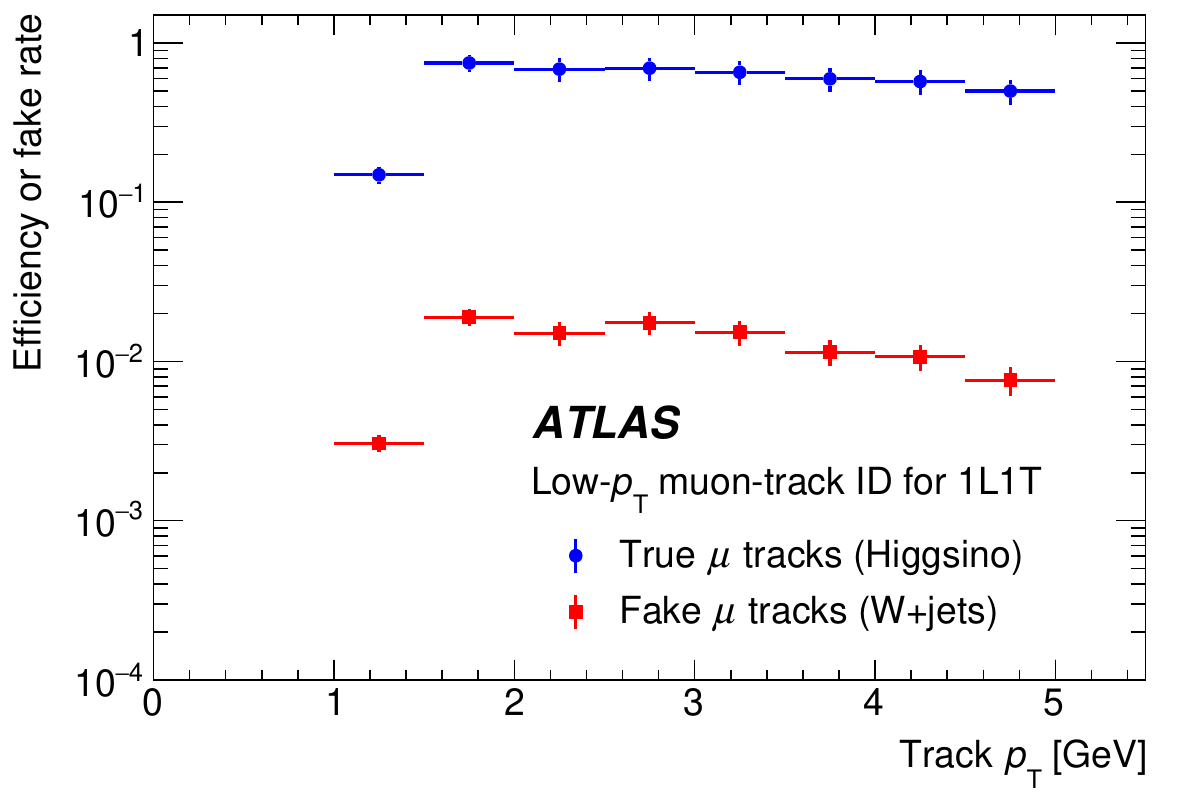}
}
\caption{\label{fig:lowptlep_eff} Simulated true-track efficiency and fake-track rate as a function of track \pt for (a) the low-\pt electron-track ID and (b) the low-\pt muon-track ID\@. The scale factors obtained from the calibration methods are applied with their corresponding uncertainty. The total uncertainty including the statistical and systematic uncertainties are shown, where the former uncertainty is found to be negligible. The true track is defined as a track matched to a lepton originating from $\ninotwo\to\ninoone\ell^+\ell^-$ decay in signal events satisfying the event selections defined later in Section~\ref{sec:selection}. The fake track is defined as a track in $W^\pm+$jets events that satisfies the same selection, excluding the track associated with the lepton from the $W^\pm$ decay.}
\end{figure}

The scale factors and their uncertainty for low-\pt lepton-tracks are obtained using individual calibration methods for the electron and muon channels. The low-\pt electron-track ID is calibrated using a `tag-and-probe method'~\cite{EGAM-2021-01} focusing on the distribution of $\Delta R$ between the low-\pt electron-track and the photon in $Z\rightarrow e^+e^-\gamma$ processes, particularly in events where an electron and the photon take away most of the energy, leaving the other electron to have a low \pt. In such event topologies, real low-\pt electrons from $Z\rightarrow e^+e^-\gamma$ tend to be distributed close to the photon and have small $\Delta R$, whereas fake misidentified electrons are randomly distributed and have broader $\Delta R$ distributions. This shape difference in $\Delta R$ between the low-\pt electron-track and the photon is used to obtain the identification efficiency in data.
The low-\pt muon-track ID is calibrated using a tag-and-probe method using $J/\psi\rightarrow\mu^+\mu^-$ processes, similar to that used in the calibration of standard muon identification~\cite{MUON-2018-03}. The invariant mass distribution of the dimuon system is fitted using crystal-ball and quadratic functions. In both cases, the scale factors are evaluated by taking the ratio between the measured efficiency in data and the simulated efficiency in MC. They are only applied to the signal MC samples and the minor SM backgrounds of the 1$\ell$1T search that are evaluated using MC samples.


%

%
\section{Event selection}
\label{sec:selection}

This section describes the event selections for the displaced track and 1$\ell$1T searches, each focusing on different mass splitting regions, and thus relying on distinct strategies exploiting NN models. For both of the searches, signal regions are defined to isolate the signal processes. As later described in Section~\ref{sec:background}, control regions are defined to normalise the background processes and validation regions adjacent to the signal regions are defined to validate the background estimate.

\subsection{Displaced track event selection}
\label{sec:selection_DT}

The displaced track search targets scenarios where $\Delta m(\chinoonepm,\ninoone)\approx0.3\textrm{--}1.0~\GeV$ and the SM decay products of the moderately long-lived ($\mathcal{O}(0.1\text{--}10~\text{mm})$) \chinoonepm and \ninotwo particles can be reconstructed as low-\pt tracks with significant $d_0$.
Events are selected using an \met trigger and are first required to satisfy the event-level selections listed in Table~\ref{tab:evsel_DT}. Although the \met trigger is fully efficient above 200~\GeV, the reconstructed \met is required to be larger than 275~\GeV to enhance signal sensitivity. For signal events in which the \ninoone-pair significantly contributes to the invisible momentum, the high \met requirement indicates a phase space in which the higgsino system is recoiling against hadronic activity in the form of ISR jets. Thus, all events are required to have at least one central jet with $\pt>250~\GeV$. Up to four jets are allowed in the event, each satisfying $\Delta\phi(\ptmiss,\mathbf{j})>0.4$. For the signal region (SR), the number of preselected leptons in each event is required to be zero.

\begin{table}[tbp]
\centering
\caption{\label{tab:evsel_DT} Event-level selections applied to all events used for the displaced track search. The single-lepton trigger is used only for events in control and validation regions. The $\Delta\phi$ variables are defined to be in the range of $[0,\pi)$. The requirement on the number of tracks refers to the tracks defined in Section~\ref{sec:reconstruction}.}
\renewcommand{\arraystretch}{1.1}
\begin{tabular}{ll}
\toprule
Variable               & Requirement \\
\midrule
Trigger & \met or single-lepton \\
\met[GeV] & \(> 275\) \\
Number of jets with $\pt>30~\GeV$ & $[1,4]$ \\
Leading jet \pt[GeV] & $>250$ \\
Leading jet $|\eta|$ & $< 2.4$ \\
\(\min \Delta\phi(\ptmiss, \mathbf{j})\) & \(> 0.4\) \\
Number of preselected leptons ($N_{\textrm{pre-lep}}$) & $=0$ (only for SR) \\
Number of tracks & $\geq 1$\\
\bottomrule
\end{tabular}
\end{table}

To exploit the available information and to maximise the signal sensitivity, a machine learning based approach using NNs is adopted for further event selection. In particular, two independent NNs are trained to discriminate between signal and background processes: one using event-level information and the other using track-level information. Both of the models are fully connected networks implemented in \texttt{PyTorch}~\cite{PyTorch_Citation} and are optimised using binary cross-entropy loss. They each adopt an optimised structure of three hidden layers with 50, 30, and 15 nodes, respectively, and rectified linear unit (ReLU) activation. Both of the output scores range from 0 to 1, with higher scores corresponding to more signal-like events or tracks. The use of two independent NNs is adopted for its high flexibility in designing and validating the background estimation method later mentioned in Section~\ref{sec:background_DT}.

The lists of input variables for both of the networks are given in Table~\ref{tab:DT_NNinputs}. The event-level NN uses the event topology information based on \met and hadronic jets.
The inputs that provide the strongest discrimination power between signal and background processes are \met and the effective mass, $m_\text{eff}$, defined as the scalar sum of \pt of all jets with $\pt>30~\GeV$ and \met.
The track-level NN exploits a variety of track information, such as the kinematics, impact parameters, isolation variables, and angular difference between the track and the missing transverse momentum.
In addition, an inclusive secondary vertexing algorithm~\cite{ATL-PHYS-PUB-2019-013} is used to identify tracks with $\pt > 500~\MeV$ that are consistent with being produced at vertices that are displaced from the primary vertex.
The invariant mass of the reconstructed secondary vertex, calculated under the assumption that each track is a charged pion, is used to discriminate between background tracks originating from the decays of long-lived hadrons (e.g. $K_{S}^{0}$, $\Lambda^{0}$) and those due to $\ninotwo \to \pi^+\pi^- \ninoone$ decays.
For reconstructed secondary vertices with exactly two associated tracks, at least one of the tracks is required to satisfy $|d_0| > 2.0~\textrm{mm}$.
The track $d_{0}$, $S(d_{0})$, and isolation variables have the highest importance as they contribute to the separation from low-\pt QCD tracks.

\begin{table}[tbp]
\centering
\caption{\label{tab:DT_NNinputs} Summary of input variables of (top) event-level NN and (bottom) track-level NN\@. Events may not contain more than four jets, following the requirement in Table~\ref{tab:evsel_DT}.}
\renewcommand{\arraystretch}{1.1}
\maxsizebox*{\textwidth}{\textheight}{
\begin{tabular}{lll}
\toprule
& Input variable & Description \\
\midrule
\vtlab{\large Event-level NN}{7}
& \met & Magnitude of the missing transverse momentum \\
& Leading jet \pt & \multirow{2}{*}{Kinematics of the leading jet} \\
& Leading jet $\eta$ & \\
& $N_{\textrm{jet}}^{30}$ & Number of jets with $\pt>30~\GeV$ \\
& $\min\Delta\phi(\ptmiss,\mathbf{j})$ & Minimum azimuthal angular difference between \ptmiss and all jets in the event \\
& $m_{\textrm{eff}}$ & Scalar sum of \pt of all jets with $\pt>30~\GeV$ and \met \\
& $H_{\textrm{T}}$ & Scalar sum of \pt of all jets $\pt>30~\GeV$ \\
\midrule
\vtlab{\large Track-level NN}{10}
& Track \pt & \multirow{2}{*}{Kinematics of the track} \\
& Track $\eta$ & \\
& Track $d_{0}$ & Transverse impact parameter \\
& Track $S(d_{0})$ & Significance of transverse impact parameter $|d_{0}/\sigma(d_{0})|$ \\
& Track $z_{0}\sin\theta$ & Longitudinal impact parameter \\
& Track $\pt^{\textrm{cone40}}$ & Scalar sum of \pt of tracks within $\Delta R<0.4$ (excluding itself) \\
& Track $\pt^{\textrm{cone100}}$ &  Scalar sum of \pt of tracks within $\Delta R<1.0$ (excluding itself) \\
& $\Delta\phi(\mathbf{trk},\ptmiss)$ & Azimuthal angular difference between the track and \ptmiss \\
& Tight Primary & Binary to indicate if the track satisfies the \textit{Tight Primary} track quality working point \\
& Secondary vertex matching & Binary to indicate if the track is matched to a secondary vertex \\
& Secondary vertex mass & Mass of secondary vertex reconstructed by associated tracks, $-1$~\GeV if unmatched \\
\bottomrule
\end{tabular}
}
\end{table}

The event-level and track-level NNs for the displaced track search are trained using signal and background MC events satisfying the selection criteria in Table~\ref{tab:evsel_DT}.
The signal samples used for the training of each NN span values of $m(\chinoonepm)$ from $85~\GeV$ to $225~\GeV$ and $\Delta m(\chinoonepm,\ninoone)$ from $0.25~\GeV$ to $1.5~\GeV$. The individual signal samples are weighted according to their production cross-sections during the training.
The simulated signal and background events are split into five batches, of which 60\%, 20\%, and 20\% are allocated to training, validation, and testing sets, respectively. The training is performed using a five-fold cross-validation configuration with early stopping to minimise overfitting. A total of approximately $2.0\times10^{5}$ and $1.0\times10^{5}$ events are used for signal and background samples, respectively. The training configurations of the event-level and track-level NNs are summarised in Table~\ref{tab:NN_config} in Appendix~\ref{sec:app_NN}.

Events satisfying the selection criteria in Table~\ref{tab:evsel_DT} are assigned an event score by the event-level NN. A track score is also assigned by calculating the output of the track-level NN for each track and selecting the track with the highest track-level NN score in the event. As listed in Table~\ref{tab:FCNN_4SR_Def_BCE}, a set of four mutually exclusive SRs are then defined in the event score vs track score 2D plane.
No veto on the presence of b-tagged jets is applied in the SRs, as such a requirement is not found to significantly reduce the SM background or increase the expected sensitivity.
The score requirements that define the four SRs are the result of an optimisation procedure that maximises the average expected significance among all signal MC samples that were used to train the NNs. The significance metric used for the optimisation is based on the profile likelihood method and described in Ref.~\cite{ATL-PHYS-PUB-2020-025}. As a result of the NN training and score optimisation procedures, the SRs provide good coverage across a range of mass splitting scenarios, though no individual SR is tuned to provide sensitivity to a specific mass splitting.

\begin{table}[tbp]
\centering
\caption{Definition of the displaced track SRs in the event score vs track score plane. Tighter SRs such as SR-DT-1 have higher sensitivity, while looser SRs such as SR-DT-4 capture additional signal with lower purity that did not fall in the SRs with lower index.}
\label{tab:FCNN_4SR_Def_BCE}
\renewcommand{\arraystretch}{1.1}
\begin{tabular}{llll}
\toprule
SR name & Event score & Track score & Additional requirement \\
\midrule
SR-DT-1 & $> 0.712$ & $> 0.788$ & -- \\
SR-DT-2 & $> 0.645$ & $> 0.799$ & $\notin\textrm{SR-DT-1}$ \\
SR-DT-3 & $> 0.645$ & $> 0.716$ & $\notin(\textrm{SR-DT-1}\cup\textrm{SR-DT-2})$\\
SR-DT-4 & $> 0.582$ & $> 0.705$ & $\notin(\textrm{SR-DT-1}\cup\textrm{SR-DT-2}\cup\textrm{SR-DT-3})$\\
\bottomrule
\end{tabular}
\end{table}


\subsection{1$\ell$1T event selection}
\label{sec:selection_1L1T}

All events used for the 1$\ell$1T search are selected by an \met trigger and are first required to satisfy the kinematic event selections listed in Table~\ref{tab:evsel_1L1T}. The reconstructed \met is required to be larger than 200~\GeV, corresponding to the lowest value at which the \met trigger is fully efficient and chosen to ensure sufficient data statistics. For signal events satisfying this high-\met requirement, the higgsino system typically recoils against hadronic activity, namely ISR jets. To explicitly select events with this topology, events are required to have at least one jet with $\pt>100~\GeV$. Moreover, events with one or more $b$-tagged jets with $\pt>20~\GeV$ are vetoed to suppress background contributions from \ttbar production.

\begin{table}[tbp]
\centering
\caption{\label{tab:evsel_1L1T} Kinematic event selections applied to all events used for the 1$\ell$1T search. The requirement on the number of tracks refers to the preselected tracks defined in Section~\ref{sec:lowPT_ID}.}
\renewcommand{\arraystretch}{1.1}
\begin{tabular}{lcc} \toprule
Variable & Electron channel & Muon channel \\
\midrule
Trigger & \multicolumn{2}{c}{\met} \\
\met[GeV] & \multicolumn{2}{c}{$> 200$} \\
Number of jets with $\pt>20~\GeV$ & \multicolumn{2}{c}{$\geq 1$} \\
Leading jet \pt[GeV] & \multicolumn{2}{c}{$> 100$} \\
Number of $b$-tagged jets ($N_{b\textrm{-jet}}^{20}$) & \multicolumn{2}{c}{$=0$} \\
Number of tight leptons ($N_{\textrm{tight-lep}}$) & $=1$ (electron) & $=1$ (muon) \\
Lepton \pt[GeV] & [4.5, 10] & [3, 10] \\
Lepton $|z_0\sin\theta|$ [mm] & \multicolumn{2}{c}{$<0.5$} \\
Lepton $\left|d_0/\sigma(d_0)\right|$ & $<5$ & $<3$ \\
Number of tracks & \multicolumn{2}{c}{$\geq 1$} \\
\bottomrule
\end{tabular}
\end{table}

Events are required to have exactly one tight lepton and are further categorised into two channels depending on the lepton flavour. The tight lepton requirement ensures that there is no overlap between the 1$\ell$1T analysis regions and the SRs of the displaced track search. The \pt of the lepton is required to be below 10~\GeV and above the lower bound of lepton identification methods, which are $4.5~\GeV$ and $3~\GeV$ for electrons and muons, respectively. As the decay length of \ninotwo is $\mathcal{O}(0.01\textrm{--}0.1~\mathrm{mm})$ for the targeted region, the leptons emerging from the $\ninotwo\rightarrow\ninoone\ell^+\ell^-$ decay are expected to be prompt. Thus, requirements on $z_{0}$ and $d_{0}$ are also imposed for the lepton. Finally, at least one additional track satisfying the criteria in Section~\ref{sec:lowPT_ID} other than the reconstructed lepton is required in each event, which correspond to lepton-track candidates also originating from \ninotwo for signal higgsino events.

After the kinematic event selections are applied, preselection criteria listed in Table~\ref{tab:trksel_1L1T} are applied to tracks to identify the candidates for the lepton-track. The charge of the track is required to be the opposite of that of the lepton, as they are expected to originate from $Z^{*}$ decay. The \pt and $|\eta|$ requirements on the track follow the requirements of the low-\pt lepton-track identification methods described in Section~\ref{sec:lowPT_ID}. The tracks are also required to be prompt with upper bounds on $|z_0\sin\theta|$ and $|d_0|$. The upper bound on $\pt^{\textrm{cone}30}$ requires there to be little activity surrounding the track, where the tight lepton is excluded from the determination of this variable in order to retain signal efficiency. The invariant mass $m(Z^{*})$ reconstructed from the lepton and the candidate lepton-track is required to be smaller than 5~\GeV and outside the range 3.0--3.2~\GeV, where the latter requirement vetoes $J/\psi$ mass resonances. Contributions from other SM resonances are found to be negligible in the selected range of $m(Z^{*})$. As the SUSY system is boosted, the lepton and the lepton-track have trajectories that are increasingly collinear. Thus, $\Delta R(\boldsymbol{\ell},\mathbf{trk})$ in $[0.1, 1.4]$ is required; the upper bound is set to 1.4 to veto most background tracks, while the lower bound is set to 0.1 to veto tracks from photon conversions nearby the lepton. Upper thresholds on the impact parameter differences between the lepton and the lepton-track are optimised to further suppress background tracks, as specified in Table~\ref{tab:trksel_1L1T}, where the requirements are looser for the electron channel to accommodate effects from bremsstrahlung.

\begin{table}[tbp]
\centering
\caption{\label{tab:trksel_1L1T} Track preselections applied to candidates of low-\pt lepton-tracks for the 1$\ell$1T search.}
\renewcommand{\arraystretch}{1.1}
\begin{tabular}{lcc} \toprule
Variable & Electron channel & Muon channel \\
\midrule
Track charge & \multicolumn{2}{c}{Opposite of lepton charge} \\
Track \pt[GeV] & $[0.5, 5]$ & $[1, 5]$\\
Track $|\eta|$ & $<2.5$ & $<1.6$ \\
Track $|z_0\sin\theta|$ [mm] & \multicolumn{2}{c}{$<0.5$} \\
Track $|d_{0}|$ [mm] & \multicolumn{2}{c}{$<0.5$} \\
Track $\pt^{\mathrm{cone30}}/\pt$ & \multicolumn{2}{c}{$<0.06$} \\
$m(Z^{*})$ [GeV] & \multicolumn{2}{c}{$<3$ or [3.2, 5]} \\
$\Delta R(\boldsymbol{\ell}, \mathbf{trk})$ & \multicolumn{2}{c}{$[0.1, 1.4]$} \\
$\Delta d_{0}(\ell, \mathrm{trk})$ [mm] & $<0.25$ & $<0.10$ \\
$\Delta z_{0}(\ell, \mathrm{trk})$ [mm] & $<0.30$ & $<0.26$ \\
\bottomrule
\end{tabular}
\end{table}

As the event topology of signal higgsino events depends sensitively on the mass splitting $\Delta m(\ninotwo,\ninoone)$, the main event selection of the 1$\ell$1T search is performed using a pNN. For very small $\Delta m(\ninotwo,\ninoone)$, the leptons emitted from the $\ninotwo\rightarrow\ninoone Z^{*}$ decay have very low \pt, and the two \ninoone tend to become more collimated, leading to higher \met values. In this case, variables such as $\Delta R(\boldsymbol{\ell}, \mathbf{trk})$ and $\Delta\phi(\ptmiss, \mathbf{Z^{*}})$ have very small values. For larger $\Delta m(\ninotwo,\ninoone)$, the leptons have higher \pt, leading to lower \met, larger $\Delta R(\boldsymbol{\ell}, \mathbf{trk})$, and larger $\Delta\phi(\ptmiss,\mathbf{Z^{*}})$. The pNN is introduced to consider these dependencies and to classify signal and background events in a dynamic manner. By choosing $\Delta m(\ninotwo,\ninoone)$ as the physics parameter $\theta$ (referred to as pNN-$\theta$), it is possible to realise $\Delta m(\ninotwo,\ninoone)$-dependent event selection with a single network model. This also simplifies the training procedure, as the pNN has the ability to smoothly interpolate between different $\Delta m(\ninotwo,\ninoone)$ values used for training, thus giving the possibility to explore various mass splitting scenarios with a limited number of training samples. Regions of the 1$\ell$1T search for a specific pNN-$\theta$ value are labelled with suffixes of `-dM\{pNN-$\theta$\}'.

Figure~\ref{fig:pnn_model_1L1T} shows the overall structure of the pNN model developed for the event selection used for both the electron and muon channels. It is implemented with \texttt{TensorFlow}~\cite{tensorflow}, using the \texttt{Keras}~\cite{chollet2015keras} frontend. Based on $\Delta m(\ninotwo,\ninoone)$ corresponding to the pNN-$\theta$ and 16 kinematic variables, the algorithm outputs a score characterizing how signal-like a given event is (referred to as pNN score) ranging from 0 to 1. The 16 kinematic variables are listed in Table~\ref{tab:pNNInputs}. Besides pNN-$\theta$, input variables such as lepton \pt, track \pt, \met, leading jet \pt, $m(Z^{*})$, $\Delta R(\boldsymbol{\ell}, \mathbf{trk})$, $H(\ell, \mathrm{trk})/\met$, $m_{\mathrm{T}}(\boldsymbol{\ell},\ptmiss)$, $\Delta \phi(\ptmiss, \mathbf{Z^{*}})$, and $\Delta \phi(\mathbf{j}_{1}, \mathbf{Z^{*}})$ have significant dependence on $\Delta m(\ninotwo,\ninoone)$, thus having high importance.\footnote{$H(\ell, \mathrm{trk})$ represents the scalar \pt sum of the lepton and the track. $m_{\mathrm{T}}(\boldsymbol{\ell},\ptmiss)$ represents the transverse mass reconstructed from the lepton and the missing transverse momentum and is defined as $\sqrt{2\pt^{\ell}\met\big\{1-\Delta\phi(\boldsymbol{\ell},\ptmiss)\big\}}$, where $\pt^{\ell}$ is the \pt of the lepton, and $\Delta\phi(\boldsymbol{\ell},\ptmiss)$ is the azimuthal angular difference between the lepton and the missing transverse momentum.} The topology differences between signal and background events tend to become more significant for smaller $\Delta m(\ninotwo,\ninoone)$ as the lepton, track, and \met are more likely to be in close proximity in the transverse plane for signal processes compared with the dominant $W^\pm+$jets background processes. For improved signal sensitivity, L1 and L2 regularisations~\cite{Ng:2004} are applied to each hidden layer to suppress the contributions of background events in high pNN-score regions.

\begin{figure}[tbp]
\centering
\includegraphics[width=0.95\textwidth]{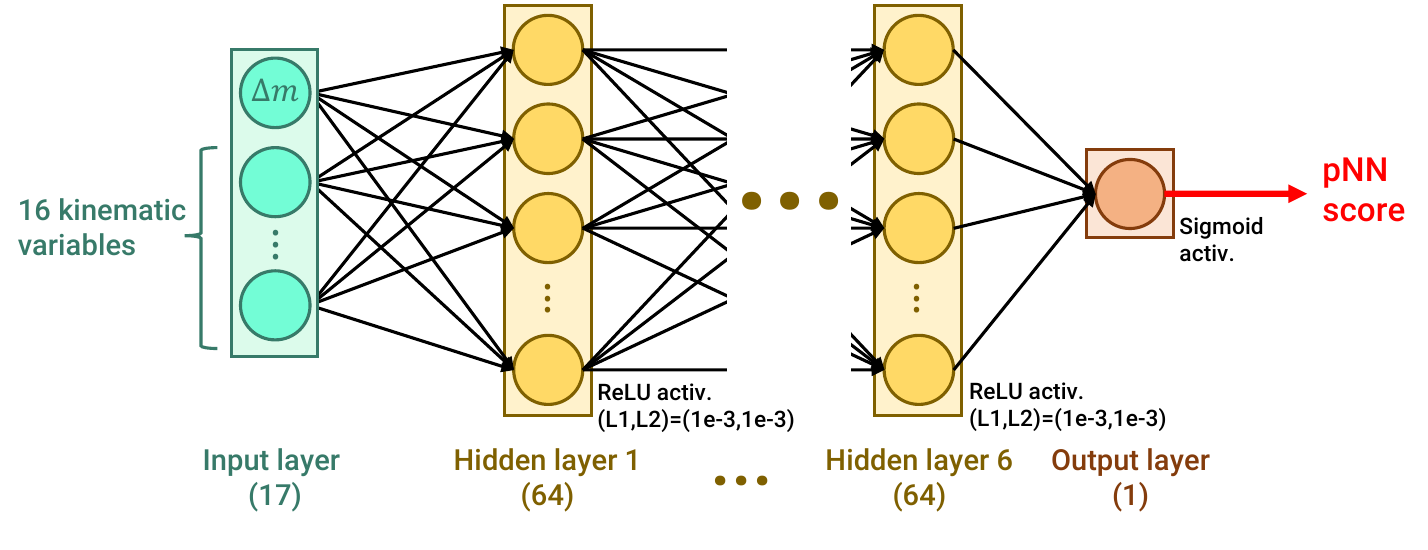}
\caption[Overall structure of the pNN model used for event selection]{Overall structure of the pNN model used for event selection. The $\Delta m(\ninotwo,\ninoone)$ and 16 kinematic variables are used as inputs, and there are six hidden layers with 64 nodes each. The nodes are fully connected. L1 and L2 regularisations are applied to each layer for increased signal sensitivity. The output pNN score ranges from 0 to 1, with 1 meaning that the input event is signal event-like.}
\label{fig:pnn_model_1L1T}
\end{figure}

\begin{table}[tbp]
\fontsize{10.5pt}{12.5pt}\selectfont
\centering
\caption{\label{tab:pNNInputs} Summary of input variables of the pNN\@. The same set of variables is used for the electron and muon channels.}
\renewcommand{\arraystretch}{1.1}
\begin{tabular}{ll}
\toprule
Input variable & Description \\
\midrule
$\Delta m(\tilde{\chi}^{0}_{2},\tilde{\chi}^{0}_{1})$ & pNN-$\theta$, mass difference between $\tilde{\chi}^{0}_{2}$ and $\tilde{\chi}^{0}_{1}$ \\
Lepton \pt, $\eta$, $\phi$ & Kinematics of the standard lepton \\
Track \pt, $\eta$, $\phi$ & Kinematics of the track \\
\met, $\phi(\ptmiss)$ & Kinematics of the missing transverse momentum \\
Leading jet \pt, $\phi$ & Kinematics of the leading jet \\
$m(Z^{*})$ & Mass of $Z^{*}$ reconstructed from the lepton and the track \\
$\Delta R(\boldsymbol{\ell}, \mathbf{trk})$ & Distance between the lepton and the track \\
$H(\ell, \mathrm{trk})/\met$ & Ratio of scalar sum of lepton \pt and track \pt to \met \\
$m_{\mathrm{T}}(\boldsymbol{\ell},\ptmiss)$ & Transverse mass reconstructed from the lepton and the missing transverse momentum \\
$\Delta \phi(\ptmiss, \mathbf{Z^{*}})$ & Azimuthal angular difference between the missing transverse momentum and $Z^{*}$ \\
$\Delta \phi(\mathbf{j}_{1}, \mathbf{Z^{*}})$ & Azimuthal angular difference between the leading jet and $Z^{*}$ \\
\bottomrule
\end{tabular}
\end{table}

Since the kinematic requirements for leptons and tracks differ for the electron and muon channels, the training of the pNN is done separately for each channel using the same pNN model structure. To train the pNN, each of the inputs are standardised into a Gaussian distribution. Higgsino production and $W^\pm+$jets MC events are used as signal and background training samples, respectively. To ensure enough statistics, higgsino production MC events are used inclusively in terms of mass points and production channels. The pNN is trained with pNN-$\theta=$ 1, 1.5, 2, 3, 4, and $5~\GeV$ and the same values are used to obtain six non-orthogonal search results for each of the lepton flavours. The pNN-$\theta$ of background training samples are assigned randomly so that their distribution matches that of the signal training samples. 70\% of the simulated signal and background events are used for training and validation of the network, and the remaining 30\% are used for testing. All training samples are required to satisfy the event and track preselections described in Table~\ref{tab:evsel_1L1T} and Table~\ref{tab:trksel_1L1T}. For the electron channel, about $2.2\times10^{4}$ signal events and $6.3\times10^{5}$ background events are used for the training and its validation, where 20\% are assigned as validation samples. For the muon channel, about $5.1\times10^{4}$ signal events and $3.8\times10^{5}$ background events are used for the training and its validation, where 20\% are assigned as validation samples. The training configurations of the pNN, common to both electron and muon channels, are summarised in Table~\ref{tab:NN_config} in Appendix~\ref{sec:app_NN}. Figure~\ref{fig:pNN_roc} shows the performance of the pNN for each pNN-$\theta$ as a receiver operating characteristic (ROC) curve. The separation of signal and background events is stronger for smaller $\Delta m(\ninotwo,\ninoone)$ values due to the mentioned topology differences.

\begin{figure}[tbp]
\centering
\subfloat[]{
\includegraphics[width=0.495\textwidth]{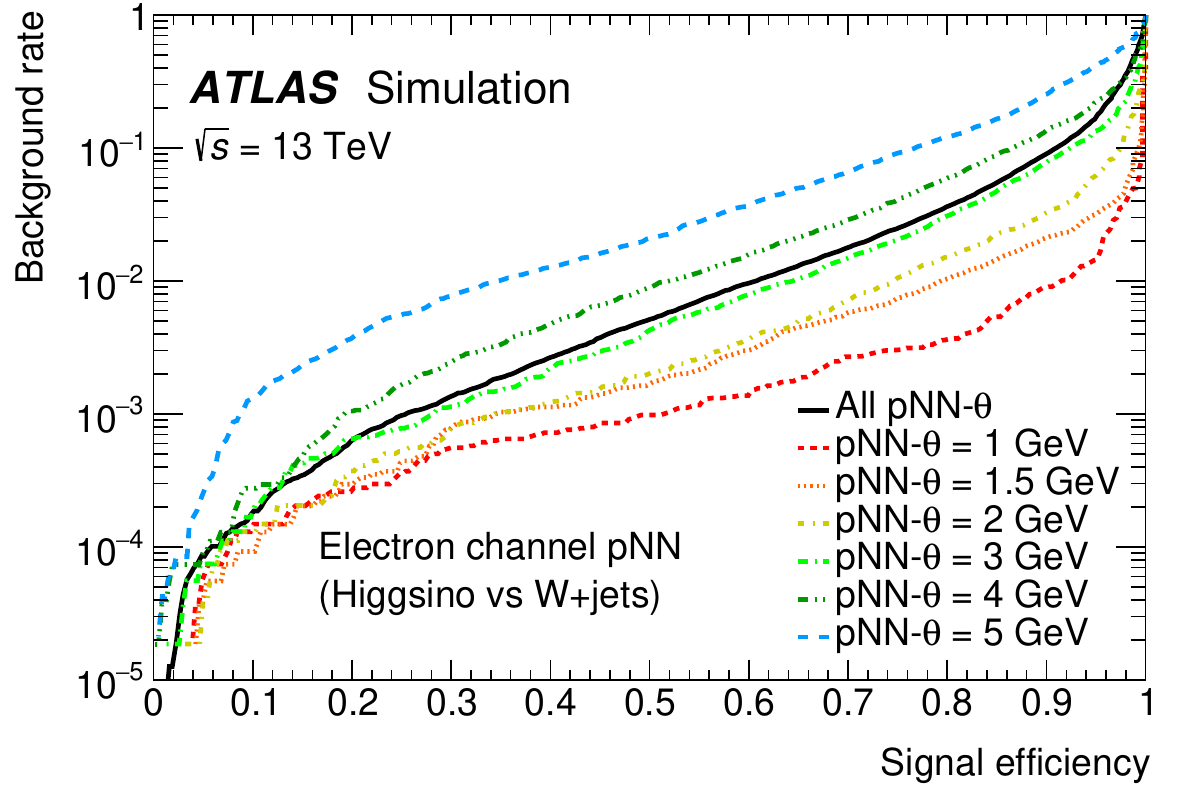}
}
\subfloat[]{
\includegraphics[width=0.495\textwidth]{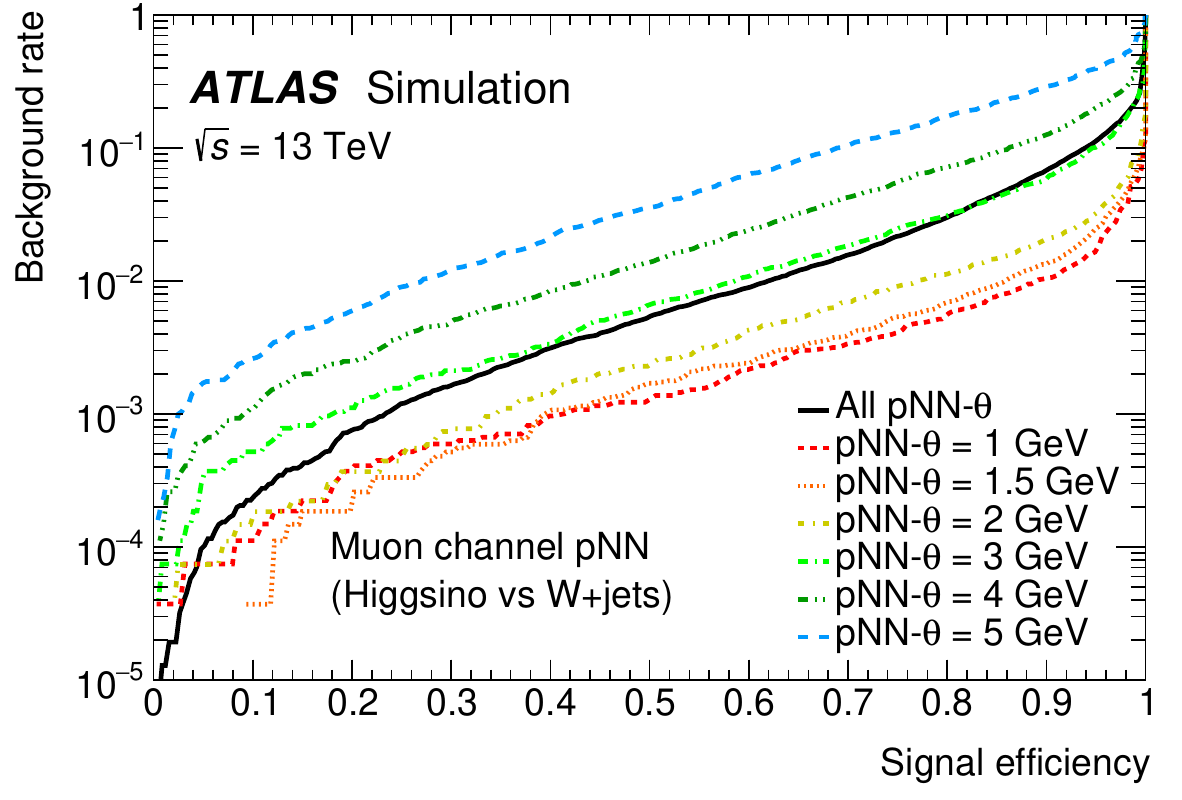}
}
\caption{\label{fig:pNN_roc} ROC curves of pNN of each pNN-$\theta$ for (a) the electron channel and (b) the muon channel. For each curve, only signal events with $\Delta m(\ninotwo,\ninoone)$ matching the pNN-$\theta$ are considered. The performance is better for smaller $\Delta m(\ninotwo,\ninoone)$ values for both channels. The steps in the curves are due to the limited statistical precision of the MC samples used.}
\end{figure}

After the application of the lepton-track preselection in Table~\ref{tab:trksel_1L1T}, the pNN score is calculated for each candidate track. The track with the highest pNN score is then selected for each event and its score is assigned as the pNN score of the event. Subsequently, selection criteria in Table~\ref{tab:SRselection_1L1T} are applied to select signal-like events with a same-flavour opposite-charge lepton and lepton-track pair to define the SR\@. As the pNN and low-\pt lepton-track ID are not directly dependent on one another, the application of the low-\pt lepton-track ID on top of the pNN score requirement significantly reduces the background contribution. The first requirement of high pNN score vetoes background events having non-signal-like event topology. Then by selecting events having a track with a large value of the low-\pt lepton-track ID score, further significant background suppression is achieved through rejection of hadronic and fake tracks, which are the main background track components in the targeted regime.

\begin{table}[tbp]
\centering
\caption{\label{tab:SRselection_1L1T} Summary of the selection criteria of the SR.}
\renewcommand{\arraystretch}{1.1}
\begin{tabular}{lcc}
\toprule
Variable & SR-1$e$1T & SR-1$\mu$1T \\
\midrule
$\Delta\phi(\ptmiss, \mathbf{Z^{*}})$ & \multicolumn{2}{c}{$<1.0$} \\
pNN score & \multicolumn{2}{c}{$>0.95$} \\
Low-\pt lepton-track ID score & $>0.8$ & $>0.6$ \\
\bottomrule
\end{tabular}
\end{table}



%

%
\section{Background estimation}
\label{sec:background}

Following the event selection criteria, SM backgrounds for the displaced track and 1$\ell$1T searches consist of events with a high-\pt jet, high \met, and low-\pt tracks.
This section presents the background estimate methods for both searches, each exploiting semi-data-driven methods to estimate contributions from low-\pt tracks for such events. The control and validation regions used in these methods are found to have negligible contributions from potential signal processes.

For the background estimation and the interpretation of final results, three configurations of fits to the data are used. The first configuration, referred to as the background-only fit, does not include any contributions from the signal models and uses only the control regions to extract the necessary normalisation and reweighting factors. Its results are then extrapolated to the validation and signal regions to compare the post-fit SM predictions with the observed data without allowing the observed data in those regions to bias the fitted background event yields. The second configuration, referred to as the discovery fit, includes the control regions and one target signal region to derive a $p$-value representing the compatibility of the observed data with the background-only hypothesis. The third configuration uses both the control and signal regions to determine the compatibility of the observed data with the target signal model. In cases where there is no significant excess in data, this configuration is often referred to as the exclusion fit.

\subsection{Displaced track background estimation}
\label{sec:background_DT}

The leading SM background contributions in the SRs for the displaced track search originate from processes that result in high \met from neutrinos, such as $Z(\to\nu\bar{\nu})$+jets and $W^\pm(\to\tau^\pm\nu_\tau)$+jets. Subleading contributions arise from diboson production and $W^\pm(\to\ell^\pm\nu_\ell)$+jets. Studies using the background MC samples show that approximately 40\% of the signal candidate tracks selected by the track-level NN in the SRs arise from pile-up interactions. These backgrounds are estimated by using a semi-data-driven approach that combines simulation-based predictions with reweighting corrections derived in control regions (CRs). Validation regions (VRs) orthogonal to the SRs are used to validate these corrections. Other processes such as top quark production and multijet processes are found to have negligible impact in the SRs and are estimated directly from MC simulation.

Each process of the leading and subleading background contributions is normalised by a corresponding normalisation factor obtained by simultaneously fitting kinematic distributions in the mutually orthogonal CRs listed in Table~\ref{tab:CR_DT_list}, defined to enhance each of the processes. No requirement is made on the track-level NN score in these CRs. Normalisation factors for $W^\pm(\to\ell^\pm\nu_\ell)$+jets and $W^\pm(\to\tau^\pm_{\textrm{lep}}\nu_\tau)$+jets in which the $\tau$-lepton decays leptonically are constrained by fitting the distribution of the transverse mass $m_{\textrm{T}}$ in CRs requiring exactly one tight lepton, defined as CR-DT-1$\ell$ ($\ell=e,\mu$). As $m_{\textrm{T}}$ tends to be smaller for $W^\pm(\to\tau^\pm_{\textrm{lep}}\nu_\tau)$+jets processes, CR-DT-1$\ell$ regions are separated into two bins with $m_{\textrm{T}}$ in $[20, 40)$ and $[40, 120)~\GeV$ to simultaneously constrain the corresponding factors. The normalisation factors obtained for $W^\pm(\to\ell^\pm\nu_\ell)$+jets and $W^\pm(\to\tau^\pm_{\textrm{lep}}\nu_\tau)$+jets are $1.20\pm0.01$ and $1.2\pm0.1$, respectively.

\begin{table}[tbp]
\centering
\caption{ \label{tab:CR_DT_list} Requirements of the CRs used in the displaced track search that are mutually orthogonal due to the lepton requirements. There are no overlapping events between the CRs and the SRs of the displaced track search. \metnolep denotes \met with the leptons treated as invisible.}
\renewcommand{\arraystretch}{1.1}
\maxsizebox*{\textwidth}{\textheight}{
\begin{tabular}{llllll}
\toprule
CR name & Trigger & $N_{\textrm{tight-lep}}$ & $N_{b\textrm{-jet}}^{20}$ & \met[GeV] & \metnolep[GeV] \\
\midrule
CR-DT-1$\ell$ & Single-lepton & 1$e$/1$\mu$ & $=0$ & -- & $>200$ \\
CR-DT-2$\ell$ & Single-lepton & 2$e$/2$\mu$ & $=0$ & -- & $>200$ \\
CR-DT-1$e$1$\mu$ & Single-lepton & 1$e$+1$\mu$ & $=0$ & -- & $>200$ \\
CR-DT-lowMET & \met & $=0$ & -- & [275, 325] & -- \\
\bottomrule
\end{tabular}
}
\end{table}

Normalisation factors for $Z(\to\nu\bar{\nu})$+jets and $Z(\to\ell^+\ell^-)$+jets processes are presumed to be the same and are constrained by fitting the dilepton mass in CRs with two same-flavour tight leptons, defined as CR-DT-2$\ell$.
While diboson processes make subleading contributions to CR-DT-2$\ell$, they are primarily constrained by the simultaneous fit in a CR that requires two different-flavour tight leptons, defined as CR-DT-1$e$1$\mu$.
The leptonic CRs CR-DT-1$\ell$, CR-DT-2$\ell$, and CR-DT-1$e$1$\mu$ are triggered with single-lepton triggers and require high \met with the leptons treated as invisible particles, denoted by \metnolep, which helps in keeping the event topology similar to the targeted process.
The $\pt$ threshold of the single-electron (single-muon) triggers varied between 24~\GeV and 26~\GeV (20~\GeV and 26~\GeV) during the 2015--2018 data-taking period.
The offline \pt of the electrons and muons are required to be above these trigger thresholds in CR-DT-1$\ell$ to ensure that the trigger efficiencies are in their plateau regions. This requirement also enforces orthogonality between CR-DT-1$\ell$ and the SRs of the $1\ell1$T search, which require the lepton $\pt$ to be below $10~\GeV$.
The number of $b$-jets is required to be zero to suppress contributions from top quark processes.
The normalisation factors obtained for $Z$+jets and diboson processes are $1.29\pm0.01$ and $1.4\pm0.2$, respectively.

Finally, CR-DT-lowMET provides constraints on the normalisation factor for $W^\pm(\to\tau^\pm_{\textrm{had}}\nu_\tau)$+jets processes in which the $\tau$-lepton decays hadronically. In this region, \met is required to be in $[275, 325]~\GeV$ and the distribution of $\min\Delta\phi(\ptmiss, \mathbf{j})$, defined as the minimum azimuthal separation between the missing transverse momentum and any jet, is used for the fit. The normalisation factor obtained for $W^\pm(\to\tau^\pm_{\textrm{had}}\nu_\tau)$+jets processes is $1.37\pm0.04$.

In addition to the event-level normalisation, track-level reweighting is applied to correct for any mismodelling in the kinematics of the tracks. The track-level reweighting consists of three sets of factors which are also obtained by a simultaneous fit later described in Section~\ref{sec:results_CRVR}. The first set of track reweighting factors is applied to consider the modelling differences between tracks originating from primary vertices and those that do not, dominantly tracks from pile-up. The reweighting factors are process-independent and are determined using CR-DT-1$\mu$, having enough events. CR-DT-1$\mu$ is categorised into two regions by $|z_{0}\sin\theta|$: a region with $|z_{0}\sin\theta|<0.1~\textrm{mm}$ dominated by tracks associated with the primary collision and a region with $|z_{0}\sin\theta|>0.5~\textrm{mm}$ dominated by pile-up tracks. The factors are binned in track \pt and are only applied to tracks with $\pt<3.2~\GeV$, where such mismodelling is prominent. The obtained first set of reweighting factors ranges in $1\textrm{--}1.4$ with a relative uncertainty below 4\%. The validity of these factors for different processes is confirmed in other CRs, such as CR-DT-1$e$ and CR-DT-lowMET.

The second set of reweighting factors corrects for the mismodelling of the track isolation variable, resulting in fewer isolated tracks in data. Similarly to the first set, this set of factors is obtained in CR-DT-1$\mu$ categorised into two regions using the same $|z_{0}\sin\theta|$ requirements and is applied separately. The factors are binned in $\pt^{\textrm{cone40}}$ and their obtained values are in the range of $0.8\textrm{--}1.1$ with a relative uncertainty below 3\%. The validity of these factors has also been confirmed in other CRs to ensure process independence.

The third reweighting factor corrects for the mismodelling of leptonic or hadronic tracks originating from $\tau$-lepton decays under the high-\met preselection with a lower bound of $275~\GeV$, which is not applied in the first and second sets of factors. A single normalisation factor is obtained in CR-DT-lowMET with additional `$\tau$-track' criteria as defined in Table~\ref{tab:DT_tautrack} to enhance the contributions from the targeted $W^\pm(\to\tau^\pm_{\textrm{had}}\nu_\tau)$+jets processes. The obtained third reweighting factor is $0.64$ with a relative uncertainty below 5\%. This factor is applied only to simulated tracks originating from $\tau$-leptons.

\begin{table}[tbp]
\centering
\caption{\label{tab:DT_tautrack} Summary of the $\tau$-track criteria applied to obtain the track reweighting factor in CR-DT-lowMET.}
\renewcommand{\arraystretch}{1.1}
\begin{tabular}{ll}
\toprule
Variable & Requirement \\
\midrule
Track \pt[GeV] & $>2.5$ \\
Track $|\eta|$ & $<1.5$ \\
Track $|d_{0}|$ [mm] & $<10$ \\
Track $z_{0}\sin\theta$ [mm]  & $<1.5$ \\
Track $\pt^{\textrm{cone40}}$ [GeV] & $<1.0$ \\
$\Delta\phi(\mathbf{trk},\ptmissnolep)$ & $\leq 1.0$ \\
Lepton veto & Not matched to a lepton-associated track \\
\bottomrule
\end{tabular}
\end{table}

As shown in Table~\ref{tab:VR_DT_list}, four VR categories are constructed to validate different aspects of the background estimation method for the displaced track search. Figure~\ref{fig:VR_DT_schema} illustrates how the different VRs map onto the event-level NN score vs track-level NN score plane. Each of the categories is further divided into three or four regions, depending on their event-level and track-level NN scores. VR-DT-$\mu$ uses the single-muon trigger and all other VRs use the \met trigger.
While some VRs may overlap with CRs by definition, they represent only a small fraction of the event yields in these CRs and therefore still provide a means of validating the background estimation strategy.

\begin{table}[tbp]
\centering
\caption{ \label{tab:VR_DT_list} Definitions of the four VR categories used in the displaced track search.}
\renewcommand{\arraystretch}{1.1}
\maxsizebox*{\textwidth}{\textheight}{
\begin{tabular}{llllll}
\toprule
VR name & $N_{\textrm{pre-lep}}$ & Event-level NN score & Track-level NN score & Additional requirement \\
\midrule
VR-DT-Evt-1 & $=0$ & $[0.712,1]$ & $<0.705$ & -- \\
VR-DT-Evt-2 & $=0$ & $[0.645,0.712)$ & $<0.705$ & -- \\
VR-DT-Evt-3 & $=0$ & $[0.582,0.645)$ & $<0.705$ & -- \\
VR-DT-Trk-1 & $=0$ & $<0.582$ & $[0.799,1]$ & $\met\in[325,500]~\GeV$ \\
VR-DT-Trk-2 & $=0$ & $<0.582$ & $[0.788,0.799)$ & $\met\in[325,500]~\GeV$ \\
VR-DT-Trk-3 & $=0$ & $<0.582$ & $[0.716,0.788)$ & $\met\in[325,500]~\GeV$ \\
VR-DT-Trk-4 & $=0$ & $<0.582$ & $[0.705,0.716)$ & $\met\in[325,500]~\GeV$ \\
VR-DT-$\mu$-1 & $=1$ (tight muon) & \multicolumn{2}{c}{Same as SR-DT-1} & Single-muon trigger,  \metnolep used \\
VR-DT-$\mu$-2 & $=1$ (tight muon) & \multicolumn{2}{c}{Same as SR-DT-2} & Single-muon trigger,  \metnolep used \\
VR-DT-$\mu$-3 & $=1$ (tight muon) & \multicolumn{2}{c}{Same as SR-DT-3} & Single-muon trigger,  \metnolep used \\
VR-DT-$\mu$-4 & $=1$ (tight muon) & \multicolumn{2}{c}{Same as SR-DT-4} & Single-muon trigger,  \metnolep used \\
VR-DT-$\tau$-1 & $=0$ & $[0.712,1]$ & -- & $\tau$-track criteria with track \pt>8~\GeV \\
VR-DT-$\tau$-2 & $=0$ & $[0.645,0.712)$ & -- & $\tau$-track criteria with track \pt>8~\GeV \\
VR-DT-$\tau$-3 & $=0$ & $[0.582,0.645)$ & -- & $\tau$-track criteria with track \pt>8~\GeV \\
\bottomrule
\end{tabular}
}
\end{table}

\begin{figure}[tbp]
\centering
\includegraphics[width=0.83\linewidth]{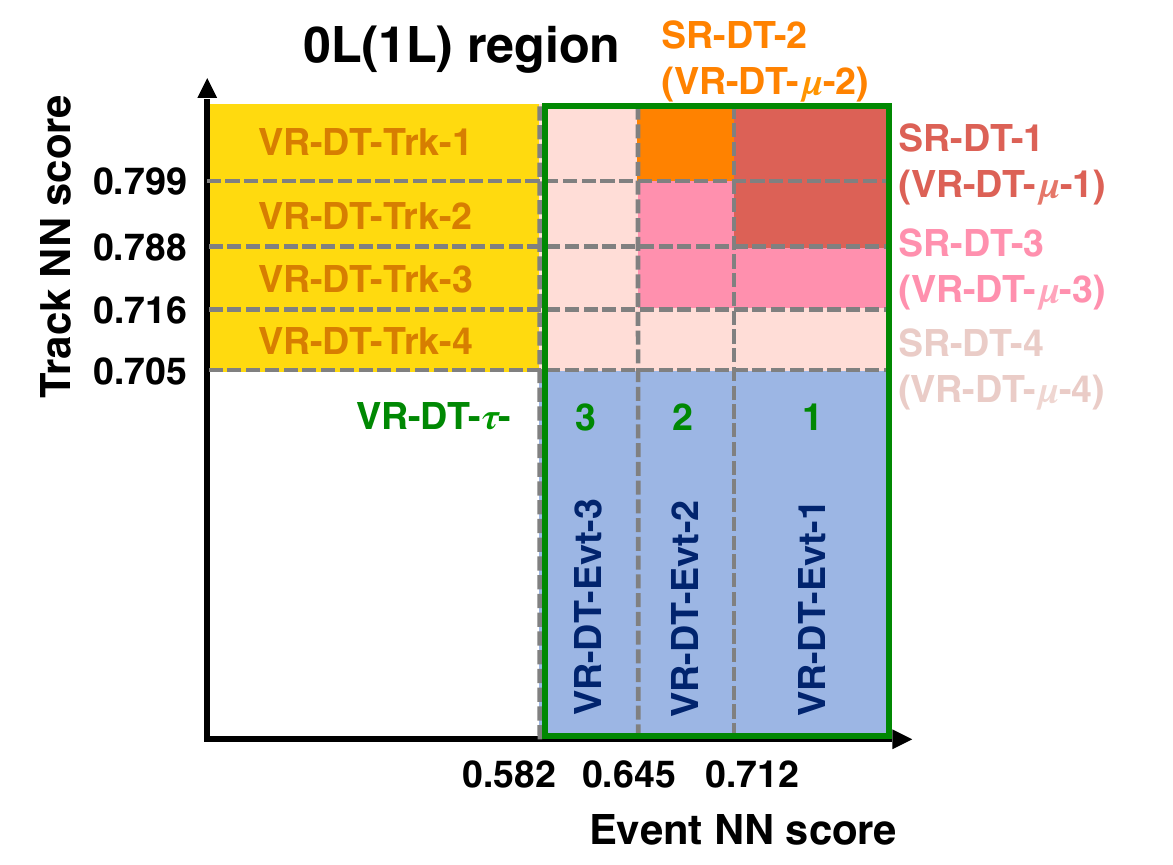}
\caption{Illustration of the four validation region categories defined in the event- vs. track-level NN score plane. Each VR is further divided into three or four slices defined by the event and track NN scores. VR-DT-$\mu$ regions have the same score requirements as the SR-DT regions but with the 1L requirement and are denoted in parentheses.}
\label{fig:VR_DT_schema}
\end{figure}

VR-DT-Evt is defined to have high event-level NN scores, but vetoes events having tracks with high track-level NN scores; it verifies the validity of the event-level NN model and the event-level normalisation factor for each background process. On the other hand, VR-DT-Trk is defined to have a high track-level NN score, but vetoes events with high event-level NN score. It is further required to have moderate \met in the range of $325\textrm{--}500~\GeV$ to ensure sufficiently low signal contamination and retain orthogonality with CR-DT-lowMET. VR-DT-Trk helps in verifying the validity of track-level NN model and the track-level corrections using reweighting factors.

To validate the estimation method in a region where the event-level and track-level networks both yield high scores, VR-DT-$\mu$ is defined by requiring one tight muon. For this region, the single-muon trigger set is used and \met is substituted by \metnolep to mimic the kinematics in the SRs excluding leptons. All of the event-level and track-level selections for this region are performed with this substitution. Furthermore, VR-DT-$\tau$ is defined to validate the modelling of $W^\pm(\to\tau^\pm\nu_\tau)$+jets processes in regions with high event-level NN scores. In VR-DT-$\tau$, tracks are required to satisfy the $\tau$-track criteria used for CR-DT-lowMET but with track $\pt>8~\GeV$. Compared to the signal processes of interest, the track \pt tends to be larger in $W^\pm(\to\tau^\pm\nu_\tau)$+jets events and this track \pt requirement is found to suppress signal contamination while retaining sufficient statistics for validating the modelling of the $W^\pm(\to\tau^\pm\nu_\tau)$+jets background.


%

\subsection{1$\ell$1T background estimation}
\label{sec:background_1L1T}

SM background processes with a high-\pt jet and a boosted system with neutrinos in the final state could satisfy the leading jet and \met selection of the 1$\ell$1T search. The identified lepton is usually a real lepton, whereas the tracks identified with the low-\pt lepton-track identification methods have significant contributions from fake leptons. Since these tracks are required to be prompt, most of them are expected to correspond to low-\pt hadrons originating from the underlying event. From these object-based considerations, the most dominant SM background process is $W^\pm+$jets with a leptonic $W^\pm$ decay. In addition to these dominant backgrounds, there are small contributions from events in which the low-\pt lepton-track corresponds to a real lepton. Such backgrounds are contributions from processes that feature two or more leptons in their final states, including $Z^{(*)}+$jets and diboson. Thus, the SM backgrounds for the 1$\ell$1T search are categorised into two components: `fake lepton-track' backgrounds from events in which the low-\pt lepton-track corresponds to a fake lepton, and `true lepton-track' backgrounds in which the low-\pt lepton-track corresponds to a real lepton. The dominant fake lepton-track backgrounds are estimated by a data-driven method described below, whereas the small true lepton-track backgrounds are estimated directly using MC simulation.

As MC simulation is not expected to accurately model processes with fake lepton-tracks, a data-driven method is used to estimate the contributions of fake lepton-track backgrounds. The method is performed by defining control regions, subtracting the true lepton-track contributions from each, and obtaining the ratio of yields of the fake lepton-track contributions between those regions. Regions are defined by applying the event and object criteria in Tables~\ref{tab:evsel_1L1T} and~\ref{tab:trksel_1L1T} and imposing requirements on the pNN score and $\Delta\phi(\ptmiss, \mathbf{Z^{*}})$ as illustrated in Figure~\ref{fig:kine_plane}.
The $\alpha$-region is defined by pNN score $>0.95$ and $\Delta\phi(\ptmiss, \mathbf{Z^{*}})<1.0$, and the $\gamma$-region is defined by pNN score $<0.4$. Both of these are further categorised into `pass' and `fail' regions following Figure~\ref{fig:bkgEstSchematic}, based on whether the low-\pt lepton candidate track passes or fails the ID-score requirements in Table~\ref{tab:SRselection_1L1T}. Therefore, the pass region within the $\alpha$-region corresponds to the SR\@. Within the $\gamma$-region, dominated by SM backgrounds, the pass and fail regions are denoted by CR-$1\ell$1T-A and CR-$1\ell$1T-B, respectively. `Transfer factors (TFs)' are measured by taking the ratio of the number of events belonging to CR-$1\ell$1T-A and that of CR-$1\ell$1T-B, separately for each of the track \pt bins in [0.5, 1.0), [1.0, 1.5), [1.5, 2.0), [2.0, 3.0), and [3.0, 5.0]~\GeV for the electron channel and [1.0, 1.5), [1.5, 2.0), [2.0, 3.0), and [3.0, 5.0]~\GeV for the muon channel.
The measured TFs are then applied to the number of events in the fail region within the $\alpha$-region, defined as CR-$1\ell$1T-C. This \pt-binned method accounts for the effects of indirect correlations between the low-\pt lepton-track ID and the event selection pNN through the track kinematics and improves the estimate of the pNN-score distribution in the SR\@. The TF values range in $\mathcal{O}(10^{-4}\textrm{--}10^{-2})$, typically having larger values in higher track \pt bins.

\begin{figure}[tbp]
\centering
\includegraphics[width=0.6\textwidth]{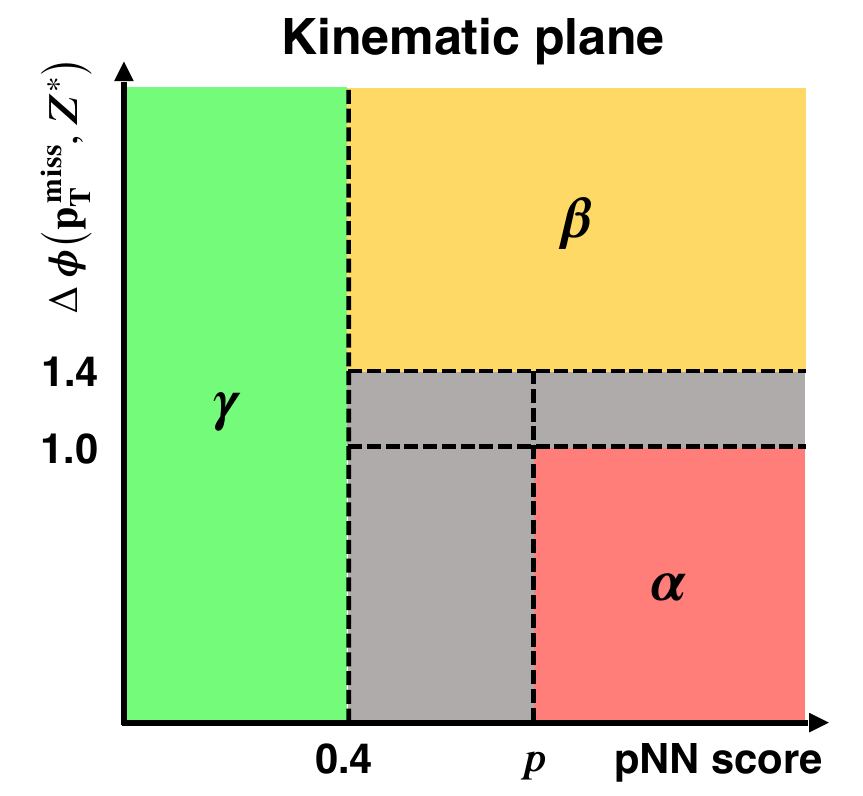}
\caption{\label{fig:kine_plane} Kinematic plane diagram used for region definitions of the 1$\ell$1T search. In terms of kinematic variables, the SRs, CRs, and VRs are defined by imposing requirements on the pNN score and $\Delta\phi(\ptmiss, \mathbf{Z^{*}})$. The value $p$ is either 0.95 for the SRs and their corresponding CRs or 0.4 for the VRs and their corresponding CRs.}
\end{figure}

\begin{figure}[tbp]
\centering
\includegraphics[width=0.75\textwidth]{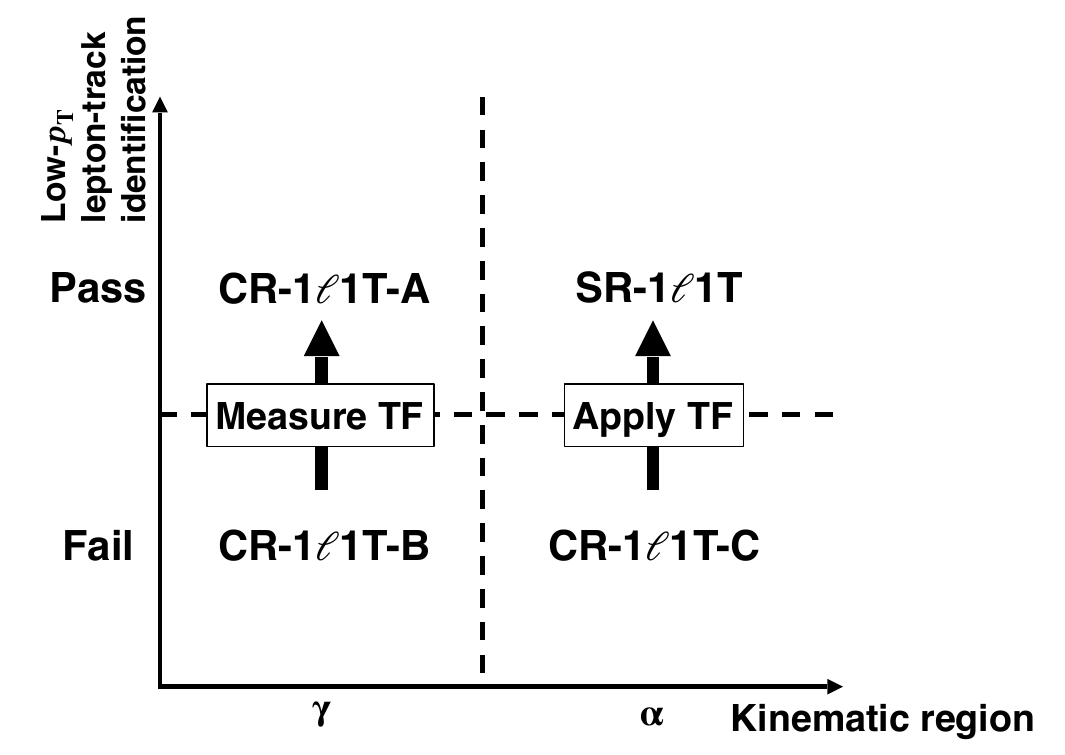}
\caption{\label{fig:bkgEstSchematic} Schematic diagram of the background estimation method for the $1\ell$1T search. CR-1$\ell$1T-A and CR-1$\ell$1T-B are used to measure the transfer factors (pass-to-fail ratio) of the low-\pt lepton-track identification. The transfer factors are binned in track \pt to correctly estimate the pNN-score shape in the SR\@. The measured transfer factors are then applied to CR-1$\ell$1T-C to estimate the background in SR-1$\ell$1T.}
\end{figure}

Three VRs are defined to validate the background estimation method of the 1$\ell$1T search. VR-1$\ell$1T-highLarge(hL) is defined as the low-\pt lepton-track ID pass region within the $\beta$-region in Figure~\ref{fig:kine_plane} with pNN score higher than 0.4 and $\Delta\phi(\ptmiss, \mathbf{Z^{*}})$ larger than 1.4, to test the validity of applying the TFs measured in low pNN-score regions to high pNN-score regions. The TFs used to estimate the background in the SR are applied to the fail region within $\beta$-region, labelled as CR-1$\ell$1T-hL, to estimate the yield in VR-1$\ell$1T-hL.

VR-1$\ell$1T-DF and VR-1$\ell$1T-SS are used to validate the overall estimation method. VR-1$\ell$1T-DF is defined by inverting the flavour requirement of the tight lepton (e.g. a muon and a low-\pt electron track are required for VR-1$e$1T-DF). To avoid possible overlaps with the SRs, the track is explicitly required to fail the `other' low-\pt lepton-track ID criterion indicated in Table~\ref{tab:SRselection_1L1T} (e.g. the track in VR-1$e$1T-DF is required to have low-\pt muon-track ID score below 0.6). VR-1$\ell$1T-SS is defined by requiring the same charge between the lepton and the low-\pt lepton-track. Except for the loosened pNN-score requirement of $>0.4$ to ensure sufficient statistics, kinematic requirements are the same as those for the SR\@. For these two VRs, the yields are obtained with dedicated CRs and TFs, similar to the method illustrated in Figure~\ref{fig:bkgEstSchematic}. All of the SR, CRs, and VRs used in the 1$\ell$1T search are listed in Table~\ref{tab:CRVR_list_1L1T}.

\begin{table}[tbp]
\centering
\caption{\label{tab:CRVR_list_1L1T} List of SR, CRs, and VRs used in the 1$\ell$1T search. OS/SS stands for opposite/same sign, and SF/DF stands for same/different flavour, of the lepton and the lepton-track. The pNN-score thresholds for $\alpha$-regions for SS-SF and OS-DF are lowered to 0.4 to increase event yields. For DF regions, as described in the text, there are additional requirements on the low-\pt lepton-track ID scores to avoid possible region overlaps with the SR in the other channel with different lepton flavour.}
\renewcommand{\arraystretch}{1.1}
\maxsizebox*{\textwidth}{\textheight}{
\begin{tabular}{llllllll}
\toprule
\multirow{2}{*}{Region name} & \multirow{2}{*}{Sign} & \multirow{2}{*}{Flavour} & \multirow{2}{*}{pNN score} & \multirow{2}{*}{$\Delta\phi(\ptmiss, \mathbf{Z^{*}})$} &  \multirow{2}{*}{Region type} & Low-\pt lepton- & \multirow{2}{*}{Description} \\
& & & & & & track ID & \\
\midrule
SR-1$\ell$1T & OS & SF & $>0.95$ & $<1.0$ & $\alpha$ & Pass & SR with OS-SF \\
\midrule
CR-1$\ell$1T-A & OS & SF & $<0.4$ & -- & $\gamma$ & Pass & TF measurement numerator \\
CR-1$\ell$1T-B & OS & SF & $<0.4$ & -- & $\gamma$ & Fail & TF measurement denominator \\
CR-1$\ell$1T-C & OS & SF & $>0.95$ & $<1.0$ & $\alpha$ & Fail & TF application CR for SR \\
CR-1$\ell$1T-hL & OS & SF & $>0.4$ & $>1.4$ & $\beta$ & Fail & TF application CR for VR-1$\ell$1T-hL \\
CR-1$\ell$1T-SS-A & SS & SF & $<0.4$ & -- & $\gamma$ & Pass & TF-SS measurement numerator \\
CR-1$\ell$1T-SS-B & SS & SF & $<0.4$ & -- & $\gamma$ & Fail & TF-SS measurement denominator \\
CR-1$\ell$1T-SS-C & SS & SF & $>0.4$ & $<1.0$ & $\alpha$ & Fail & TF-SS application CR for VR-SS \\
CR-1$\ell$1T-DF-A & OS & DF & $<0.4$ & -- & $\gamma$ & Pass & TF-DF measurement numerator \\
CR-1$\ell$1T-DF-B & OS & DF & $<0.4$ & -- & $\gamma$ & Fail & TF-DF measurement denominator \\
CR-1$\ell$1T-DF-C & OS & DF & $>0.4$ & $<1.0$ & $\alpha$ & Fail & TF-DF application CR for VR-DF \\
\midrule
VR-1$\ell$1T-hL & OS & SF & $>0.4$ & $>1.4$ & $\beta$ & Pass & VR within OS-SF with large $\Delta\phi(\ptmiss, \mathbf{Z^{*}})$ \\
VR-1$\ell$1T-SS & SS & SF & $>0.4$ & $<1.0$ & $\alpha$ & Pass & VR with SS-SF \\
VR-1$\ell$1T-DF & OS & DF & $>0.4$ & $<1.0$ & $\alpha$ & Pass & VR with OS-DF \\
\bottomrule
\end{tabular}
}
\end{table}



%

%
\section{Systematic uncertainty}
\label{sec:systematics}

Systematic uncertainty is evaluated for all signal samples and background processes. The sources of uncertainty are common to both the displaced track and 1$\ell$1T searches unless noted.

\begin{figure}[tbp]
\centering
\subfloat[]{
\includegraphics[width=0.87\textwidth]{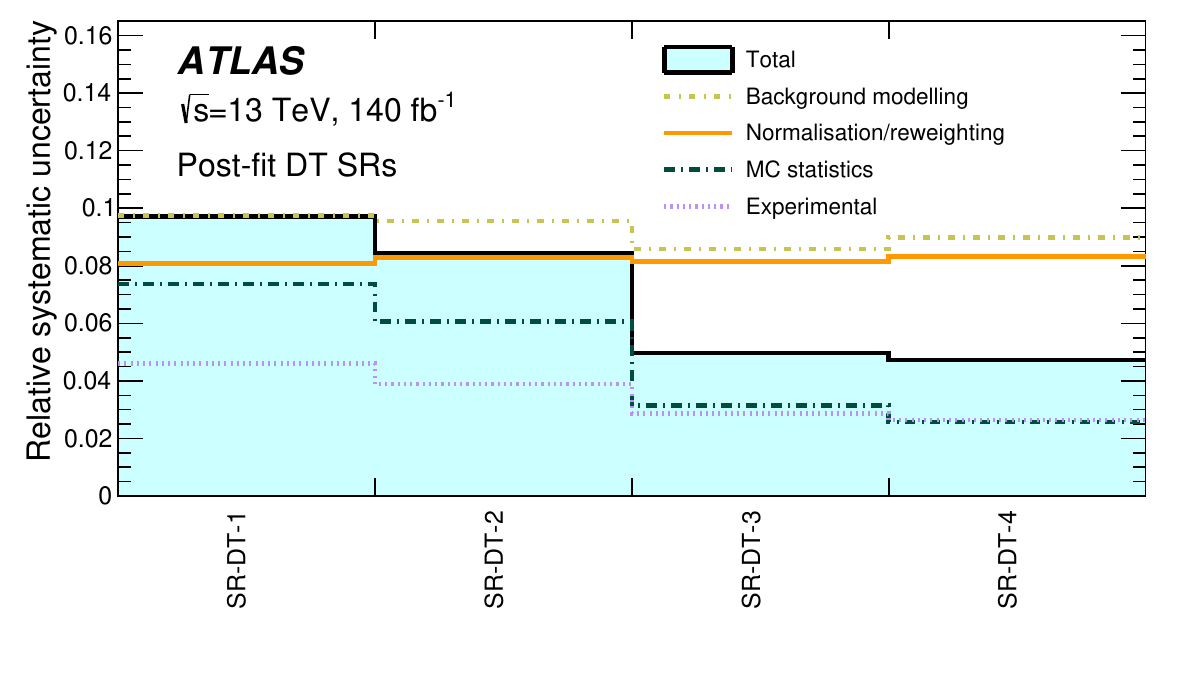}
}
\\
\vspace{20pt}
\subfloat[]{
\includegraphics[width=0.87\textwidth]{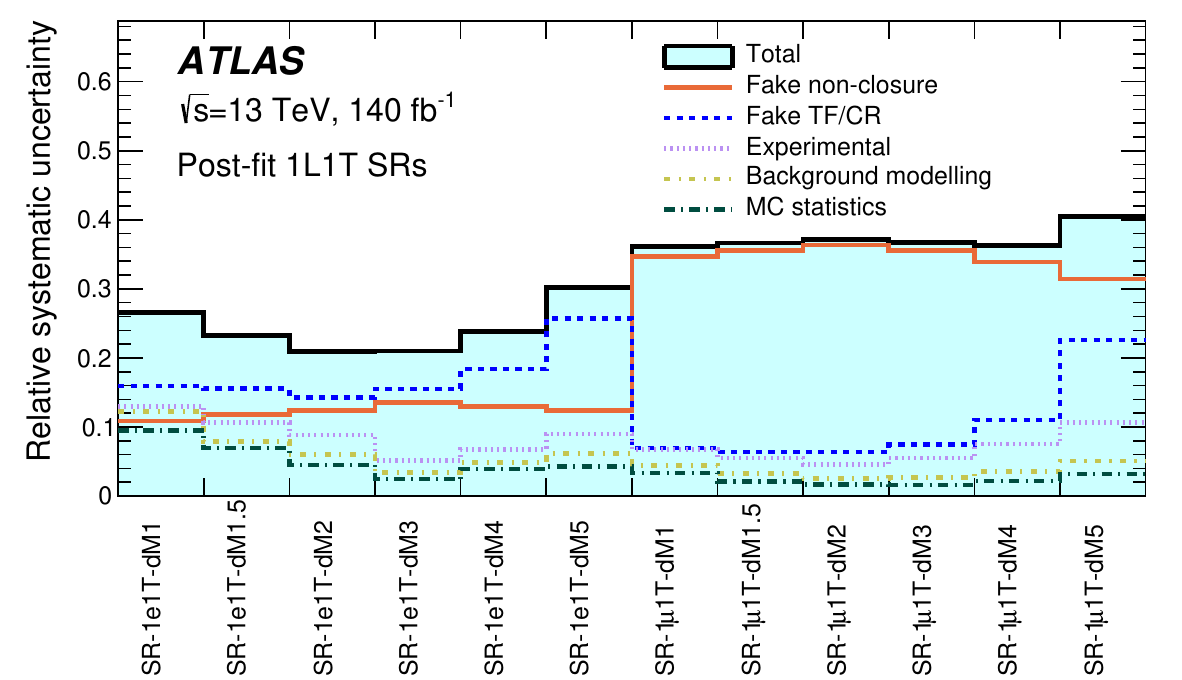}
}
\caption{\label{fig:SRsys} The relative systematic uncertainty (post-fit) in the fitted SM background event yields in the SRs obtained from a background-only fit to the CRs for (a) the displaced track search and (b) the 1$\ell$1T search. The `Normalisation/reweighting' category for the displaced track search includes all sources of uncertainty relevant to the normalisation and reweighting factors used in the background estimation. The `Fake non-closure' category for the 1$\ell$1T search denotes the non-closure uncertainty assigned to the fake lepton-track background component. The `Fake TF/CR' category for the 1$\ell$1T search includes all sources of uncertainty related to the transfer factors and statistical uncertainty in the CRs for the data-driven estimate of the fake lepton-track component. Sources of uncertainty related to the reconstruction and identification of jets, \met, leptons, and lepton-tracks are included in the `Experimental' category. The `Background modelling' category includes the theory uncertainty in the background processes. The `MC Statistics' uncertainty originates from the limited size of the MC samples. The individual sources of uncertainty can be correlated and therefore do not necessarily sum in quadrature to the total uncertainty.}
\end{figure}

Figure~\ref{fig:SRsys} shows the relative systematic uncertainty in the predicted SM background yields in the SRs derived from background-only fits to the CRs. Sources of experimental uncertainty include those related to the reconstruction and selection of objects relevant to the analysis. As both the displaced track and 1$\ell$1T searches target boosted topologies with a high-\pt leading jet and high \met, sources of uncertainty related to these objects are found to be dominant experimental uncertainty. Sources of uncertainty related to jets, which mainly involve the jet energy scale (JES) and jet energy resolution (JER), are evaluated as a function of jet \pt and $\eta$. This uncertainty is determined from data and MC studies such as the measurement of jet-\pt balance in dijet and $Z/\gamma$+jets processes~\cite{JETM-2018-05}. Sources of uncertainty related to $b$-jet identification and vertex tagging are also considered. Uncertainty in the modelling of \met in the MC simulation is estimated by propagating the energy and momentum scale of the objects entering the calculation of \met and the variations of the resolution and scale of the soft-term. Sources of systematic uncertainty for leptons include those in the reconstruction, identification and isolation efficiencies, and those in the momentum/energy resolution and scale~\cite{EGAM-2018-01,MUON-2018-03}. Other sources of experimental uncertainty originate from pile-up modelling, tracking efficiency, and track impact parameter resolution.
For the background MC samples of the 1$\ell$1T search, the experimental uncertainty only affects the subdominant true lepton-track component, hence to avoid fit instabilities, a single 50\% envelope uncertainty is assigned, which is found to cover all of the individual sources of experimental uncertainty, the largest of which is 40\%. Moreover, the 1$\ell$1T search considers sources of uncertainty in low-\pt lepton-track identification performance, such as those originating from the calibration.

For the displaced track search, uncertainty in the semi-data driven background estimation is considered. Uncertainty in the event-level normalisation factors for each background processes and track-level reweighting factors, originating from the data and MC statistics in the CRs, is taken into account by performing a simultaneous fit with all sources of uncertainty included. The relative uncertainty values of these factors are typically below 10\%.

For the 1$\ell$1T search, uncertainty in the transfer factors and statistical uncertainty in the corresponding CRs is incorporated into a simultaneous fit. Moreover, a non-closure uncertainty is applied to the fake lepton-track component to consider sources of uncertainty originating from the data-driven estimation method itself, such as the choice of variables used to define the CRs. The non-closure uncertainty is defined in such a way as to ensure compatibility between the total background prediction and the observed data in all VRs of the 1$\ell$1T search, taking into account the statistical uncertainty in the data. A non-closure uncertainty of 15\% and 40\% is assigned to the fake lepton-track components of the 1$e$1T and 1$\mu$1T channels, respectively.

Theory modelling uncertainty is also considered for MC samples of dominant background processes.
For the displaced track search, these sources of uncertainty are considered for all background processes. For the 1$\ell$1T search, the fake lepton-track background is estimated using data-driven methods and therefore no theory uncertainty is assigned to this component. The remaining 10--20\% of the total background in this search arises from processes that produce true lepton-tracks and is estimated using MC samples. Theory uncertainty is therefore considered for the $V$+jets and diboson processes that dominate the true lepton-track background.

Common to both of the searches, uncertainty originating from the choice of QCD renormalisation and factorisation scales is evaluated by varying the corresponding MC generator parameters up and down by a factor of two around their nominal values. Uncertainty associated with the choice of PDF set~~\cite{Ball:2012cx,Ball:2014uwa} and the strong coupling constant $\alpha_{\textrm{s}}$ is also considered. Furthermore for the displaced track search, uncertainty in the resummation and matching scale between matrix elements and parton showers is considered for $V$+jets samples by varying up and down the corresponding parameters in \textsc{Sherpa}.

For signal samples, uncertainty originating from multi-parton interactions and parton showers is considered on top of the theory uncertainty mentioned above. This uncertainty is evaluated from the change of yields with different tune variations using a few of the samples. Five types of variations are considered according to the A14 tune of \PYTHIA~\cite{ATL-PHYS-PUB-2014-021}: one for underlying event effects, one for jet structure effects, and three for different aspects of extra jet production. The sum in quadrature of the five variations is used as the overall uncertainty. As this uncertainty is found to have little dependence on the SUSY particle masses, uncertainties of 6.4\% and 15\% are assigned to all signal samples for the displaced track and 1$\ell$1T searches, respectively.

As shown in Figure~\ref{fig:SRsys}, the systematic uncertainty of the displaced track search is dominated by the uncertainty related to the normalisation and track reweighting factors, and the background modelling uncertainty.
The impact of a given category of uncertainty is estimated by taking the results from the nominal background-only likelihood fit to the CRs, fixing the parameters associated with the other sources of uncertainty to their best-fit values, and propagating the variations of the remaining floating parameters to the background prediction in the SRs using the covariance matrix from the nominal fit~\cite{Baak:2014wma}.
The profile likelihood method used for the fits is described in more detail in Section~\ref{sec:results}.
Since the individual sources of uncertainty in different categories can be correlated, the uncertainties from each category do not necessarily sum in quadrature to the total systematic uncertainty using this decomposition method, and can even be larger than the total in the case of the displaced track SRs.
This feature arises mainly from anti-correlations between the theoretical uncertainties and those related to the normalisation and reweighting factors for the dominant $Z(\rightarrow \nu\bar{\nu})$+jets and $W^\pm(\rightarrow \tau^\pm \nu_\tau)$+jets backgrounds, leading to a partial cancellation when they are included simultaneously in the likelihood fit.
For the 1$\ell$1T search, the non-closure uncertainty in the fake lepton-track contribution is the most significant, especially for 1$\mu$1T.
For 1$e$1T, uncertainty in the transfer factors and CR statistical uncertainty become more dominant, due to limited data.


%

%
\section{Results}
\label{sec:result}

%

%
\label{sec:results}
Data in the CRs, VRs, and SRs are compared with SM predictions using a profile likelihood method~\cite{Cowan:2010js}. The sources of systematic uncertainty are treated as nuisance parameters with Gaussian-distributed constraints in the likelihood. The correlations among the sources of uncertainty are considered in this method. Following the comparisons in the control and validation regions in Section~\ref{sec:results_CRVR}, results in the signal regions are shown in Section~\ref{sec:results_SR}. Section~\ref{sec:results_disc} and Section~\ref{sec:results_excl} show the results of model-independent and model-dependent interpretations, respectively.

\subsection{Control and validation regions}
\label{sec:results_CRVR}
Background-only fits of the CRs are performed using only the CRs to constrain the fit parameters. For the displaced track search, a simultaneous fit is performed using all CRs to constrain all event normalisation and track reweighting factors. For the 1$\ell$1T search, data in CR-1$\ell$1T-A, CR-1$\ell$1T-B, and CR-1$\ell$1T-C are fitted simultaneously to constrain the TFs and to obtain the background contribution in the SRs. The same procedure is performed to estimate the SM background yields in the VRs, using the CRs defined in Section~\ref{sec:background_1L1T}.

The SM background prediction obtained from the background-only fit is first compared with data in the VRs to verify the validity and accuracy of the background estimation method. Figure~\ref{fig:pullplots_VR} shows a comparison between the data and estimated background event yields in the VRs, where all differences are below the 2$\sigma$ level. The non-closure uncertainty in the fake lepton-track components mentioned in Section~\ref{sec:systematics} is applied to the 1$\ell$1T search. Figures~\ref{fig:kine_VR_DT} and~\ref{fig:kine_VR_1L1T} show examples of kinematic distributions in the VRs after a background-only fit using the CRs.

\begin{figure}[tbp]
\centering
\subfloat[]{
\includegraphics[width=0.93\textwidth]{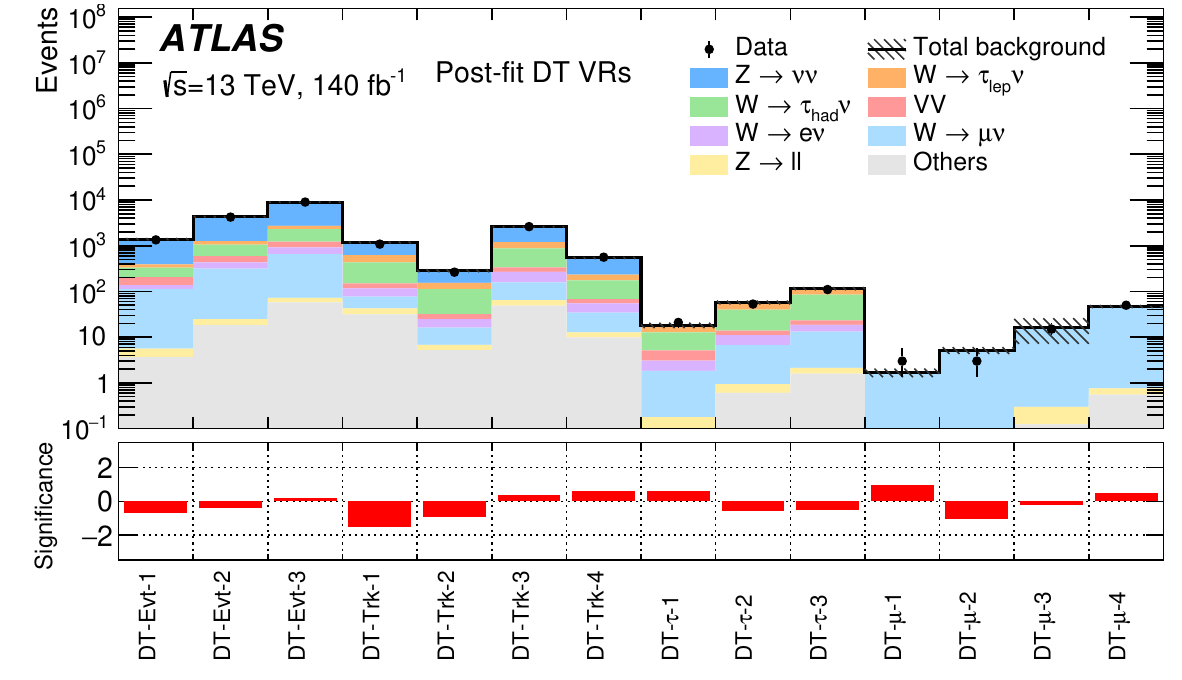}
}
\\
\vspace{20pt}
\subfloat[]{
\includegraphics[width=0.93\textwidth]{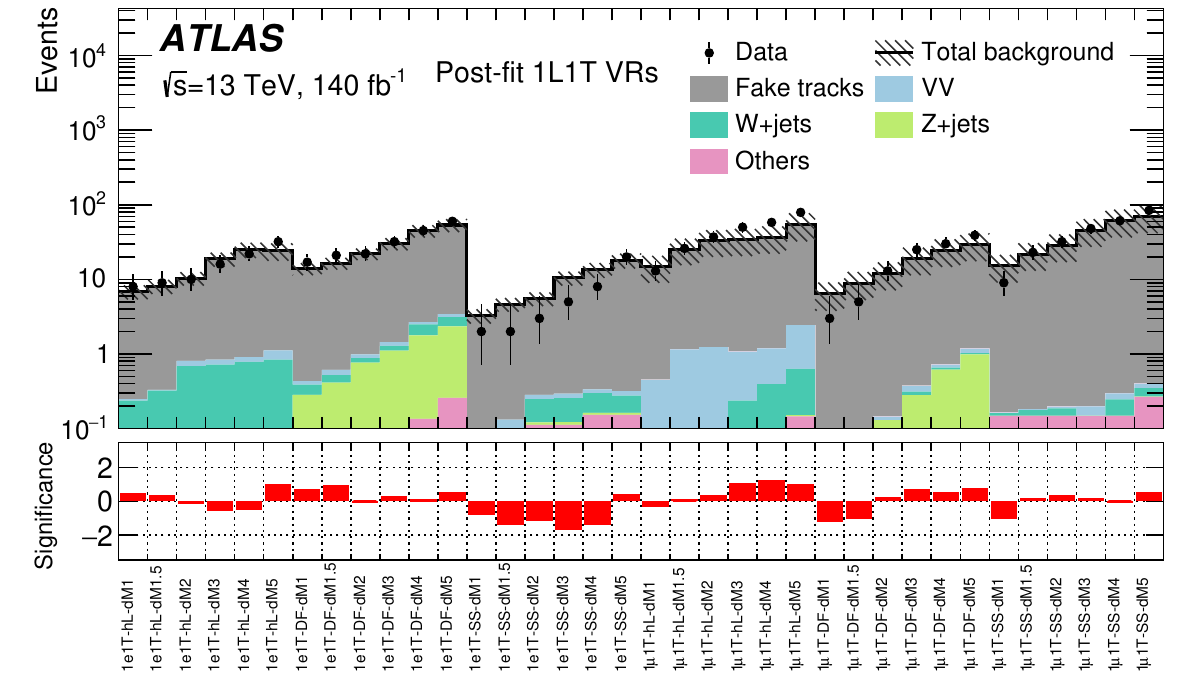}
}
\caption{\label{fig:pullplots_VR} Comparison of the observed and expected event yields in all VRs of (a) the displaced track search and (b) the 1$\ell$1T search. The bottom panel in each plot shows the significance of the difference between the observed and the predicted yields, following the profile likelihood method of Ref.~\cite{Cousins:2007bmb} in the case where the observed yield exceeds the prediction and using the same expression with an overall minus sign when the observed yield is below the prediction. (a) All slices are shown for each of the four VR categories: VR-DT-Evt, VR-DT-Trk, VR-DT-$\tau$, and VR-DT-$\mu$. (b) On the left of the plot are VRs for 1$e$1T and on the right are VRs for 1$\mu$1T. All six pNN-$\theta=$1, 1.5, 2, 3, 4, and 5~\GeV slices are shown for each of the three VRs: VR-hL, VR-DF, and VR-SS\@. The category `Others' contains background processes including top quarks, multijets and $\gamma$+jets. Uncertainty in the estimated background includes both the statistical and systematic uncertainties.}
\end{figure}

\begin{figure}[tbp]
\centering
\subfloat[]{
\includegraphics[width=0.49\textwidth]{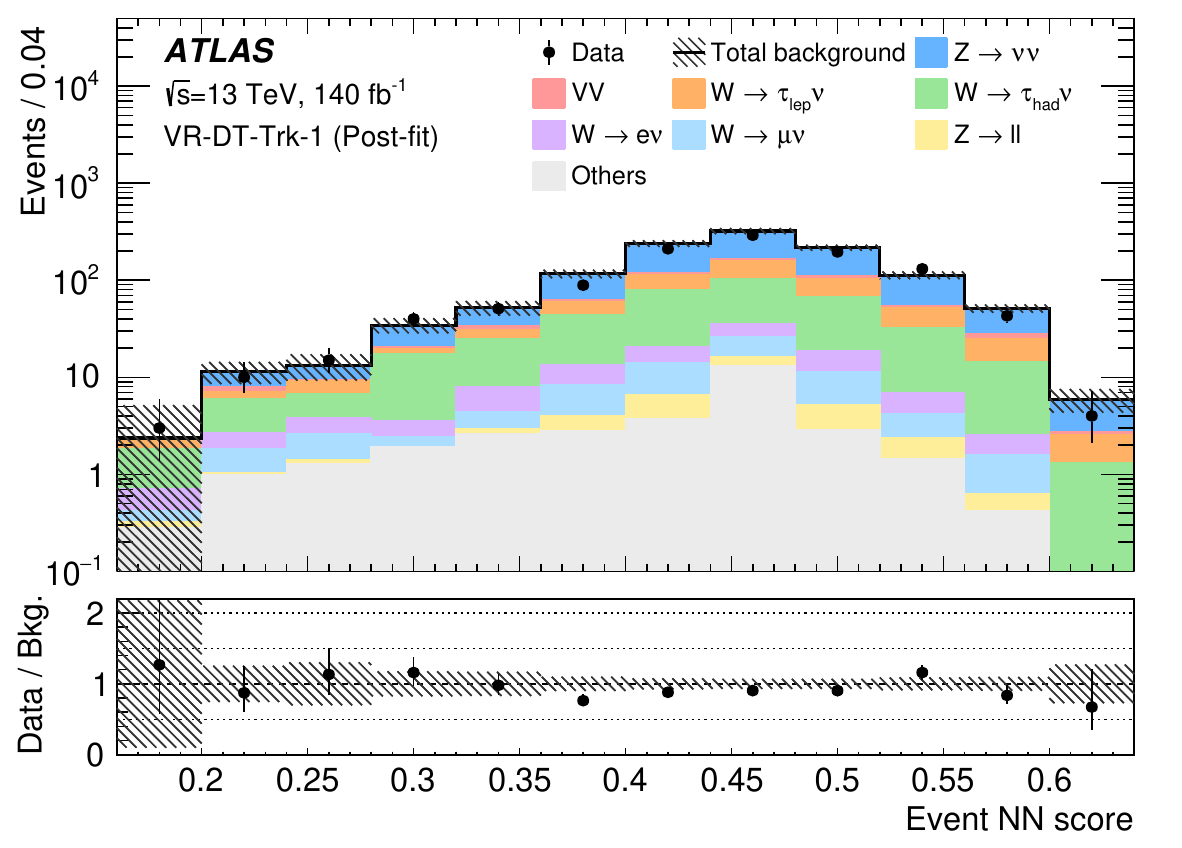}
}
\subfloat[]{
\includegraphics[width=0.49\textwidth]{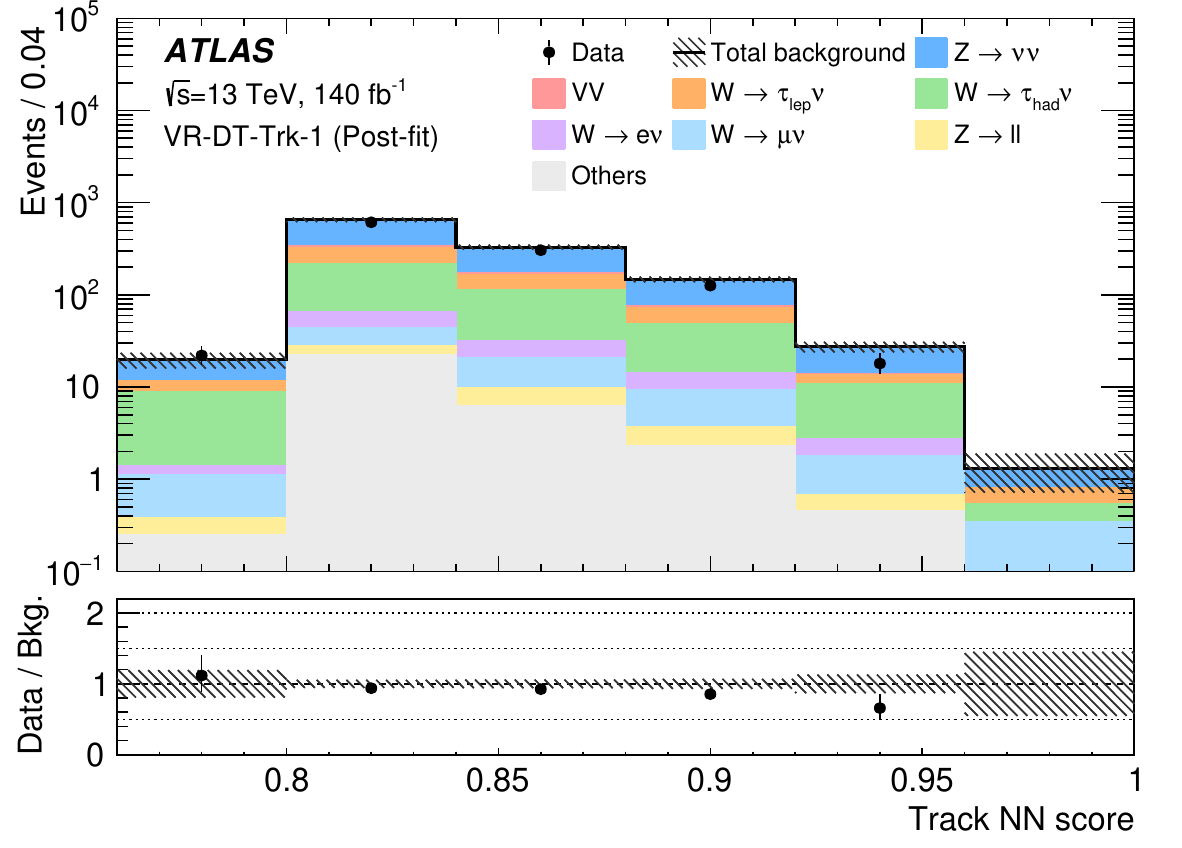}
}
\\
\vspace{15pt}
\subfloat[]{
\includegraphics[width=0.49\textwidth]{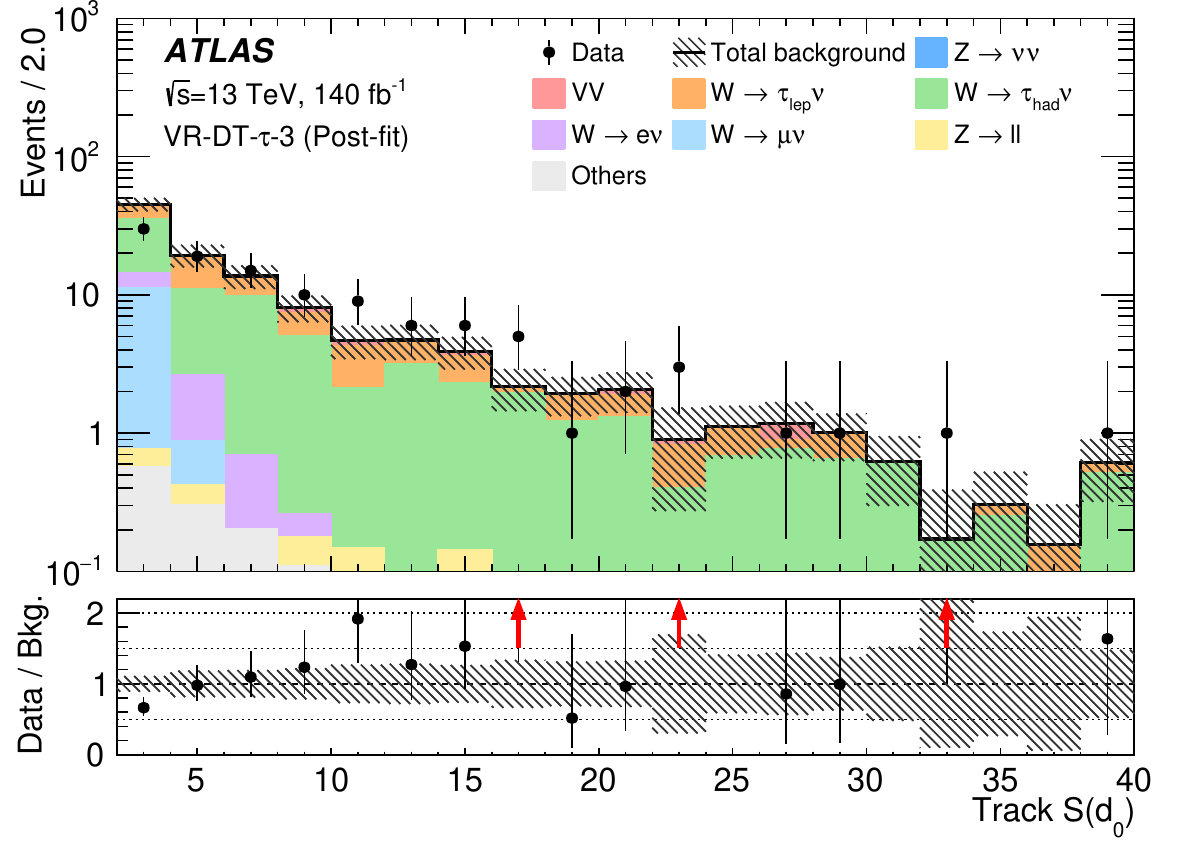}
}
\subfloat[]{
\includegraphics[width=0.49\textwidth]{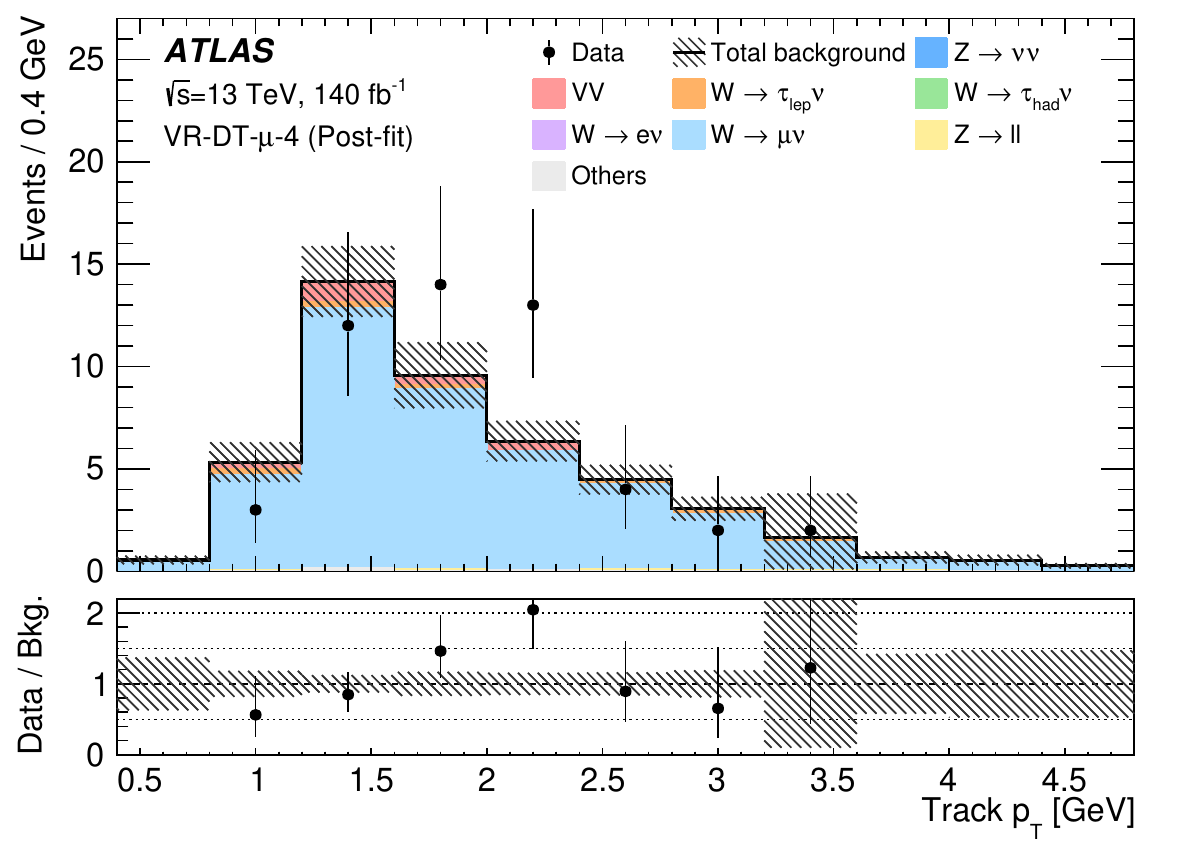}
}

\caption{\label{fig:kine_VR_DT} Examples of kinematic distributions in the VRs of the displaced track search after the background-only fit using the CRs. (a) and (b) show distributions of event-level and track-level NN scores in VR-DT-Trk-1, respectively. (c) shows the distribution of transverse impact parameter significance, $S(d_{0})$, of the track in VR-DT-$\tau$-3. (d) shows the distribution of track \pt in VR-DT-$\mu$-4. The category `Others' contains background processes including top quarks, multijets and $\gamma$+jets. The uncertainty bands on the expected background include all statistical and systematic uncertainties. Upper arrows in the lower panel indicate data points outside the vertical range shown.}
\end{figure}

\begin{figure}[tbp]
\centering
\subfloat[]{
\includegraphics[width=0.49\textwidth]{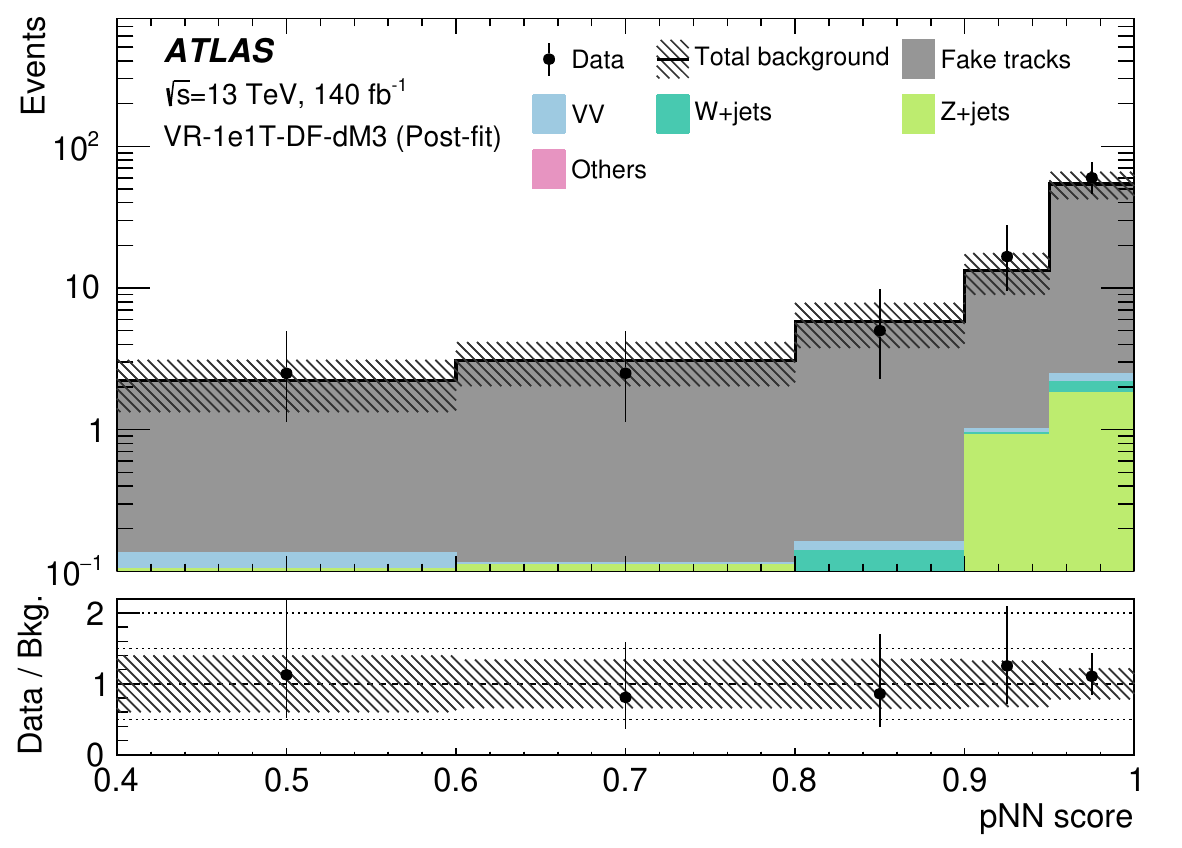}
}
\subfloat[]{
\includegraphics[width=0.49\textwidth]{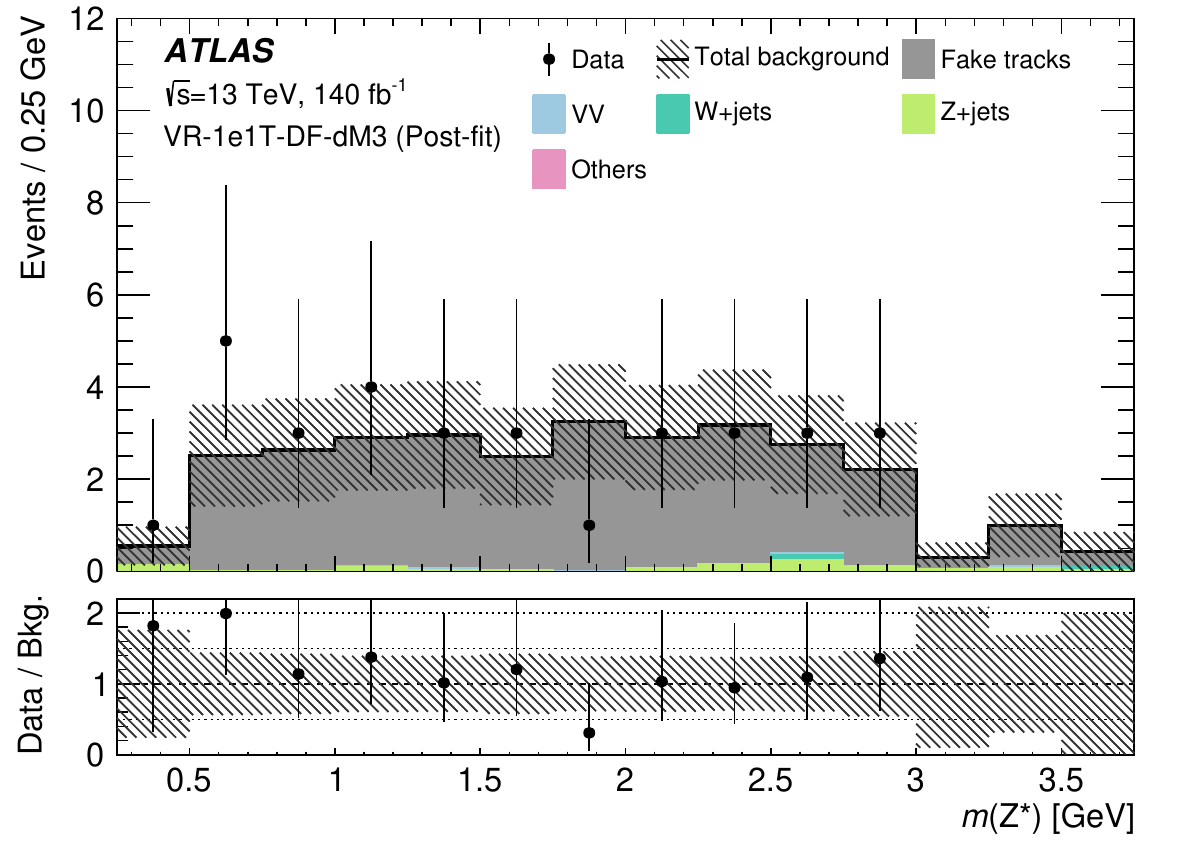}
}
\\
\vspace{15pt}
\subfloat[]{
\includegraphics[width=0.49\textwidth]{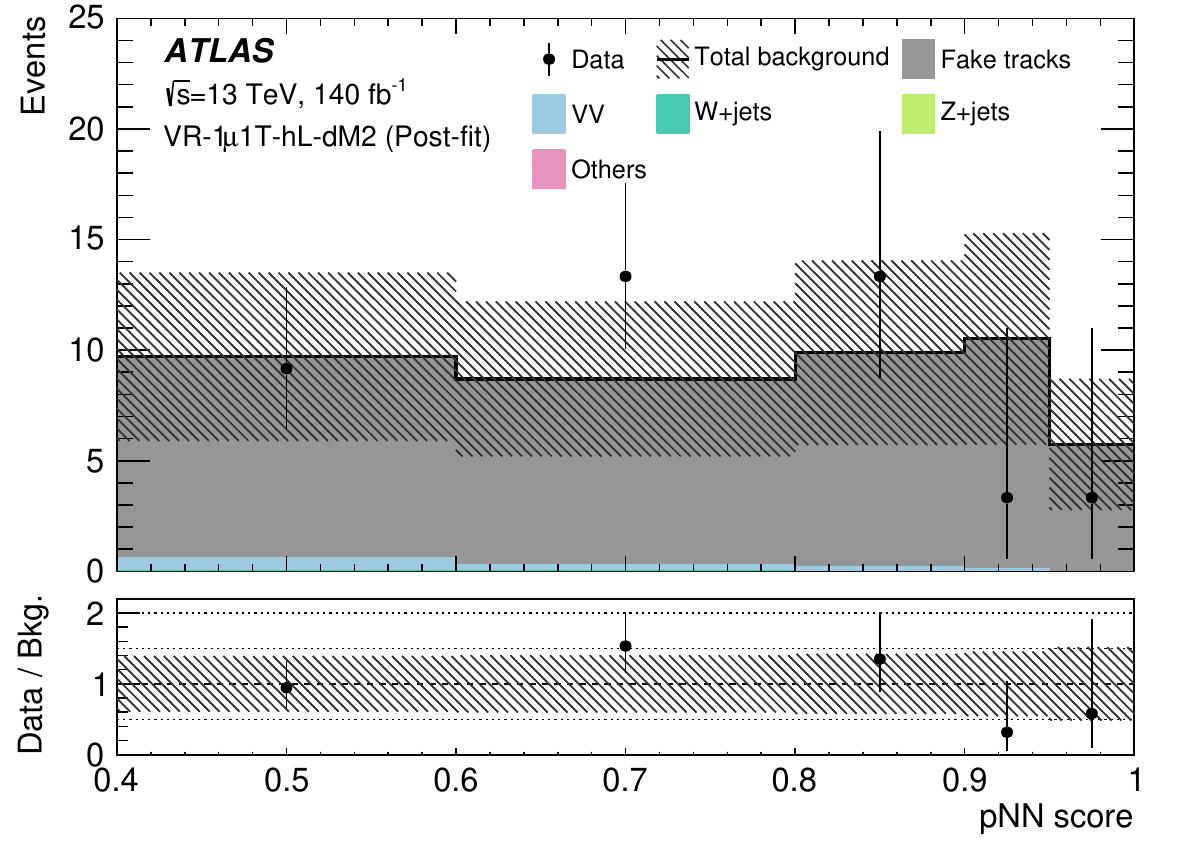}
}
\subfloat[]{
\includegraphics[width=0.49\textwidth]{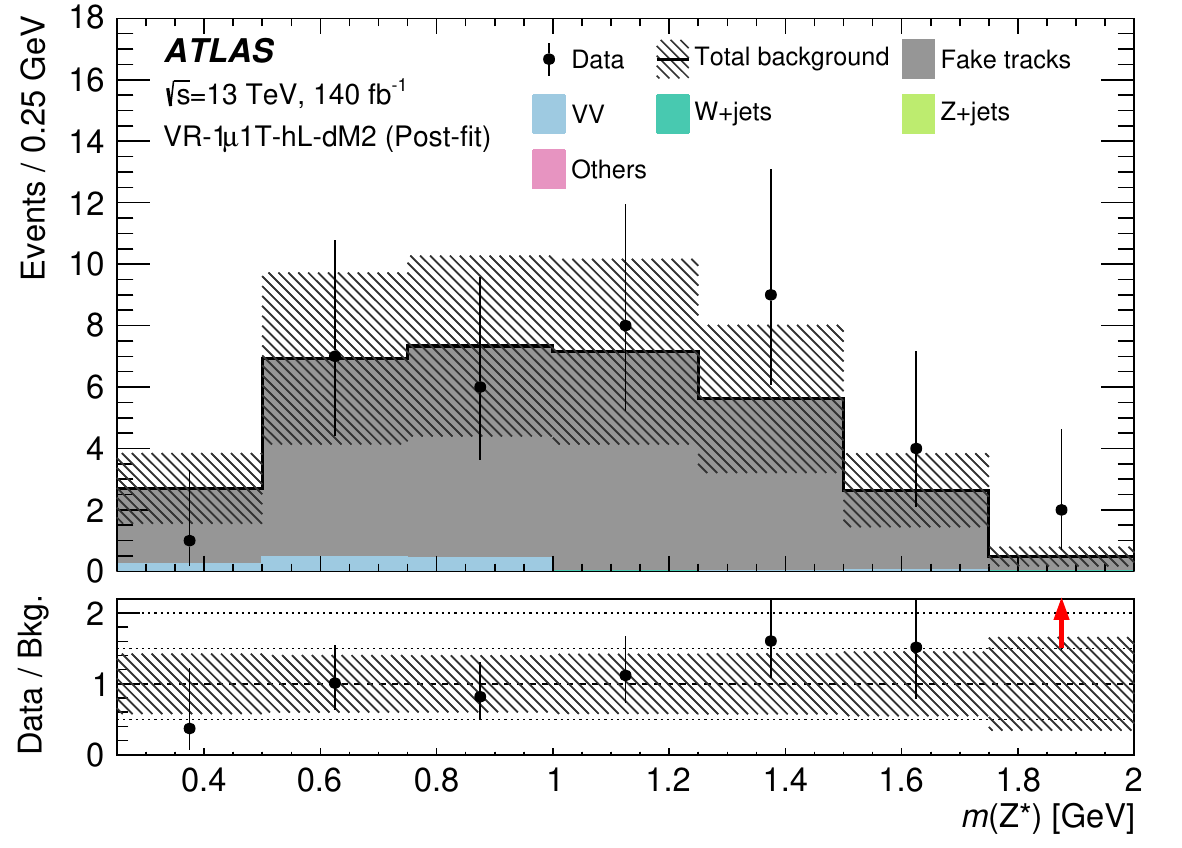}
}
\caption{\label{fig:kine_VR_1L1T} Examples of kinematic distributions in the VRs of the $1\ell$1T search after the background-only fit using the CRs. (a) and (b) show distributions of pNN score and $m(Z^{*})$ in VR-1$e$1T-DF-dM3, respectively. (c) and (d) show distributions of pNN score and $m(Z^{*})$ in VR-1$\mu$1T-hL-dM2. The category `Others' contains background processes including top quarks, multijets and $\gamma$+jets. The uncertainty bands on the expected background include all statistical and systematic uncertainties. Upper arrows in the lower panel indicate data points outside the vertical range shown.}
\end{figure}

\subsection{Event yields in signal regions}
\label{sec:results_SR}
The data and the SM background prediction obtained from the background-only fit are then compared in the SRs. Such comparisons are shown in Figure~\ref{fig:pullplots_SR} and Table~\ref{tab:SR_DT_1L1T_combined}. Deviations in all regions are within 2$\sigma$. The largest deviations for the displaced track and 1$\ell$1T searches are found to be $-1.3\sigma$ in SR-DT-4 and $1.5\sigma$ in SR-1$\mu$1T-dM5, respectively. Figures~\ref{fig:kine_SR_DT} and~\ref{fig:kine_SR_1L1T} show examples of kinematic distributions in the SRs after a background-only fit using the CRs. For the 1$\ell$1T search, the $m(Z^{*})$ distributions for the signal and background processes in Figure~\ref{fig:kine_SR_1L1T} demonstrate that the pNN is successfully learning to select events based on the reconstructed $Z^{*}$ mass, which is expected to have a kinematic cutoff at $\Delta m(\ninotwo,\ninoone)$ for signal processes.

\begin{figure}[tbp]
\centering
\subfloat[]{
\includegraphics[width=0.93\textwidth]{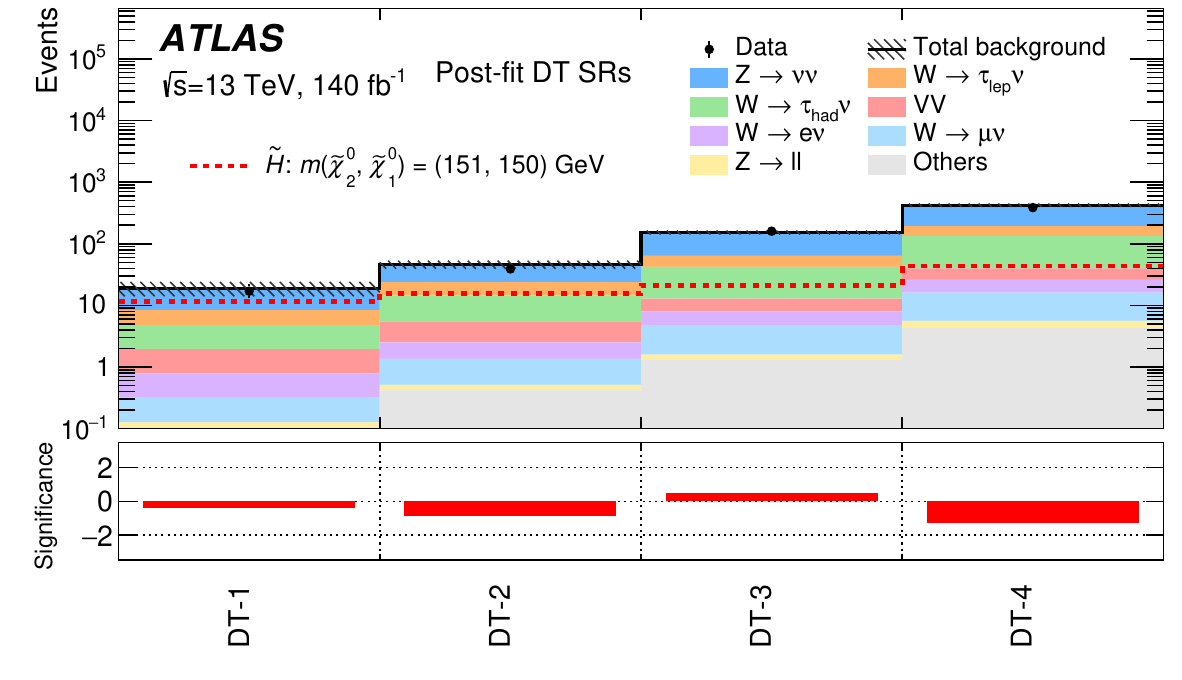}
}
\\
\vspace{20pt}
\subfloat[]{
\includegraphics[width=0.93\textwidth]{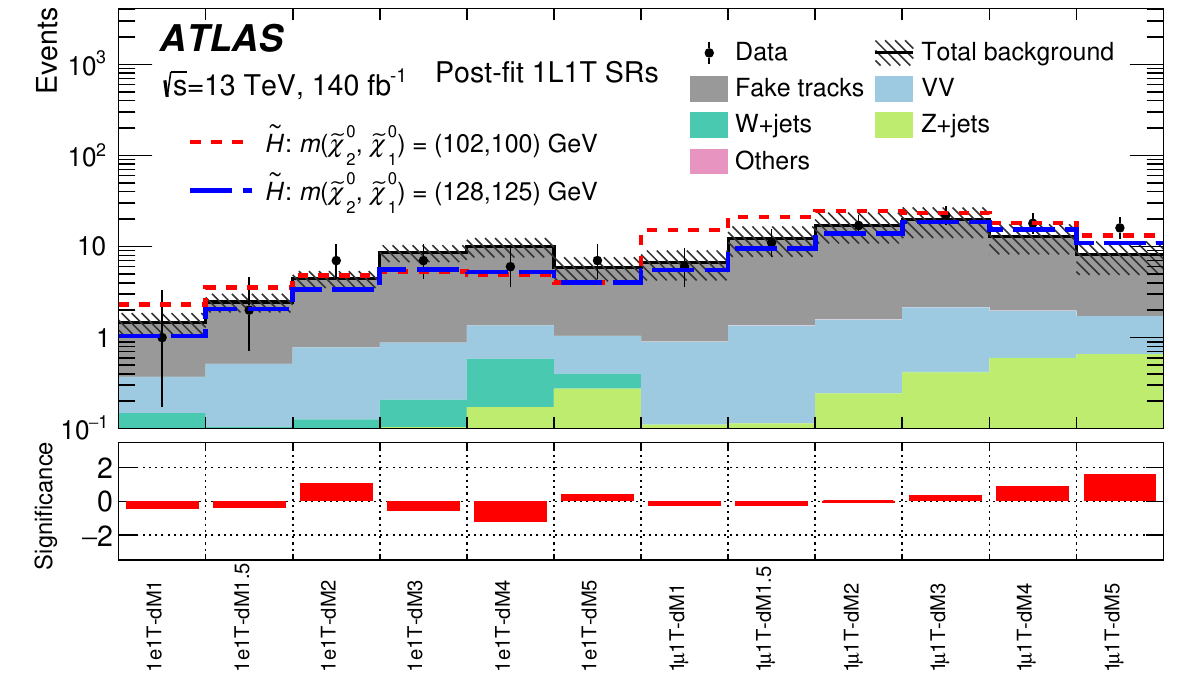}
}
\caption{\label{fig:pullplots_SR} Comparison of the observed and expected event yields in all SRs of (a) the displaced track search and (b) the 1$\ell$1T search. The bottom panel in each plot shows the significance of the difference between the observed and the predicted yields, following the profile likelihood method of Ref.~\cite{Cousins:2007bmb} in the case where the observed yield exceeds the prediction and using the same expression with an overall minus sign when the observed yield is below the prediction. (a) All four SRs are shown for the displaced track search. (b) On the left of the plot are the SRs for 1$e$1T and on the right are the SRs for 1$\mu$1T. All six pNN-$\theta=$1, 1.5, 2, 3, 4, and 5~\GeV slices are shown. The category `Others' contains background processes including top quarks, multijets and $\gamma$+jets. Uncertainty in the estimated background includes both statistical and systematic uncertainties. Predicted yields for some signal mass points are overlaid for reference.}
\end{figure}

\clearpage
\begin{table}[htbp]
\centering
\caption{\label{tab:SR_DT_1L1T_combined} Observed and predicted event yields in the SRs from a background-only fit using the CRs for (top) the displaced track search and (bottom) the 1$\ell$1T search. The category `Others' contains background processes including top quarks, multijets and $\gamma$+jets. Uncertainty in the predicted background yields includes statistical and systematic uncertainties.}
\renewcommand{\arraystretch}{1.1}
\maxsizebox*{\textwidth}{\textheight}{
\begin{tabular}{l rrrr}
\toprule
SR name & SR-DT-1 & SR-DT-2 & SR-DT-3 & SR-DT-4 \\
\midrule
Observed & $17$ & $39$ & $160$ & $386$ \\
\midrule
Fitted SM events & $19.0 \pm 1.9$ & $46.2 \pm 3.9$ & $152.9 \pm 7.6$ & $424 \pm 20$ \\
\midrule
$Z\rightarrow\nu\bar{\nu}$ & $10.5 \pm 1.5$ & $21.9 \pm 3.2$ & $87 \pm 10$ & $226 \pm 29$  \\
$W^\pm\rightarrow \tau^\pm_{\textrm{had}}\nu_\tau$ & $3.0 \pm 0.7$ & $9.6 \pm 3.0$ & $31.6 \pm 6.0$ & $99 \pm 23$ \\
$W^\pm\rightarrow \tau^\pm_{\textrm{lep}}\nu_\tau$ & $3.6 \pm 1.1$ & $9.5 \pm 2.0$ & $21.4 \pm 5.6$ & $60 \pm 12$ \\
$VV$ & $1.2 \pm 0.4$ & $2.9 \pm 0.5$ & $4.8 \pm 1.0$ & $13.4 \pm 2.5$ \\
$W^\pm\rightarrow e^\pm\nu_e$ & $0.5 \pm 0.2$ & $1.1 \pm 0.5$ & $3.2 \pm 1.7$ & $9.8 \pm 4.1$ \\
$W^\pm\rightarrow \mu^\pm\nu_\mu$ & $0.20 \pm 0.05$ & $0.84 \pm 0.08$ & $3.2 \pm 0.3$ & $10.5 \pm 1.0$ \\
$Z\rightarrow \ell^+\ell^-$ & $0.04 \pm 0.02$ & $0.1 \pm 0.1$ & $0.30 \pm 0.10$ & $1.3 \pm 0.4$ \\
Others & $0.1 \pm 0.1$ & $0.4 \pm 0.2$ & $1.3 \pm 0.4$ & $4.3 \pm 1.5$ \\
\bottomrule
\end{tabular}
}

\bigskip
\renewcommand{\arraystretch}{1.2}
\maxsizebox*{\textwidth}{\textheight}{
\begin{tabular}{l l rrrrrr}
\toprule
& SR pNN-$\theta$ [GeV]& 1 & 1.5 & 2 & 3 & 4 & 5 \\
\toprule
\vtlab{\large SR-1$e$1T}{7}
& Observed & 1 & 2 & 7 & 7 & 6 & 7 \\
\cmidrule{2-8}
& Fitted SM events & $1.5 \pm 0.4$ & $2.5 \pm 0.6$ & $4.4 \pm 0.9$ & $8.5 \pm 1.8$ & $10.0 \pm 2.4$ & $5.8 \pm 1.8$ \\
\cmidrule{2-8}
& Fake tracks & $1.1 \pm 0.4$ & $1.9 \pm 0.6$ & $3.7 \pm 1.1$ & $7.6 \pm 2.2$ & $8.6 \pm 2.5$ & $4.8 \pm 1.8$ \\
& $VV$ & $0.22 \pm 0.17$ & $0.41 \pm 0.29$ & $0.66 \pm 0.43$ & $0.67 \pm 0.43$ & $0.78 \pm 0.53$ & $0.64 \pm 0.44$ \\
& $W^\pm+$jets & $0.10^{+0.13}_{-0.10}$ & $0.10^{+0.13}_{-0.10}$ & $0.10^{+0.13}_{-0.10}$ & $0.10^{+0.13}_{-0.10}$ & $0.41^{+0.43}_{-0.41}$ & $0.12^{+0.14}_{-0.12}$ \\
& $Z+$jets & $0.04^{+0.09}_{-0.04}$ & $<0.01$ & $0.02^{+0.09}_{-0.02}$ & $0.10^{+0.16}_{-0.10}$ & $0.17^{+0.20}_{-0.17}$ & $0.27^{+0.28}_{-0.27}$ \\
& Others & $<0.01$ & $<0.01$ & $<0.01$ & $<0.01$ & $<0.01$ & $<0.01$ \\
\midrule
\vtlab{\large SR-1$\mu$1T}{7}
& Observed & 6 & 11 & 17 & 22 & 18 & 16 \\
\cmidrule{2-8}
& Fitted SM events & $6.6 \pm 2.2$ & $12.2 \pm 4.5$ & $17.2 \pm 6.4$ & $19.6 \pm 7.2$ & $12.9 \pm 4.7$ & $8.1 \pm 3.3$ \\
\cmidrule{2-8}
& Fake tracks & $5.7 \pm 2.5$ & $10.8 \pm 4.6$ & $15.6 \pm 6.6$ & $17.5 \pm 7.5$ & $10.9 \pm 4.9$ & $6.4 \pm 3.5$ \\
& $VV$ & $0.79 \pm 0.52$ & $1.2 \pm 0.8$ & $1.3 \pm 0.8$ & $1.7 \pm 1.0$ & $1.4 \pm 0.8$ & $1.1 \pm 0.7$ \\
& $W^\pm+$jets & $<0.01$ & $<0.01$ & $<0.01$ & $<0.01$ & $<0.01$ & $<0.01$ \\
& $Z+$jets & $0.11 \pm 0.09$ & $0.11 \pm 0.09$ & $0.24 \pm 0.18$ & $0.41 \pm 0.28$ & $0.59 \pm 0.36$ & $0.66 \pm 0.41$ \\
& Others & $<0.01$ & $<0.01$ & $<0.01$ & $<0.01$ & $<0.01$ & $<0.01$ \\
\bottomrule
\end{tabular}
}
\end{table}

\clearpage

\begin{figure}[tbp]
\centering
\subfloat[]{
\includegraphics[width=0.49\textwidth]{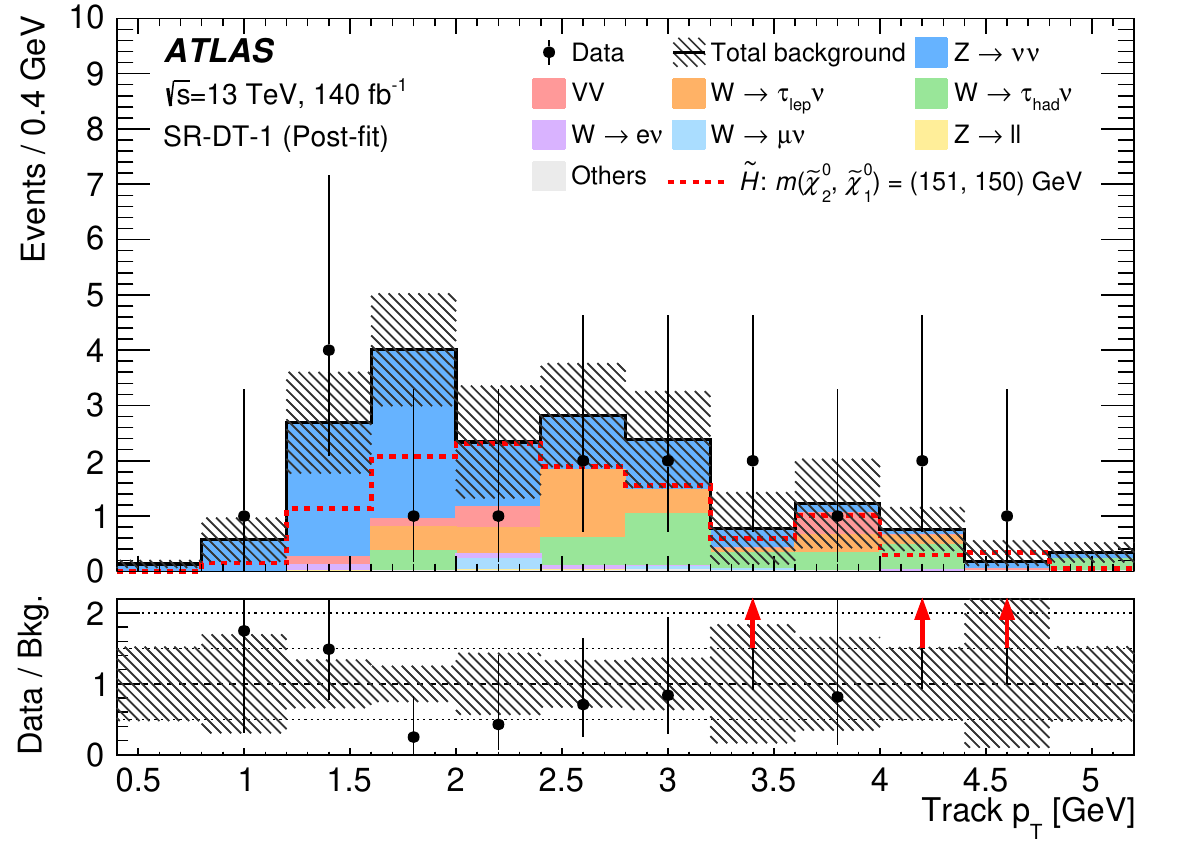}
}
\subfloat[]{
\includegraphics[width=0.49\textwidth]{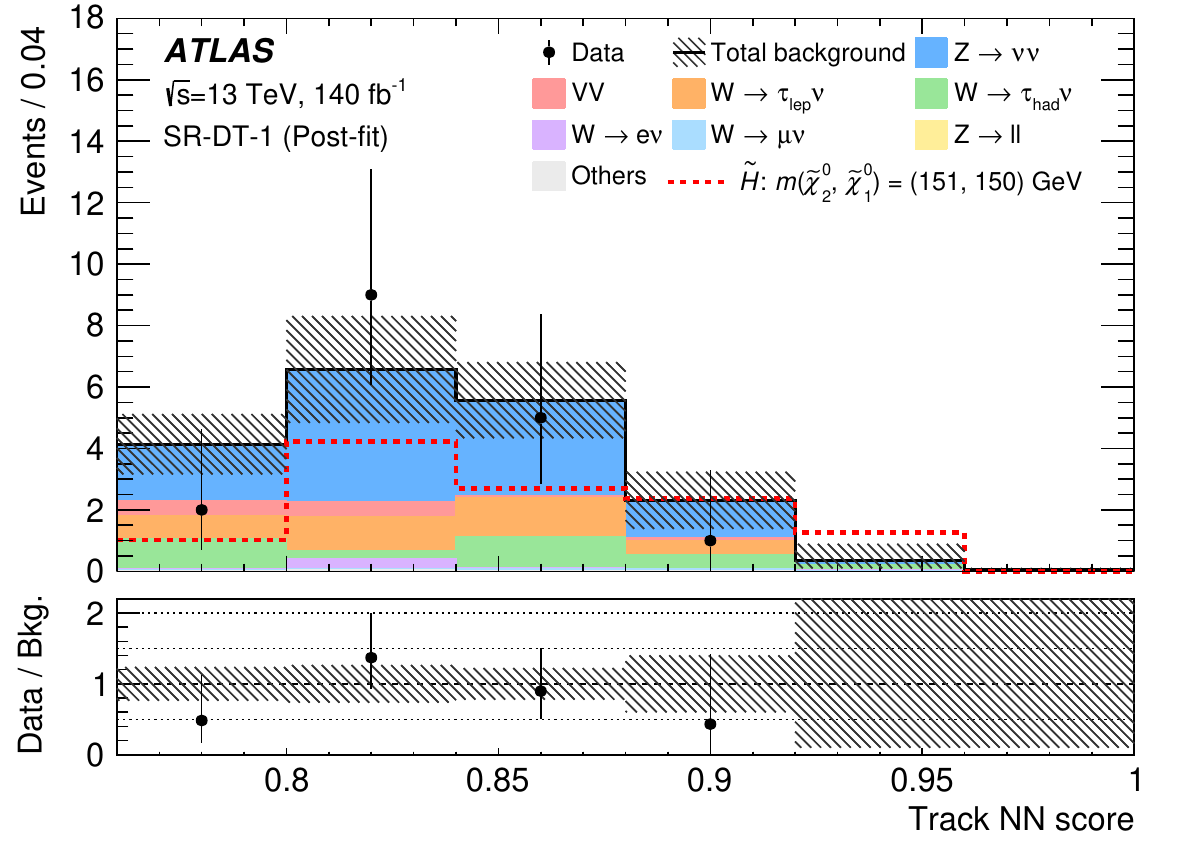}
}
\\
\vspace{15pt}
\subfloat[]{
\includegraphics[width=0.49\textwidth]{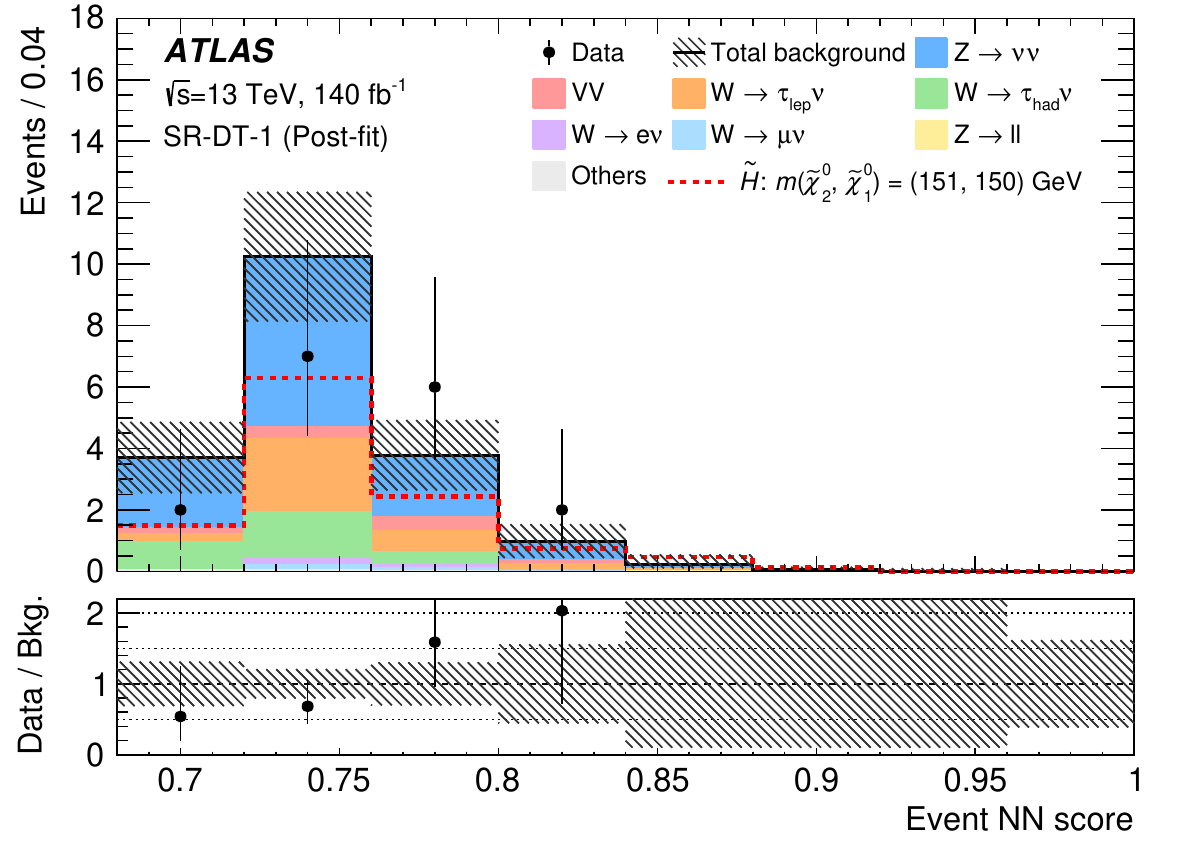}
}
\subfloat[]{
\includegraphics[width=0.49\textwidth]{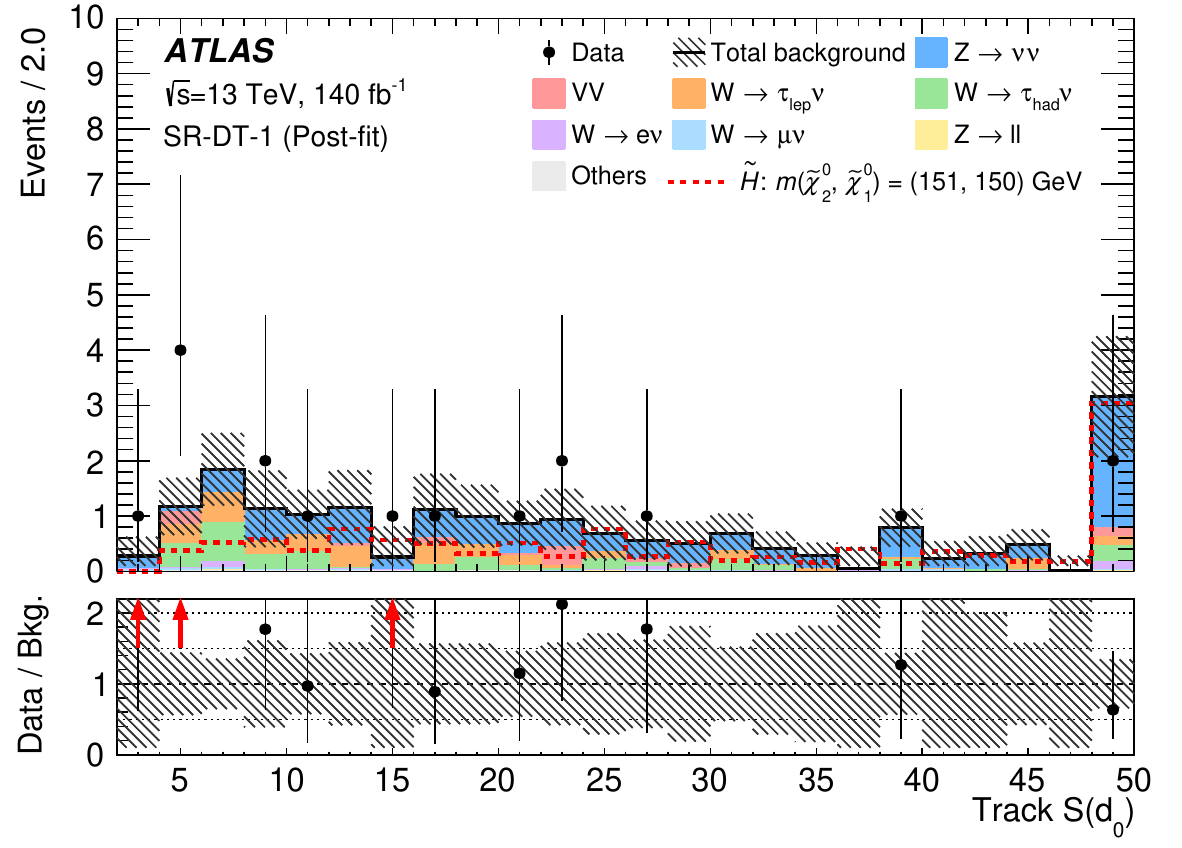}
}
\caption{\label{fig:kine_SR_DT} Examples of kinematic distributions in the SRs of the displaced track search after the background-only fit using the CRs. Overflow events are included in the last bin of the track $S(d_0)$ distribution.The category `Others' contains background processes including top quarks, multijets and $\gamma$+jets. The uncertainty bands on the expected background include all statistical and systematic uncertainties. Upper arrows in the lower panel indicate data points outside the vertical range shown. Predicted yields for a signal mass point of $m(\ninotwo, \chinoonepm, \ninoone)=(151, 150.5, 150)~\GeV$ are overlaid for reference.}
\end{figure}

\begin{figure}[tbp]
\centering
\subfloat[]{
\includegraphics[width=0.49\textwidth]{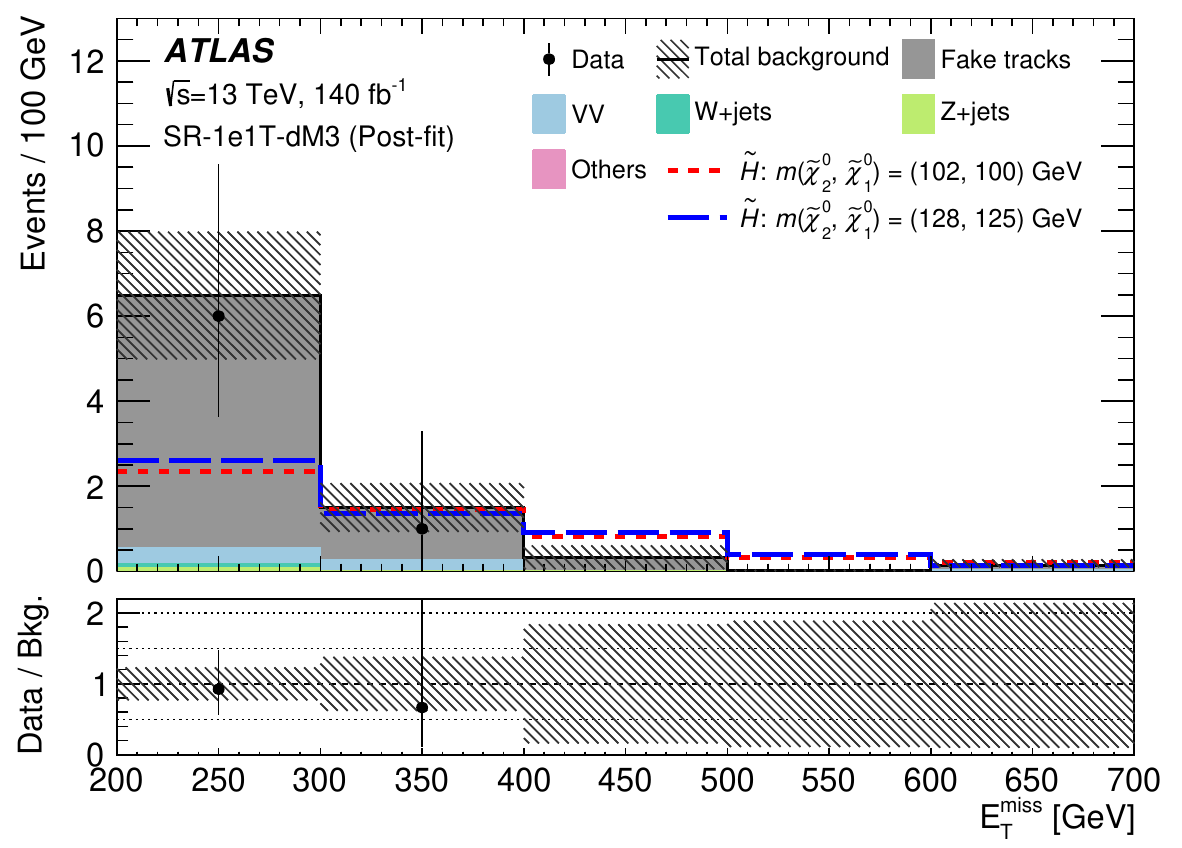}
}
\subfloat[]{
\includegraphics[width=0.49\textwidth]{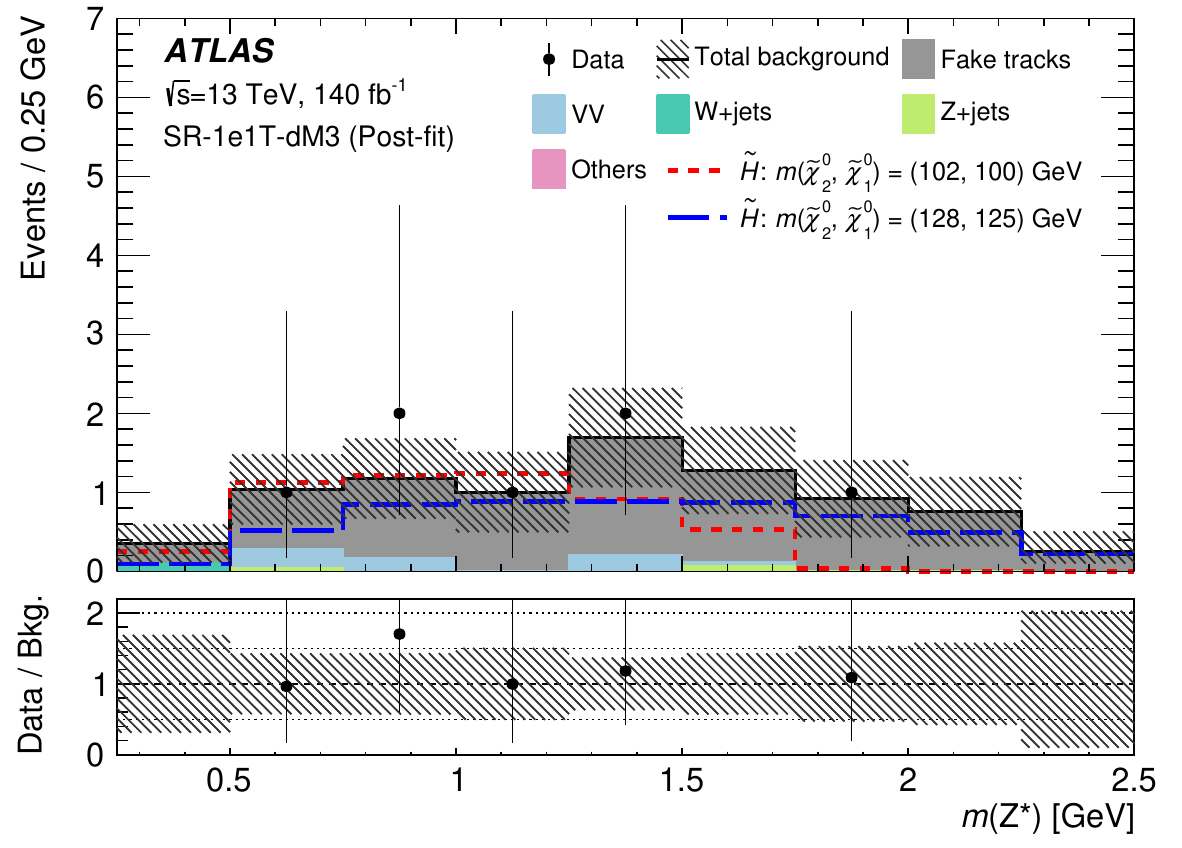}
}
\\
\vspace{15pt}
\subfloat[]{
\includegraphics[width=0.49\textwidth]{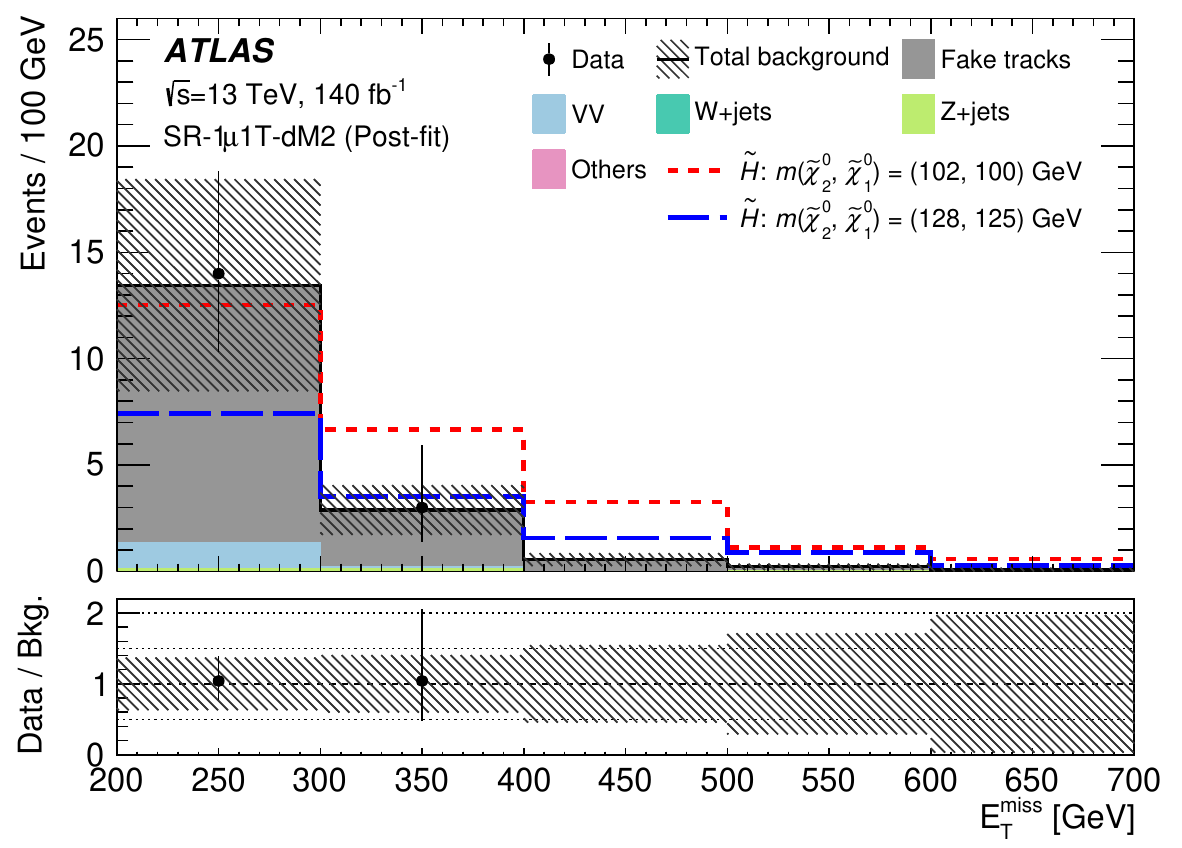}
}
\subfloat[]{
\includegraphics[width=0.49\textwidth]{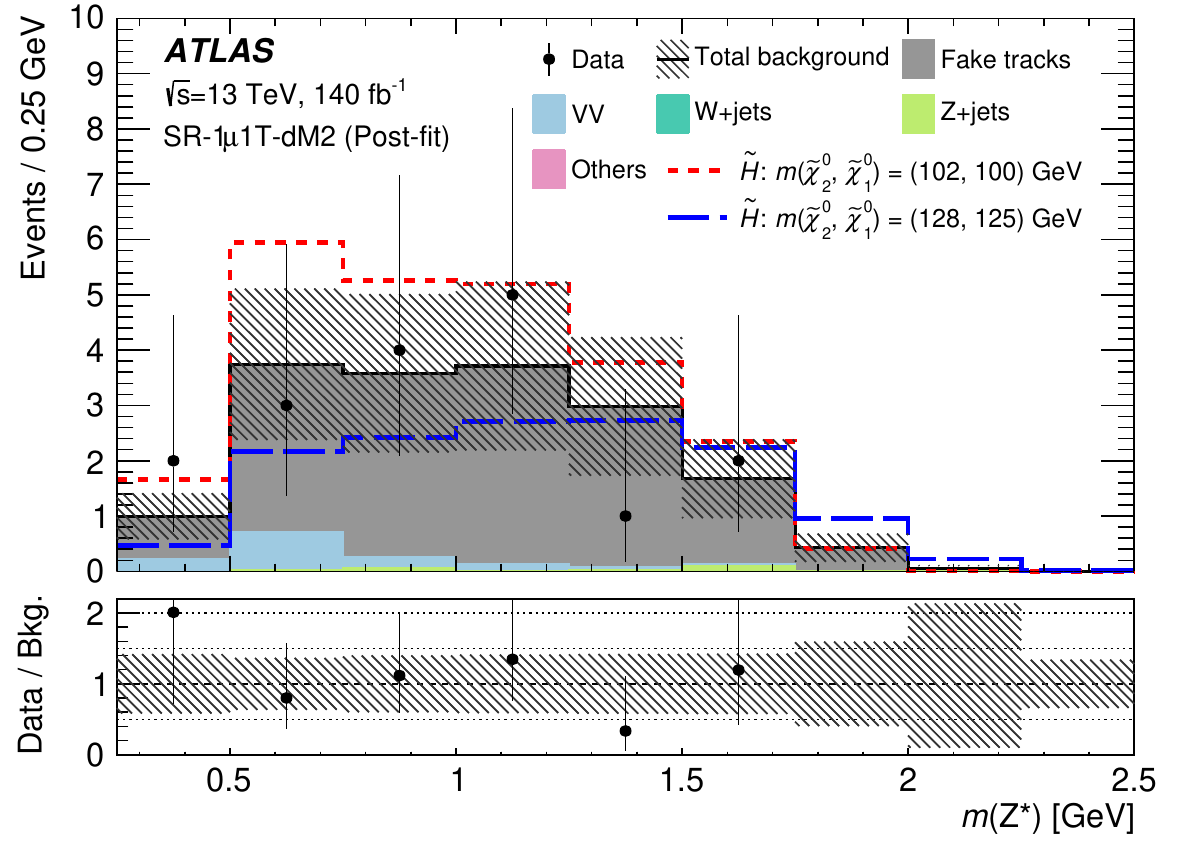}
}
\caption{\label{fig:kine_SR_1L1T} Examples of kinematic distributions in the SRs of the $1\ell$1T search after the background-only fit using the CRs. (a) and (b) show distributions of \met and $m(Z^{*})$ in SR-1$e$1T-dM3, respectively. (c) and (d) show distributions for \met and $m(Z^{*})$ in  SR-1$\mu$1T-dM2, respectively. The category `Others' contains background processes including top quarks, multijets and $\gamma$+jets. The uncertainty bands on the expected background include all statistical and systematic uncertainties. Predicted yields for signal mass points of $m(\ninotwo, \ninoone)=(102, 100)$ and $(128, 125)~\GeV$ are overlaid for reference.}
\end{figure}

\subsection{Model-independent interpretation}
\label{sec:results_disc}
For the 1$\ell$1T search, given the similarity of the phase space relevant to each of the SRs, SR-1$\ell$1T-dM1, SR-1$\ell$1T-dM3, and SR-1$\ell$1T-dM5 are selected to test the excesses of events in the SRs and to set upper limits on the visible cross-sections of physics beyond the SM\@. For this purpose, each SR is constructed by combining the corresponding electron and muon channel SRs into a single bin. A discovery fit configuration mentioned in Section~\ref{sec:background} is applied, where a generic signal model with an unconstrained normalisation parameter is included to estimate the contributions from any phenomena beyond the SM\@. The probability under the SM-only hypothesis is quantified by calculating the $p$-value and the corresponding significance for each SR.
For the final state targeted by the displaced track search, analogous results are reported in Ref.~\cite{SUSY-2020-04}, where the single-bin SRs are defined by a fixed set of kinematic requirements without the use of machine learning classifiers, thus enabling more
straightforward reinterpretations of the results in the context of some generic beyond the SM physics process.

Table~\ref{tab:1L1T_disc} shows the results of the model-independent interpretation for the 1$\ell$1T search. Data and predicted SM yields agree well in SR-1$\ell$1T-dM1 and SR-1$\ell$1T-dM3, resulting in 95\% confidence level (CL) upper limits of 0.05~fb and 0.13~fb on the visible cross-section of physics beyond the SM\@. Consistent with the deviation observed in SR-1$\mu$1T-dM5, a deviation of $1.5\sigma$ is observed between data and the SM predictions in SR-1$\ell$1T-dM5, resulting in a looser upper limit of 0.14~fb.

\begin{table}[tbp]
\caption{Results of the model-independent interpretation of the 1$\ell$1T search. For each SR indicated in the first column, the observed $N_{\textrm{obs}}$ number of events and predicted $N_{\textrm{exp}}$ number of SM events are shown in the second and third columns. The following columns show the observed 95\% CL upper limits on the visible cross-section, $\langle\epsilon{\mathrm{\sigma}}\rangle_{\mathrm{obs}}^{95}$, and on the observed and expected number of signal events corresponding to physics beyond the SM, denoted by $S_{\mathrm{obs}}^{95}$ and $S_{\mathrm{exp}}^{95}$, respectively. The $\pm1\sigma$ variations are also shown for $S_{\mathrm{exp}}^{95}$. The final columns show the $p$-value for the SM-only hypothesis $p(S=0)$ and the corresponding Gaussian-equivalent statistical significance $\sigma$, where $p$-values are capped at 0.5. For cases in which the data yield is smaller than expected, the $p$-value and the significance are not shown.}
\label{tab:1L1T_disc}
\renewcommand{\arraystretch}{1.1}
\centering
\begin{tabular}{lccccccc}
\toprule
SR name & $N_{\textrm{obs}}$ & $N_{\textrm{exp}}$ & $\langle\epsilon{\mathrm{\sigma}}\rangle_{\mathrm{obs}}^{95}$[fb] & $S_{\mathrm{obs}}^{95}$ & $S_{\mathrm{exp}}^{95}$ & $p(S=0)$ & $\sigma$ \\
\midrule
SR-1$\ell$1T-dM1 & 7 & $8.1\pm2.4$ & $0.05$ &  $7.4$ & $ { 8.1 }^{ +3.7 }_{ -2.2 }$ & -- & -- \\
SR-1$\ell$1T-dM3 & 29 & $28.1\pm7.4$ & $0.13$ &  $18.5$ & $ { 18.1 }^{ +6.5 }_{ -4.7 }$ & $ 0.47 $ & $0.07$ \\
SR-1$\ell$1T-dM5 & 23 & $13.9\pm3.7$ & $0.14$ &  $20.1$ & $ { 12.7 }^{ +5.3 }_{ -3.5 }$ & $ 0.07$ & $1.48$ \\
\bottomrule
\end{tabular}
\end{table}

\subsection{Model-dependent interpretation}
\label{sec:results_excl}

Exclusion fits are performed to set limits on the higgsino model under study. For the displaced track search, all CRs and SRs are fitted simultaneously with a parameter of interest corresponding to the signal strength, which scales the signal yields in all regions. For the 1$\ell$1T search, the same procedure is performed for each pNN-$\theta$. The $\textrm{CL}_{\textrm{s}}$ prescription~\cite{Read:2002hq} is used to perform hypothesis tests on the target higgsino model. Exclusion limits at 95\% CL are presented in a two-dimensional plane with the horizontal axis given by the mass of \chinoonepm and the vertical axis given by the mass difference between \chinoonepm and \ninoone, where the simplified higgsino model assumes $\Delta m(\ninotwo,\ninoone)=2\Delta m(\chinoonepm,\ninoone)$.

Figure~\ref{fig:combined_limit} shows the exclusion contours for the displaced track and 1$\ell$1T searches overlaid on other search results. For the displaced track search, as slight deficits are observed in SR-DT-1, 2, and 4, the observed limit extends beyond the expected, reaching up to about $m(\chinoonepm)=200~\GeV$ for $\Delta m(\chinoonepm,\ninoone)=0.6~\GeV$. Excluded $\Delta m(\chinoonepm,\ninoone)$ ranges from 0.3 to 1.2~\GeV.

The exclusion contours of the 1$\ell$1T search are drawn by taking an envelope of the six contours obtained from the six pNN-$\theta$ configurations. This procedure is performed by selecting the pNN-$\theta$ with the best expected $\textrm{CL}_{\textrm{s}}$ for each mass of the signal samples. The expected sensitivity of the 1$\ell$1T search originates largely from 1$\mu$1T, as the muon-identification efficiency is higher compared with that of electrons in the very low-\pt regime. The effects of a slight deficit in data observed in SR-1$e$1T-dM4 and slight excess observed in SR-1$\mu$1T-dM5 cancel out to yield an observed limit close to the expected limit. The observed limit extends up to $m(\chinoonepm)=130~\GeV$ at $\Delta m(\chinoonepm,\ninoone)=2~\GeV$.

The two searches complementarily explore simplified higgsino models in $\Delta m(\chinoonepm,\ninoone)=0.3\textrm{--}3~\GeV$, surpassing existing limits set by the LEP experiment~\cite{LEPlimits} and ATLAS searches~~\cite{SUSY-2018-16,SUSY-2020-04,SUSY-2018-09}.

\begin{figure}[tbp]
\centering
\includegraphics[width=0.8\textwidth]{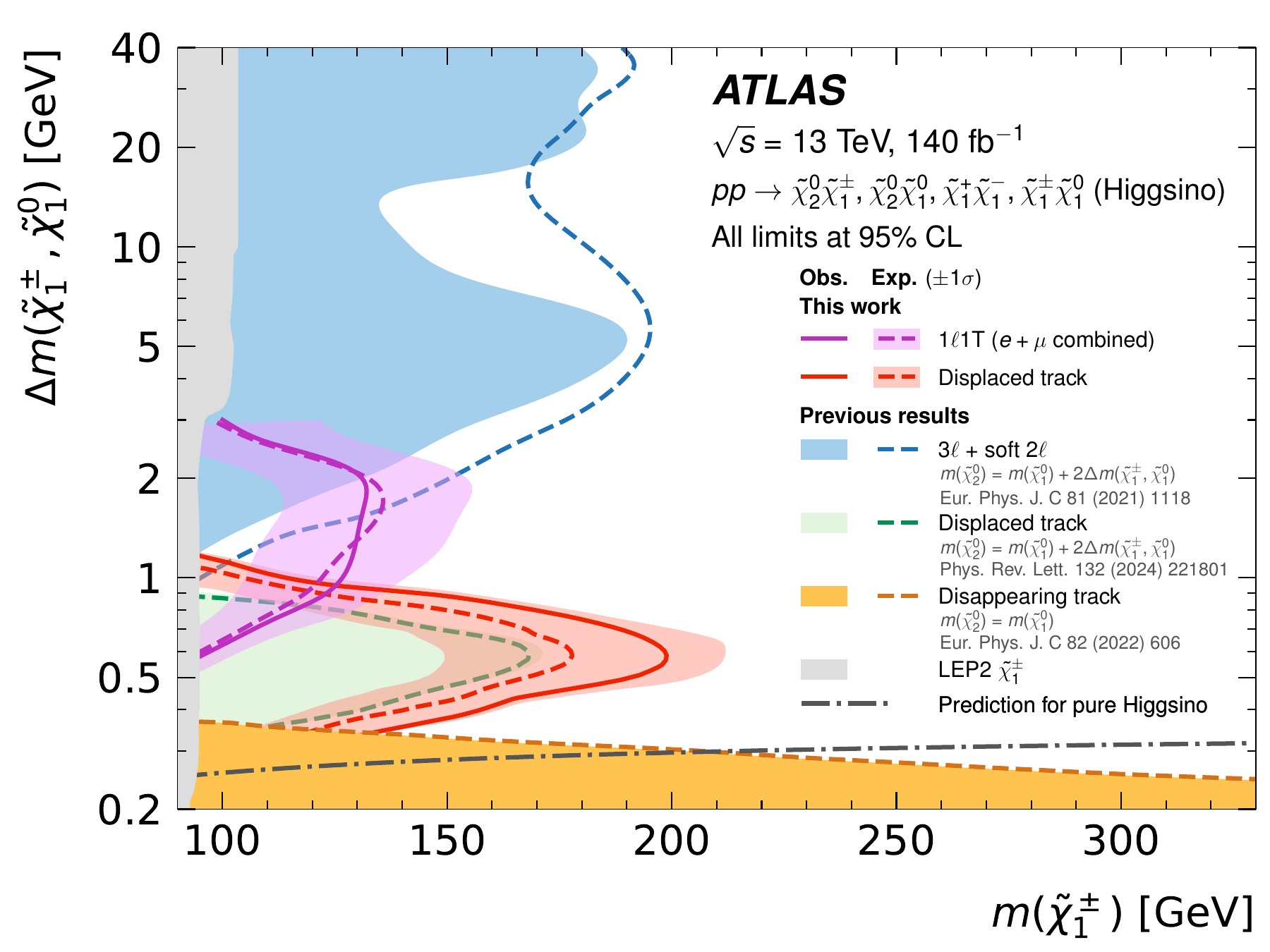}
\caption{\label{fig:combined_limit} Observed (solid line) and expected (dashed line) limits at 95\% CL for simplified higgsino models for 1$\ell$1T (purple) and displaced track (red) searches. The limits are drawn in the $\Delta m(\chinoonepm,\ninoone)$ vs $m(\chinoonepm)$ plane, with the assumption of $\Delta m(\ninotwo,\ninoone)=2\Delta m(\chinoonepm,\ninoone)$. The grey region denotes the mass exclusion limits from the LEP experiment~\cite{LEPlimits}. The blue, light green, and yellow regions indicate the limits from ATLAS soft lepton~\cite{SUSY-2018-16,SUSY-2019-09}, displaced track~\cite{SUSY-2020-04}, and disappearing track~\cite{SUSY-2018-19} searches.}
\end{figure}


%

%
%
\FloatBarrier

\section{Conclusion}
\label{sec:conclusion}

Two searches for the direct production of higgsinos with compressed mass spectra are presented, based on
$140~\ifb$ of $\sqrt{s}=13~\TeV$ $pp$ collision data collected by the ATLAS experiment at the LHC.
Events are required to contain missing transverse momentum, a high-$\pt$ jet, and at least one soft charged particle
consistent with the decay of a higgsino-like \ninotwo or \chinoonepm into a nearly mass-degenerate higgsino-like \ninoone.
For models with $\Delta m(\chinoonepm, \ninoone) \lesssim 1~\GeV$, the charginos are expected to have non-negligible lifetimes
and low-$\pt$, and displaced tracks from their potential decays are identified to reduce contributions from SM background processes.
For models with larger mass splittings where the higgsino decays are expected to be prompt, events are selected that feature
two low-$\pt$ electrons or muons with opposite charge, where one of the leptons is identified using novel soft-lepton taggers.
Both of the searches use NNs to provide discrimination between signal and SM background processes.

In each search, the data are found to be consistent with the SM predictions and the results are used to set
lower limits on the higgsino masses at 95\% CL in a simplified model where $\Delta m(\ninotwo, \ninoone) = 2 \Delta m(\chinoonepm, \ninoone)$.
The 1$\ell$1T search is able to further constrain the region where $0.8~\GeV \lesssim \Delta m(\chinoonepm, \ninoone) \lesssim 2.0~\GeV$,
extending the limits established by other experiments up to a maximum of
$m(\chinoonepm) = 132~\GeV$ for $\Delta m(\chinoonepm, \ninoone) = 1.8~\GeV$.
Additionally, upper limits on the visible cross-section are set at 95\% CL for generic beyond the SM physics processes in the 1$\ell$1T SRs that are optimised for 1, 3, and 5~\GeV mass splittings.
The displaced track search is able to extend the limits on $m(\chinoonepm)$ previously set by the ATLAS experiment in this final state by approximately $30~\GeV$,
with a maximum exclusion set at $m(\chinoonepm) = 199~\GeV$ for $\Delta m(\chinoonepm, \ninoone) = 0.6~\GeV$.
Taken together, the searches presented here exclude \chinoonepm masses below $126~\GeV$ at 95\% CL over the targeted range of $\Delta m(\chinoonepm, \ninoone)$ within the simplified higgsino model.
The limits set by the LEP experiments are now superseded by ATLAS in all ranges of $\Delta m(\chinoonepm, \ninoone)$ by these searches, providing additional scrutiny on the existence of light higgsinos with compressed mass spectra.


%

%
%
%
%
%
\section*{Acknowledgements}
%

%

%
%

%
%

We thank CERN for the very successful operation of the LHC and its injectors, as well as the support staff at
CERN and at our institutions worldwide without whom ATLAS could not be operated efficiently.

The crucial computing support from all WLCG partners is acknowledged gratefully, in particular from CERN, the ATLAS Tier-1 facilities at TRIUMF/SFU (Canada), NDGF (Denmark, Norway, Sweden), CC-IN2P3 (France), KIT/GridKA (Germany), INFN-CNAF (Italy), NL-T1 (Netherlands), PIC (Spain), RAL (UK) and BNL (USA), the Tier-2 facilities worldwide and large non-WLCG resource providers. Major contributors of computing resources are listed in Ref.~\cite{ATL-SOFT-PUB-2026-001}.

We gratefully acknowledge the support of ANPCyT, Argentina; YerPhI, Armenia; ARC, Australia; BMWFW and FWF, Austria; ANAS, Azerbaijan; CNPq and FAPESP, Brazil; NSERC, NRC and CFI, Canada; CERN; ANID, Chile; CAS, MOST and NSFC, China; Minciencias, Colombia; MEYS CR, Czech Republic; DNRF and DNSRC, Denmark; IN2P3-CNRS and CEA-DRF/IRFU, France; SRNSFG, Georgia; BMFTR, HGF and MPG, Germany; GSRI, Greece; RGC and Hong Kong SAR, China; ICHEP and Academy of Sciences and Humanities, Israel; INFN, Italy; MEXT and JSPS, Japan; CNRST, Morocco; NWO, Netherlands; RCN, Norway; MNiSW, Poland; FCT, Portugal; MNE/IFA, Romania; MSTDI, Serbia; MSSR, Slovakia; ARIS and MVZI, Slovenia; DSI/NRF, South Africa; MICIU/AEI, Spain; SRC and Wallenberg Foundation, Sweden; SERI, SNSF and Cantons of Bern and Geneva, Switzerland; NSTC, Taipei; TENMAK, T\"urkiye; STFC/UKRI, United Kingdom; DOE and NSF, United States of America.

Individual groups and members have received support from BCKDF, CANARIE, CRC and DRAC, Canada; CERN-CZ, FORTE and PRIMUS, Czech Republic; COST, ERC, ERDF, Horizon 2020, ICSC-NextGenerationEU and Marie Sk{\l}odowska-Curie Actions, European Union; Investissements d'Avenir Labex, Investissements d'Avenir Idex and ANR, France; DFG and AvH Foundation, Germany; Herakleitos, Thales and Aristeia programmes co-financed by EU-ESF and the Greek NSRF, Greece; BSF-NSF and MINERVA, Israel; NCN and NAWA, Poland; La Caixa Banking Foundation, CERCA Programme Generalitat de Catalunya and PROMETEO and GenT Programmes Generalitat Valenciana, Spain; G\"{o}ran Gustafssons Stiftelse, Sweden; The Royal Society and Leverhulme Trust, United Kingdom; United States of America.

In addition, individual members wish to acknowledge support from CERN: European Organization for Nuclear Research (CERN DOCT); Chile: Agencia Nacional de Investigaci\'on y Desarrollo (ANID FONDECYT reg. 1230987, FONDECYT 1230812, FONDECYT 1240864, Fondecyt 3240661, Fondecyt Regular 1240721); China: Chinese Ministry of Science and Technology (MOST-2023YFA1605700, MOST-2023YFA1609300), National Natural Science Foundation of China (NSFC - 12175119, NSFC 12275265); Czech Republic: Czech Science Foundation (GACR - 24-11373S), Ministry of Education Youth and Sports (ERC-CZ-LL2327, FORTE CZ.02.01.01/00/22\_008/0004632), PRIMUS Research Programme (PRIMUS/21/SCI/017); EU: H2020 European Research Council (ERC - 101002463); European Union: European Research Council (BARD No. 101116429, ERC - 948254, ERC 101089007), European Regional Development Fund (HE COFUND GA No.101081355, ERDF), Horizon 2020 Framework Programme (MUCCA - CHIST-ERA-19-XAI-00), European Union, Future Artificial Intelligence Research (FAIR-NextGenerationEU PE00000013), Italian Center for High Performance Computing, Big Data and Quantum Computing (ICSC, NextGenerationEU); France: Agence Nationale de la Recherche (ANR-21-CE31-0013, ANR-21-CE31-0022, ANR-22-EDIR-0002, ANR-24-CE31-0504-01); Germany: Baden-Württemberg Stiftung (BW Stiftung-Postdoc Eliteprogramme), Deutsche Forschungsgemeinschaft (DFG - 469666862, DFG - CR 312/5-2); China: Research Grants Council (GRF); Italy: Istituto Nazionale di Fisica Nucleare (ICSC, NextGenerationEU), Ministero dell'Università e della Ricerca (NextGenEU 153D23001490006 M4C2.1.1, NextGenEU I53D23000820006 M4C2.1.1, NextGenEU I53D23001490006 M4C2.1.1, SOE2024\_0000023); Japan: Japan Society for the Promotion of Science (JSPS KAKENHI JP22H01227, JSPS KAKENHI JP22H04944, JSPS KAKENHI JP22KK0227, JSPS KAKENHI JP24K23939, JSPS KAKENHI JP25H00650, JSPS KAKENHI JP25H01291, JSPS KAKENHI JP25K01023); Norway: Research Council of Norway (RCN-314472); Poland: Ministry of Science and Higher Education (IDUB AGH, POB8, D4 no 9722), Polish National Science Centre (NCN 2021/42/E/ST2/00350, NCN OPUS 2023/51/B/ST2/02507, NCN OPUS nr 2022/47/B/ST2/03059, NCN UMO-2019/34/E/ST2/00393, UMO-2022/47/O/ST2/00148, UMO-2023/49/B/ST2/04085, UMO-2023/51/B/ST2/00920, UMO-2024/53/N/ST2/00869); Portugal: Foundation for Science and Technology (FCT); Spain: Agencia de Gestión de Ayudas Universitarias y de Investigación (AGAUR - 2023 BP 00141), Generalitat Valenciana (ASFAE/2022/008), Ministry of Science and Innovation (MCIN \& NextGenEU PCI2022-135018-2, MICIN \& FEDER PID2021-125273NB, RYC2019-028510-I, RYC2020-030254-I, RYC2021-031273-I, RYC2022-038164-I), Ministerio de Ciencia, Innovación y Universidades/Agencia Estatal de Investigaci\'on (PID2022-142604OB-C22); Sweden: Carl Trygger Foundation (Carl Trygger Foundation CTS 22:2312), Swedish Research Council (Swedish Research Council 2023-04654, VR 2021-03651, VR 2022-03845, VR 2022-04683, VR 2023-03403, VR 2024-05451), Knut and Alice Wallenberg Foundation (KAW 2018.0458, KAW 2022.0358, KAW 2023.0366); Switzerland: Swiss National Science Foundation (SNSF - PCEFP2\_194658); United Kingdom: Leverhulme Trust (Leverhulme Trust RPG-2020-004), Royal Society (NIF-R1-231091); United States of America: U.S. Department of Energy (ECA DE-AC02-76SF00515), Neubauer Family Foundation.

%
%


%
\clearpage
\appendix
\part*{Appendix}
\addcontentsline{toc}{part}{Appendix}
%

\FloatBarrier
\section{NN training configurations}
\label{sec:app_NN}

Table~\ref{tab:NN_config} shows the training configurations of the event-level and track-level NNs used in the displaced track search presented in Section~\ref{sec:selection_DT} and the pNN used in the 1$\ell$1T search presented in Section~\ref{sec:selection_1L1T}. The configuration of the pNN is common to both electron and muon channels.

\begin{table}[htbp]
\centering
\caption{\label{tab:NN_config} Summary of the training configurations of the event-level and track-level NNs for the displaced track (DT) search and the event selection pNN for the 1$\ell$1T search.}
\renewcommand{\arraystretch}{1.1}
\begin{tabular}{lccc}
\toprule
Hyperparameter & Event-level (DT) & Track-level (DT) & pNN (1$\ell$1T)    \\
\midrule
Number of learnable parameters & 2411 & 2611 & 21957 \\
EarlyStopping~\cite{PyTorch_Citation,tensorflow,chollet2015keras} & \multicolumn{3}{c}{Enabled} \\
Initial learning rate  &  \multicolumn{2}{c}{0.01} & 0.0001 \\
Loss type & \multicolumn{3}{c}{Binary cross-entropy} \\
Optimiser & \multicolumn{3}{c}{Adam~\cite{Kingma:2014vow}} \\
\bottomrule
\end{tabular}
\end{table}


\clearpage
\printbibliography
\clearpage
 
\begin{flushleft}
\hypersetup{urlcolor=black}
{\Large The ATLAS Collaboration}

\bigskip

\AtlasOrcid[0000-0002-6665-4934]{G.~Aad}$^\textrm{\scriptsize 104}$,
\AtlasOrcid[0000-0001-7616-1554]{E.~Aakvaag}$^\textrm{\scriptsize 17}$,
\AtlasOrcid[0000-0002-5888-2734]{B.~Abbott}$^\textrm{\scriptsize 123}$,
\AtlasOrcid[0000-0002-0287-5869]{S.~Abdelhameed}$^\textrm{\scriptsize 119a}$,
\AtlasOrcid[0000-0002-1002-1652]{K.~Abeling}$^\textrm{\scriptsize 55}$,
\AtlasOrcid[0000-0001-5763-2760]{N.J.~Abicht}$^\textrm{\scriptsize 49}$,
\AtlasOrcid[0000-0002-8496-9294]{S.H.~Abidi}$^\textrm{\scriptsize 30}$,
\AtlasOrcid[0009-0003-6578-220X]{M.~Aboelela}$^\textrm{\scriptsize 45}$,
\AtlasOrcid[0000-0002-9987-2292]{A.~Aboulhorma}$^\textrm{\scriptsize 36e}$,
\AtlasOrcid[0000-0001-5329-6640]{H.~Abramowicz}$^\textrm{\scriptsize 157}$
% ...
\bigskip
\\

% ...
$^{17}$Department for Physics and Technology, University of Bergen, Bergen; Norway.\\
% ...
$^{30}$Physics Department, Brookhaven National Laboratory, Upton NY; United States of America.\\
% ...
$^{36}$$^{(a)}$Facult\'e des Sciences Ain Chock, Universit\'e Hassan II de Casablanca;$^{(b)}$Facult\'{e} des Sciences, Universit\'{e} Ibn-Tofail, K\'{e}nitra;$^{(c)}$Facult\'e des Sciences Semlalia, Universit\'e Cadi Ayyad, LPHEA-Marrakech;$^{(d)}$LPMR, Facult\'e des Sciences, Universit\'e Mohamed Premier, Oujda;$^{(e)}$Facult\'e des sciences, Universit\'e Mohammed V, Rabat;$^{(f)}$Institute of Applied Physics, Mohammed VI Polytechnic University, Ben Guerir; Morocco.\\
% ...
$^{45}$Physics Department, Southern Methodist University, Dallas TX; United States of America.\\
% ...
$^{49}$Fakult\"{a}t Physik, Technische Universit{\"a}t Dortmund, Dortmund; Germany.\\
% ...
$^{55}$II. Physikalisches Institut, Georg-August-Universit\"{a}t G\"ottingen, G\"ottingen; Germany.\\
% ...
$^{104}$CPPM, Aix-Marseille Universit\'e, CNRS/IN2P3, Marseille; France.\\
% ...
$^{119}$$^{(a)}$New York University Abu Dhabi, Abu Dhabi;$^{(b)}$United Arab Emirates University, Al Ain; United Arab Emirates.\\
% ...
$^{123}$Homer L. Dodge Department of Physics and Astronomy, University of Oklahoma, Norman OK; United States of America.\\
% ...
$^{157}$Raymond and Beverly Sackler School of Physics and Astronomy, Tel Aviv University, Tel Aviv; Israel.\\
% ...
$^{a}$ Also at Affiliated with an institute formerly covered by a cooperation agreement with CERN.\\
$^{b}$ Also at An-Najah National University, Nablus; Palestine.\\
$^{c}$ Also at Borough of Manhattan Community College, City University of New York, New York NY; United States of America.\\
$^{d}$ Also at Center for Interdisciplinary Research and Innovation (CIRI-AUTH), Thessaloniki; Greece.\\
$^{e}$ Also at Centre of Physics of the Universities of Minho and Porto (CF-UM-UP); Portugal.\\
$^{f}$ Also at CERN, Geneva; Switzerland.\\
% ...

\end{flushleft}


%

%
%

%

%
%
%
%

\end{document}